\documentclass[12pt]{article}
\usepackage{amsmath,amsfonts,amssymb,adjustbox, array, amsthm}
\usepackage{xcolor}
\usepackage{filecontents}
\usepackage{algorithm}
\usepackage{algorithmicx}
\floatstyle{plaintop}
\restylefloat{algorithm}
\usepackage{bbm}
\usepackage{etoolbox}
\usepackage{graphicx,psfrag,epsf}
\usepackage{enumerate}
\usepackage{natbib}
\usepackage{url} 
\usepackage{comment}
\usepackage{multirow}
\usepackage{float}
\usepackage{tabularx}

\usepackage{xr-hyper}
\usepackage{hyperref}
\makeatletter
\newcommand*{\addFileDependency}[1]{
\typeout{(#1)}
\@addtofilelist{#1}
\IfFileExists{#1}{}{\typeout{No file #1.}}
}\makeatother

\newcommand*{\myexternaldocument}[1]{%
\externaldocument{#1}%
\addFileDependency{#1.tex}%
\addFileDependency{#1.aux}%
}
\myexternaldocument{SuppPub}
\newcommand{\blind}{0}

\begin{document}

\def\spacingset#1{\renewcommand{\baselinestretch}%
{#1}\small\normalsize} \spacingset{1}

\newcommand{\field}[1]{\mathbb{#1}}
\newcommand{\R}{\field{R}}
\newcommand{\F}{\field{F}}
\newcommand{\G}{\field{G}}
\newcommand{\p}{\field{P}}
\newcommand{\N}{\field{N}}
\newcommand{\Z}{\field{Z}}
\newcommand{\E}{\field{E}}
\newcommand{\C}{\mathcal{C}}
\newcommand{\FF}{\mathcal{F}}
\newcommand{\GG}{\mathcal{G}}
\newcommand{\PP}{\mathcal{P}}
\newcommand{\MS}{\mathcal{S}}
\newcommand{\RR}{\mathcal{R}}
\newcommand{\NN}{\mathcal{N}}
\newcommand{\bv}{\mbox{\boldmath$b$}}
\newcommand{\bw}{\mathbf{w}}
\newcommand{\bx}{\mathbf{x}}
\newcommand{\betav}{\mbox{\boldmath$\beta$}}
\newcommand{\cv}{\mbox{\boldmath$c$}}
\newcommand{\etav}{\mbox{\boldmath$\eta$}}
\newcommand{\thetav}{\mbox{\boldmath$\theta$}}
\def\argmax{\mathop{\mbox{argmax}}}
\def\argmin{\mathop{\mbox{argmin}}}
\newcommand{\Cov}{\text{Cov}}
\def\var{\mbox{Var}}
\def\cov{\mbox{Cov}}
\def\x{{\bf x}}
\def\y{{\bf y}}
\def\z{{\bf z}}
\def\v{{\bf v}}
\def\c{{\bf c}}
\def\W{{\bf W}}
\def\Q{{\bf Q}}
\def\HH{{\bf H}}
\def\ZZ{{\bf Z}}

\newtheorem{theorem}{Theorem}
\newtheorem{lemma}{Lemma}
\newtheorem{corol}{Corollary}
\newtheorem{proposition}{Proposition}
\newtheorem{Conjecture}{Conjecture}

\newtheorem{definition}{Definition}
\newtheorem{condition}{Condition}
\newtheorem{example}{Example}

\newtheorem{remark}{Remark}
\newcommand{\refersize}{\fontsize{11pt}{12pt}\selectfont}

\if0\blind
{
  \title{\bf Self-convolved Bootstrap for M-regression under Complex Temporal Dynamics}
  \author{Miaoshiqi Liu\hspace{.2cm}\\
    Department of Statistical Sciences, University of Toronto\\
    and \\
    Zhou Zhou \thanks{Email: \texttt{zhou.zhou@utoronto.ca}}\hspace{.2cm}\\
    Department of Statistical Sciences, University of Toronto}
  \maketitle
} \fi

\if1\blind
{
  \bigskip
  \bigskip
  \bigskip
  \begin{center}
    {\LARGE\bf Self-convolved Bootstrap for M-regression under Complex Temporal Dynamics}
\end{center}
  \medskip
} \fi

\bigskip
\begin{abstract}
The paper considers simultaneous nonparametric inference for a wide class of M-regression models with time-varying coefficients. The covariates and errors of the regression model are tackled as a general class of 
nonstationary time series and are allowed to be cross-dependent. A novel and easy-to-implement self-convolved bootstrap procedure is proposed. With only one tuning parameter, the bootstrap facilitates a $\sqrt{n}$-consistent inference of the cumulative regression function for the M-estimators under complex temporal dynamics, even under the possible presence of breakpoints in time series. Our methodology leads to a unified framework to conduct general classes of Exact Function Tests, Lack-of-fit Tests, and Qualitative Tests for the time-varying coefficients. These tests enable one to, among many others, conduct variable selection, check for constancy and linearity, as well as verify shape assumptions, including monotonicity and convexity. As applications, our method is utilized to study the time-varying properties of global climate data and Microsoft stock return, respectively.
\end{abstract}

\noindent%
{\it Keywords:} Bahadur representation, cumulative regression function, M estimator, nonstationary time series, quantile regression, shape test.
\vfill

\newpage
\spacingset{1.45} 
\section{Introduction}
\label{sec:intro}
There has been an increasing necessity to account for temporal non-stationarity in finance and economics. For instance, financial returns are frequently non-stationary \citep{francq2022volatility}. 
In climate economics, climate change has become progressively unpredictable with extreme weather happening globally with greater intensity \citep{ebi2021extreme}. See also \cite{dette2011measure, MR3097614, GN,kreiss2015bootstrapping,dahlhaus2019towards} among others for the recent literature on non-stationary time series analysis. As deviations from temporal homogeneity become more pronounced, static models may fail to capture the dynamic features of the data. In the context of regression analysis, one successful generalization is to use the varying-coefficient model \citep{hastie1993varying}, which is capable of describing the non-constant relationship between the predictors and the response \citep{hoover1998nonparametric, fan2000simultaneous, KARMAKAR2022408}. In the present work, we shall further consider the following time-varying M-regression \citep{Huber_1964, Huber_1973} model for non-stationary time series:
\begin{eqnarray}
y_i=\x_i^{\top}\betav_i+e_i, \quad i=1,2,\cdots,n,
\label{eq:tv-model}
\end{eqnarray}
where $\{\x_i=(x_{i1},\cdots,x_{ip})^\top\}_{i=1}^n$ is a $p$-dimensional predictor process,  $\{e_i\}_{i=1}^n$ is the error process satisfying
$\E(\psi(e_i)|\x_i,\x_{i-1},\cdots,\x_1)=0$ almost surely with $\psi(\cdot)$ being the left derivative of a convex function $\rho(\cdot)$
, and $\betav_i = \betav(t_i)=$  $(\beta_0(t_i), \cdots, \beta_{p-1}(t_i))^\top$ is the coefficient $\betav(t)$ evaluated at time $t_i = i/n$.
We study model \eqref{eq:tv-model} in a general framework where $\{\x_i\}$ and $\{e_i\}$ belong to a general class of non-stationary time series models with both smoothly and abruptly time-varying dynamics. The detailed assumptions of the model are deferred to Section \ref{sec:aux}. Observe that \eqref{eq:tv-model} includes a wide range of regression models with flexible choices of $\rho(\cdot)$. Prominent examples include the least square regression with $\rho(x)=x^2$, the quantile regression with $\rho(x)=\tau x^++(1-\tau)(-x)^+$, $\tau\in(0,1)$, $x^+=\max(x,0)$, the $L^q$ regression with $\rho(x)=|x|^q$, $q>0$, the expectile regression with $\rho(x)=|I(x\le 0)-\alpha|x^2$, $\alpha\in(0,1)$, and Huber's robust regression with $\rho(x)=x^2I(|x|\le \varsigma)/2+(\varsigma|x|-\varsigma^2/2)I(|x|>\varsigma)$, $\varsigma>0$. 

This article aims to provide a unified framework for nonparametric inference of 
model \eqref{eq:tv-model}. 
Due to the complex temporal dynamics and the broad choice of loss functions, it is difficult to directly estimate the limiting distributions of the M-estimators, in which case resampling methods such as the bootstrap is typically adopted to circumvent the difficulty. 
In the context of nonparametric M-regression, however, so far there exist no valid bootstrap methods for its simultaneous inference when the regressors and errors are non-stationary time series with possible jumps in the underlying data generating mechanism. The most related work can be found in \citet{wu2017nonparametric}, who explored time-varying coefficient quantile regression with smoothly time-varying dynamics. We also refer to \citet{FRIEDRICH2022105345} who proposed an autoregressive sieve bootstrap framework, valid under $L_2$ regression with strictly stationary $\alpha$-mixing predictors. Extending to our current setup, the primary challenge in designing bootstrap methods originates from the inconsistency in estimating various key quantities around the jump points. Additionally, bootstrap inference for M-regression involves estimating a density-like function, which requires the delicate choice of a tuning parameter \citep{koenker2005quantile}. For instance, in quantile regression where $\rho(x)=\tau x^++(1-\tau)(-x)^+$, the asymptotic behavior of local estimates of $\betav(\cdot)$ is tied to a density-type quantity $\Sigma(\cdot)$ \citep{wu2017nonparametric}. 
In moderate samples, the quality of the bootstrap inference depends on the estimation accuracy of $\Sigma(\cdot)$, which substantially entails the nontrivial task of selecting the tuning parameter. Lastly, the existing literature on bootstrap inference of nonparametric M-regression, such as \citet{FRIEDRICH2022105345} and \citet{wu2017nonparametric}, often constructs the bootstrap based on residuals. In moderate samples, the performance of such residual-based bootstrap can be significantly hampered by the estimation error of the residuals as estimators of the true regression errors. 

The methodological innovation of this paper lies in proposing a novel self-convolved bootstrap for the simultaneous nonparametric inference of Model \eqref{eq:tv-model}, which shares a simple and unified construction among a broad class of M-estimators and hypothesis tests, and is consistent under very general forms of non-stationary temporal dependence and predictor-error dependence. As its name implies, the self-convolved bootstrap only necessitates the convolution of the M-estimators with i.i.d. auxiliary standard normal variables, eliminating the need for additional computation. Particularly, it avoids estimating the aforementioned density-like quantities or residuals. The construction principle applies uniformly to a wide class of M-estimators and therefore saves lots of effort when conducting inference based on various estimators simultaneously. The validity of the proposed bootstrap arises from two main observations. Firstly, by taking the second-order difference of the local M-estimators as shown in Section \ref{sec:rb}, their progressive sum can be well approximated by a weighted block sum of the data through a uniform Bahadur representation. Secondly, the weighted block sum convolved by i.i.d. auxiliary standard normal variables takes the form of a robust multiplier bootstrap \citep{zhou2013heteroscedasticity}, which can consistently mimic the probabilistic behavior of their partial sum processes.
We will show that the easy-to-implement bootstrap simulates the probabilistic behavior of the M-estimators consistently under complex temporal dynamics. Compared to the bootstrap methods developed in \citet{KARMAKAR2022408}, \citet{wu2017nonparametric} and \citet{FRIEDRICH2022105345}, the self-convolved bootstrap enjoys broader applicability yet maintains a simple implementation.

Facilitating the self-convolved bootstrap, our framework utilizes the cumulative regression function (CRF) 
\begin{eqnarray}\label{eq:crf}
\Lambda(t)=(\Lambda_1(t),\cdots,\Lambda_p(t))^\top=\int_0^t \betav(s)\,ds,\quad 0\le t\le 1,
\end{eqnarray} instead of $\betav(t)$ for inference, driven by two main reasons. First, as is suggested in Theorem \ref{thm:2}, the CRF can be estimated at a parametric rate of $\sqrt{n}$. Such merits of getting $\sqrt{n}$ convergence via integration have been demonstrated by \citet{hall1987estimation}, \citet{bickel1988estimating}, \cite{huang1999nonparametric} and many others. Recently, \citet{mies2021functional} employed the method to estimate parameters of locally stationary time series, leading to a $\sqrt{n}$ functional central limit theorem. \citet{CAI20231447} also obtained a similar result regarding rolling regression estimators, where the error process is assumed to be a martingale difference. Although the results in the existing literature have gradually adapted from the i.i.d. case to a broader non-stationary time series setup, they generally rely on a closed-form statistic, e.g., the least square estimator, and therefore do not apply to the M-regression setting. To our knowledge, no comparable research has been conducted on uniform nonparametric inference of the CRF under the time-varying M-regression framework. The second reason, possibly more salient, pertains to the complexity of the temporal dynamics considered. Specifically, it is difficult to consistently estimate the (asymptotic) distributional behavior of the M-estimators around the time points where $\x_i$ and/or $e_i$ experience abrupt changes. As a result, directly inferring the M-estimators of $\betav(\cdot)$ simultaneously over time becomes challenging. Nevertheless, we discover that aggregated M-estimators of the CRF converge to a Gaussian process uniformly over time under mild conditions, which can be effectively approximated by the self-convolved bootstrap mentioned earlier. Consequently, it is simpler and more accurate to make simultaneous inferences of $\Lambda(\cdot)$ than $\betav(\cdot)$ under complex temporal dynamics. As far as we know, no previous literature has used the method of integration or aggregation to alleviate the effect of complex dependence and heteroscedasticity in regression analysis. 

The self-convolved bootstrap facilitated by the CRF addresses different types of tests using a unified asymptotic result, which ventures into a new avenue compared to the conventional methods. To elaborate, we consider three types of hypothesis tests: Exact Function Tests, Lack-of-fit Tests, and Qualitative Tests. The first type, \textit{Exact Function Test (EFT)}, takes the form of $H_0: \betav(t) = \betav_0(t)$ for some specific $\betav_0(t)$, which, for example, enables one to check the significance of the variables by letting $\betav_0(t) = \mathbf{0}$. Secondly, the \textit{Lack-of-fit Test (LOFT)}, also referred to as diagnostic test \citep{stute1997nonparametric}, tackles the circumstance where the null hypothesis is a parametric family while the alternative remains nonparametric \citep{He_Zhu_2003}. This includes, but is not limited to, the tests of constancy and linearity of the function versus a general smooth alternative. An example in the realm of econometrics is \citep{Chen_Hong_2012}, where the LOFT is used to tackle the long-standing problem of detecting structural changes in economic relationships. Apart from the aforementioned two types of tests, one may also want to check qualitative beliefs on functions where the null hypothesis is a relatively general nonparametric class of functions \citep{komarova2019testing}. For example, many economic models use certain shape restrictions (e.g., monotonicity, convexity) as plausible restrictions \citep{chetverikov_2019,fang2021projection}. While those prespecified qualitative shape assumptions facilitate the inference and enhance performance \citep{friedman1984monotone, MATZKIN19942523}, wrong conclusions can be drawn when these assumptions are not satisfied. The necessity to conduct \textit{Qualitative Tests (QT)}, therefore, becomes axiomatic. 

Historically in the context of regression, much literature has focused on the Lack-of-fit Test on the regression mean function, serving as a key component in model checking. One popular approach is to estimate the parametric and nonparametric curves separately and compute their discrepancies as a natural test statistic, as seen in \citet{hardle1993comparing} and \cite{hong1995consistent}. Similarly, \citet{Chen_Hong_2012} proposed a generalized Hausman test for checking parameter stability in time series models, applicable to stationary data under a $L_2$ framework. Alternatively, another distinguished direction is to study the empirical process based on residuals; see \citet{stute1998model,koul1999nonparametric}. Recently, \cite{mies2021functional} utilized the integration technique for change-point detection of local parameters. In comparison, our current work is based on a new self-convolved bootstrap technique and applies to a broader range of tests and a more general M-regression setting. Different from the LOFT where the null hypothesis forms a parametric family, qualitative hypotheses often include shape constraints that are more complicated to deal with. To test the monotonicity of a regression curve, \cite{bowman1998testing} exploited \cite{silverman1981using}'s idea of critical bandwidth. \cite{gijbels2000tests} and \cite{ghosal2000testing} formulated monotonicity of the regression curve as the concordance of $X$ and $Y$. As a distinctive approach, \cite{durot2003kolmogorov} transformed the test of monotonicity into testing whether the integral coincides with its least concave majorant. A localized version of the test was later elaborated by \cite{akakpo2014testing}. 
Other contributions in nonparametric qualitative tests include \citet{hall2000testing,gijbels2017shape,komarova2019testing,fang2021projection}, among others.

The rest of the paper is structured as follows. Section \ref{sec:pre} introduces the piece-wise locally stationary time series, dependence measures, and local linear M-estimators. The estimator of the CRF is proposed, with its asymptotic results stated in Theorem \ref{thm:2}. Section \ref{sec:rb} presents the self-convolved bootstrap. In Section \ref{sec:test}, three types of hypothesis tests are discussed, where detailed algorithms and their asymptotic behaviors are provided. The finite-sample performance of our method is demonstrated in Section \ref{sec:simu}, followed by empirical illustrations using global climate data and Microsoft stock return in Section \ref{sec:data}. Regularity conditions of the model and auxiliary theoretical results about Bahadur representation are given in Section \ref{sec:aux}. Detailed proofs
are provided in the supplementary material.

\section{Preliminary}\label{sec:pre}
We start by introducing some notations. For an $m$-dimensional (random) vector $\boldsymbol{v} = (v_1, \cdots, v_m), m \ge 1$, let $|v|$ be the Euclidean norm and $|v|_{\infty} :=\max_{i = 1}^m |v_i|$. Let $I\{\cdot\}$ be the indicator function and we denote the weak convergence by $\Rightarrow$.
\subsection{Non-stationary time series models}
The following piece-wise locally stationary (PLS) time series models \citep{zhou2013heteroscedasticity} are used to describe the complex temporal dynamics of the covariates and errors. For a sequence of i.i.d. random variables $\{\eta_i\}_{i = -\infty}^{\infty}$, let $\{\eta'_i\}_{i = -\infty}^{\infty}$ be an i.i.d. copy of $\{\eta_i\}_{i = -\infty}^{\infty}$. Define $\FF_i=(\cdots,\eta_{i-1},\eta_{i})$, and $\FF_i^* = ({\cal F}_{-1},\eta_0',\eta_1,\cdots,\eta_i)$. 
\begin{definition}
We say that $\{\ZZ_i\}$ is a $d$-dimensional piece-wise locally stationary time series with $r$ break points and filtration $\{\FF_i\}$ if
\begin{eqnarray}\label{eq:pls}
\ZZ_i=\sum_{j=0}^r G_j(i/n,\FF_i)I\{i/n\in(w_j,w_{j+1}]\}, \quad i=1,2,\cdots,n,
\end{eqnarray}
where $0=w_0<w_1<\cdots<w_r<w_{r+1}=1$, 
and $G_j :$ $[w_j,w_{j+1}]\times \R^\field{Z}\rightarrow \R^{d}$, $j=0,1,\cdots,r$ are possible nonlinear filters.
\end{definition}
The breakpoints $w_1,\cdots, w_r$ are assumed to be fixed but unknown, and the number of breaks $r$ is assumed to be bounded. The PLS process can capture a broad class non-stationary behavior in practice because it allows the underlying data-generating mechanism to evolve smoothly between breakpoints (provided that the filters $G_j(t,\FF_i)$ are smooth in $t$) and permits the latter mechanism to change abruptly from $G_{j-1}(\cdot)$ to $G_j(\cdot)$ at the breakpoint $w_j$. Let $\zeta_i$ be the index $j$ such that $i/n \in(w_j,w_{j+1}]$. Then \eqref{eq:pls} is equivalent to $\ZZ_i=G_{\zeta_i}(i/n,\FF_i)$. For $q>0$ and $i \ge 0$, we define the dependence measures
\begin{eqnarray}\label{eq:dep_mea}
\delta_{\ZZ}(i, q)
 = \max_{0\le j\le r }\sup_{w_j\le t\le w_{j+1}} \|G_j({t,\cal F}_i) - G_j(t,\FF_i^*) \|_q,
\end{eqnarray}
where $\|\cdot\|_q=\{\E[|\cdot|^q]\}^{1/q}$ is the ${\cal L}^q$ norm. Write $\|\cdot\|:=\|\cdot\|_2$. Let $\delta_{\ZZ}(i,q)=0$ if $i<0$. Intuitively, $\delta_{\ZZ}(i, q)$ measures uniformly the changes in a dynamic system's output when the innovations of the system $i$-steps ahead are replaced with i.i.d. copies. Therefore the decay speed of $\delta_{\ZZ}(i, q)$ as a function of $i$ measures the strength of the system's temporal dependence. We refer the readers to \citep{zhou2013heteroscedasticity} for more discussions and examples of the PLS time series models and the associated dependence measures.

Here we consider filtrations $\FF_i=(\cdots,\eta_{i-1},\eta_{i})$ and $\GG_i = (\cdots,\tau_{i-1},\tau_{i})$, where $\{\eta_i\}_{i = -\infty}^{\infty}$ and $\{\tau_i\}_{i = -\infty}^{\infty}$ are independent. Assume piece-wise locally stationary time series $\x_i = \HH_{\zeta_i}(t_i, \GG_i)$, $e_i = G_{\zeta_i}(t_i, \FF_i, \GG_i)$. Observe that $\x_i$ and $e_i$ can be dependent as the innovations $\GG_i$ which generate $\x_i$ are also involved in $e_i$'s generating process. On the other hand, the factors influencing $e_i$ but are unrelated to $\x_i$ are captured in $\FF_i$. Without loss of generality, assume $\{\x_i\}$ and $\{e_i\}$ share the same $r$ breakpoints  $0=w_0<\cdots<w_r<w_{r+1}=1$. 

\subsection{Aggregated local estimation of the CRF}
Throughout this paper, we assume that the regression function $\betav(t) \in {\cal C}^2[0,1]$.
For any $t\in (0,1)$, since $\betav(s) \approx \betav(t) + \betav^{\prime}(t)(s-t) $ in a small neighborhood of $t$, we define the preliminary local linear M-estimates of $\betav(t)$ and $\betav^{\prime}(t)$ by 
\begin{eqnarray}\label{eq:local_linear}
\tilde{\betav}_{b_n}(t) := \left(\begin{array}{c}\hat{\betav}_{b_{n}}(t) \\ \hat{\betav}_{b_{n}}^{\prime}(t)\end{array}\right)=\argmin_{\bv_0, \bv_1\in\R^p}\sum_{i=1}^n\rho(y_i-\x_i^\top\bv_0 - \x_i^\top\bv_1(t_i-t)) K((t_i-t)/b_n),
\end{eqnarray}
where $t_i=i/n$, $b_n$ is a bandwidth satisfying $b_n\rightarrow 0$, and $\rho(\cdot)$ is a convex loss function with left derivative $\psi(\cdot)$. In this paper, we also assume robustness of the loss functions in the sense that $|\psi(x) - \psi(y)| \le M_1 + M_2|x - y|$ for $\forall x, y \in \R$ and some positive constants $M_1, M_2$. $K(\cdot)$ is a kernel function satisfying $K\in{\cal K}$ and ${\cal K}$ is collection of symmetric and ${\cal C}^1$ density functions with support $[-1,1]$.

Recall the CRF defined in \eqref{eq:crf}, then the regression function $\betav(\cdot)$ can be retrieved via 
\begin{eqnarray*}
\betav(t)=\Lambda'(t-)\quad 0<t\le 1 \quad\mbox{and}\quad \betav(0)=\Lambda'(0+),
\end{eqnarray*}
where $\Lambda'(t-)$ and $\Lambda'(t+)$ represent the left and right derivative of $\Lambda(t)$, respectively. 

As is discussed in Section \ref{sec:aux}, the M-estimator $\hat{\betav}_{b_{n}}(t)$ obtained in \eqref{eq:local_linear} involves a bias term regarding $\betav^{\prime\prime}(t)$. In order to proceed without estimating $\betav^{\prime\prime}(t)$, we shall employ a Jackknife bias-corrected estimator denoted as $\check{\betav}_{b_n}(t)$, where
\noindent
\begin{eqnarray}\label{eq:jack}
	\check{\betav}_{b_n}(t) := 2\hat{\betav}_{b_n/\sqrt{2}}(t) - \hat{\betav}_{b_n}(t).
\end{eqnarray}
The implementation of the bias-correction procedure is asymptotically equivalent to using the second-order kernel $K^*(x):= 2\sqrt{2}K(\sqrt{2}x)-K(x)$. In this way, the bias of $\check{\betav}_{b_n}(t)$ is asymptotically negligible under mild conditions. Based on the Jackknife bias-corrected estimator, we propose to estimate $\Lambda(t)$ via 
\begin{eqnarray}\label{eq:hat_crf}
\check{\Lambda}(t_j):=\sum_{i=1}^{j}\check{\betav}_{b_n}(t_i)/n,\quad 1\le j\le n,
\end{eqnarray}
and $\check{\Lambda}(0):=0$. For any $t\in[0,1]$, let $\check{\Lambda}(t)$ be the linear interpolation of the sequence $\{\check{\Lambda}(t_j)\}_{j=0}^{n}$. We omit the subscript $b_n$ of $\check{\betav}_{b_n}(t)$ hereafter, whenever no confusion caused.

Let $\mathbf{C}$ be a fixed $s\times p$, $s \le p$ full rank matrix, we are interested in testing the dynamic pattern of $\betav_{\mathbf{C}}(\cdot):=\mathbf{C} \betav(\cdot)$, which includes any linear combination of $\{\betav_j(\cdot)\}_{j=0}^{p-1}$. As a result, for the CRF ${\Lambda}_{\mathbf{C}}(t) := \mathbf{C}{\Lambda}(t)$, its estimator can be obtained via $\check{\Lambda}_{\mathbf{C}}(t) := \mathbf{C}\check{\Lambda}(t)$. Based on the Bahadur representation presented in Section \ref{sec:aux}, the following theorem establishes a $\sqrt{n}$-consistent Gaussian weak convergence result for the proposed aggregated M-estimator:
\begin{theorem}\label{thm:2}
Suppose assumptions (A1)-(A7) in Section \ref{sec:aux} hold, $\frac{nb_n ^4}{\log^8 n} \to \infty$, $n^c b_n \to 0$ for some positive constant $c$, and let $\Psi_{\mathbf{C}}(t) = (\mathbf{C}\Sigma^{-1}(t)\Psi(t)\Sigma^{-1}(t)\mathbf{C}^\top)^{\frac{1}{2}}$ with $\Sigma(t)$ and $\Psi(t)$ defined in Section \ref{sec:aux}. 
Then we have
\begin{eqnarray*}\label{eq: limiting_distribution}
	\left\{\sqrt{n}(\check{\Lambda}_{\mathbf{C}}(t) - \Lambda_{\mathbf{C}}(t))\right\}_{0<t<1} \Rightarrow\left\{\mathbf{U}(t)\right\}_{0<t< 1},
\end{eqnarray*}
where $\left\{\mathbf{U}(t)\right\}$ is a mean 0 Gaussian process with $\Cov(\mathbf{U}(t_1), \mathbf{U}(t_2)) = \int_{0}^{\min(t_1,t_2)}\Psi_{\mathbf{C}}(t)\Psi_{\mathbf{C}}^{\top}(t) \; dt$.
\end{theorem} 

\section{The Self-convolved Bootstrap}\label{sec:rb}
The Gaussian process $\mathbf{U}(t)$ expressed in Theorem \ref{thm:2} provides a theoretical foundation for the simultaneous inference of $\Lambda(t)$, but the implementation is still unaccomplished due to its complex covariance structure. To circumvent the problem, we propose a self-convolved bootstrap method to mimic the behavior of the Gaussian process $\mathbf{U}(t)$. 

Let $\mu = \int_{-2}^{2}[K^*(t-1)+K^*(t+1)-2K^*(t)]^2 \; \mathrm{d}t$, $\{R_i\}_{i = 1}^n$ be i.i.d. standard normal random variables independent of $\{(\mathbf{x}_i, y_i)\}_{i = 1}^n$. For a given bandwidth $c_n$, define the bootstrap process $\{\tilde{\boldsymbol{\Phi}}_{n}^{o}(t), t \in (0, 1)\}$ as the linear interpolation of $\{\boldsymbol{\Phi}_{n,j}^{o}\}$, where
$$
\boldsymbol{\Phi}_{n,j}^{o} = \mathbf{C}\sum_{i = \left\lceil 2 n c_n \right \rceil}^{j} \sqrt{\frac{c_n}{\mu}}(\check{\betav}_{c_n}(t_i + c_n) + \check{\betav}_{c_n}(t_i - c_n)-2\check{\betav}_{c_n}(t_i))R_i,\quad j =  \left\lceil 2 n c_n \right \rceil,  \cdots, n -  \left\lceil 2 n c_n \right \rceil.
$$
    
The self-convolved bootstrap $\tilde{\boldsymbol{\Phi}}_n^o(t)$ only requires the convolution of local M-regression estimates with i.i.d. standard normal random variables $\{R_i\}_{i=1}^n$. The heuristics stem from two fundamental observations. Firstly, the Bahadur representation of $\{\check{\betav}(t_i)\}$ in Section \ref{sec:aux} suggests that $\check{\betav}_{c_n}(t_i + c_n) + \check{\betav}_{c_n}(t_i - c_n)-2\check{\betav}_{c_n}(t_i)$ can be expressed as a weighted block sum of $\{\mathbf{x}_i\psi(e_i)\}_{i = 1}^n$, where we take the second-order difference as a bias-correction technique. Secondly, \citet{zhou2013heteroscedasticity} demonstrated that for a broad range of non-stationary time series, progressive convolutions of their block sums with i.i.d. standard normal random variables can consistently mimic the joint probabilistic behavior of their partial sum processes. Observe that $\mathbf{U}(t)$ represents the limiting behavior of a weighted partial sum process of $\{\mathbf{x}_i\psi(e_i)\}_{i = 1}^n$. By leveraging these observations, the self-convolved bootstrap achieves an accurate simulation of the limiting Gaussian process of the CRF and retains a concise form for a broad range of M-estimators, utilizing just a single tuning parameter.

In contrast, it is difficult to make inference of $\betav(t)$ directly based on $\check{\betav}_{b_n}(t)$, as the density-like quantity involved in the distributional behavior of $\check{\betav}_{b_n}(t)$ is hard to estimate in practice due to possible discontinuities and the no-trivial task of choosing smoothing parameters. As a result, the CRF considered in this paper, combined with the proposed self-convolved bootstrap, enables us to conduct inference more efficiently and accurately.

To quantify the consistency of $\{\tilde{\boldsymbol{\Phi}}_{n}^{o}(t), t \in (0, 1)\}$, we derive a direct comparison of the distribution between $\{ \boldsymbol{\Phi}_{n,j}^{o}\}_{j = \lceil 2nc_n \rceil}^{n-\lceil 2nc_n \rceil}$ and $\{\mathbf{U}(t_j)\}{_{j = \lceil 2nc_n \rceil}^{n-\lceil 2nc_n \rceil}}$ in the following theorem. 
First, we define the $ns$-dimensional vector $\mathbf{\Theta}^{B} = [(\mathbf{\Theta}_1^{B})^{\top}, (\mathbf{\Theta}_2^{B})^{\top}, \cdots,(\mathbf{\Theta}_n^{B})^{\top}]$ with
$$
    \mathbf{\Theta}_j^{B} : = \left\{\begin{array}{ll} \boldsymbol{\Phi}_{n, \left\lceil 2 n c_n \right \rceil}^{o} & \text { if } j \leq  \left\lceil 2 n c_n \right \rceil \\ \boldsymbol{\Phi}_{n,j}^{o} & \text { if }  \left\lceil 2 n c_n \right \rceil<j \leq n- \left\lceil 2 n c_n \right \rceil \\ \boldsymbol{\Phi}_{n,n- \left\lceil 2 n c_n \right \rceil}^{o}  & \text { if } n- \left\lceil 2 n c_n \right \rceil < j \leq n \end{array}\right. ,
    $$
    and similarly define the $ns$-dimensional vector $\mathbf{U} = [(\mathbf{U}_1)^{\top}, (\mathbf{U}_2)^{\top}, \cdots,(\mathbf{U}_n)^{\top}]$ with
$$
    \mathbf{U}_i := \left\{\begin{array}{ll} \mathbf{U}(t_{ \left\lceil 2 n c_n \right \rceil}) & \text { if } i \leq  \left\lceil 2 n c_n \right \rceil \\ \mathbf{U}(t_i) & \text { if }  \left\lceil 2 n c_n \right \rceil <i \leq n- \left\lceil 2 n c_n \right \rceil \\ \mathbf{U}(t_{n- \left\lceil 2 n c_n \right \rceil})  & \text { if } n- \left\lceil 2 n c_n \right \rceil < i \leq n. \end{array}\right. 
    $$  
Recall $\mathbf{U}(t)$ is defined in Theorem \ref{thm:2}.
Furthermore, we define 
\[\Delta:=\max _{1 \leq i, j \leq  ns}\left|\left[\operatorname{Cov}\left(\mathbf{U}\right)-\operatorname{Cov}(\mathbf{\Theta}^{B} \mid \{(\mathbf{x}_k,y_k)\}_{k = 1}^n)\right]_{i j}\right| \]
as a measure of the difference in covariance structure between the bootstrapped process and the limiting Gaussian process.

\begin{theorem}\label{thm:3}
Suppose assumptions (A1)-(A7) in Section \ref{sec:aux} hold, $\frac{nc_n ^4}{\log^8 n} \to \infty$, and {$nc_n^5 \to 0$}, then we have $\Delta = O_{\mathbb{P}}(\sqrt{c_n}+nc_n^5)$. 
Define the sequence of events $A_n = \{\Delta \le (\sqrt{c_n}+nc_n^5)h_n\}$ where $h_n > 0$ is a sequence diverging at an arbitrarily slow rate. Then $\mathbb{P}(A_n) = 1-o(1)$. On the event $A_n$, for any $\delta \in (0,1)$, we have
\begin{equation}\label{eq:comparison}
\begin{aligned}
     &\sup _{|x|>d_{n, p}^{\circ}}\left|\mathbb{P}\left(\max_{\lceil 2nc_n \rceil \le i \le n - \lceil 2nc_n \rceil}|\mathbf{U}(t_i)|_\infty \leq x\right)-\mathbb{P}\left(\max_{\lceil 2nc_n \rceil \le i \le n - \lceil 2nc_n \rceil}|\boldsymbol{\Phi}_{n,i}^{o}|_\infty \leq x \mid \{(\mathbf{x}_i,y_i)\}_{i = 1}^n \right)\right| \\  \lesssim & \left((\sqrt{c_n}+nc_n^5)h_n\right)^{1 / 3} (\log(n))^{1+4\delta /3} + \dfrac{1}{n(\log(n))^{1/2}}, 
\end{aligned}
\end{equation}
where
$d_{n, p}^{\circ}=C\left[\left((\sqrt{c_n}+nc_n^5)h_n\right)^{1 / 3} \log ^{\delta / 3}(n)+\log ^{-\delta}(n)\right]$ with some finite constant $C$ that does not depend on $n$. 
\end{theorem}

\begin{remark}\label{remark: intutition_of_bootstrap} 
Theorem \ref{thm:3} guarantees that the conditional behavior of the bootstrap process consistently mimics the distribution of the limiting process ${\bf U}(t)$ in ${\cal L}_\infty$ norm. The restricted range $|x|>d_{n, p}^{\circ}$ is to make the Gaussian approximation valid under the circumstance that the variances of $\boldsymbol{\Phi}_{n,i}^{o}$ have no lower bounds. Note that $d_{n, p}^{\circ}$ and the upper bound in \eqref{eq:comparison} converge to $0$ as $n \to \infty$. Accordingly, for a given $\alpha$ and when $n$ is sufficiently large, $d_{n, p}^{\circ}$ will be dominated by the $(1-\alpha)$-th percentile of $\max_{\lceil 2nc_n \rceil \le i \le n - \lceil 2nc_n \rceil}|\boldsymbol{\Phi}_{n,i}^{o}|_\infty$, thereby ensuring that the results in Theorem \ref{thm:3} asymptotically validate the bootstrap procedure. Furthermore, under the conditions of Theorem \ref{thm:3}, $c_n$ converges to $0$, hence the consistency of the bootstrap $\tilde{\boldsymbol{\Phi}}_{n}^{o}(t)$ demonstrated on $[2c_n, 1-2c_n]$ is asymptotically uniform on $t \in (0,1)$. 
\end{remark}

\section{Applications to Hypothesis Testing}\label{sec:test}
In principle, hypotheses regarding the regression function $\betav(t)$ can be expressed equivalently in terms of the CRF $\Lambda(t)$. In this Section, we shall explore the application of the CRF and the self-convolved bootstrap to three general classes of hypothesis tests of the M-regression: \textit{Exact Function Test (EFT)}, \textit{Lack-of-fit Test (LOFT)}, and \textit{Qualitative Test (QT)}.

\subsection{Exact Function Test}\label{sec:EFT}
 The Exact Function Test amounts to testing $H_0: \betav_{\mathbf{C}}(t) = f(t), t \in (0,1)$, where $f(
 \cdot)$ is a known function. It can be reformulated as $H_0: \Lambda_{\mathbf{C}}(t) = \int_{0}^t f(s) \;ds$. Define
 \begin{eqnarray}\label{eq:t_EFT}
	T_e := \sup_{t\in(0,1)}\sqrt{n}\left|\check{\Lambda}_{\mathbf{C}}(t) - \int_{0}^t f(s) \;ds\right|_{\infty}.
\end{eqnarray}
For any prespecified level $\alpha$, our goal is to find the critical value $q_{\alpha}$, such that $\mathbb{P}\left(T_e > q_\alpha \mid H_0\right) = \alpha$ asymptotically. The detailed procedures for conducting EFT are given in Algorithm \ref{alg:EFT}. Propositions \ref{prop: typeIerror_LOFT} and \ref{prop: power_LOFT} in Section \ref{sec:LOFT} validate the latter algorithm theoretically.

\begin{algorithm}
\floatname{algorithm}{\bf Algorithm}
\caption{Exact Function Test}
\vspace{4pt}
\hrule
\vspace{4pt}
\label{alg:EFT}
\begin{algorithmic}[1]
\State Select appropriate bandwidths $b_n$, $c_n$ according to the procedures in section \ref{sec:bw}, define $i_* = \max(\lceil nb_n \rceil, \lceil 2nc_n \rceil)$ and $i^* = n - i_*$.

\State Obtain the Jackknife bias-corrected M-estimators $\check{\betav}_{b_n}(t_i)$, $\check{\betav}_{c_n}(t_i)$, $\check{\betav}_{c_n}(t_i -c_n)$ and $\check{\betav}_{c_n}(t_i + c_n)$ for $i = 1, 2, \cdots, n$, compute $\{\check{\Lambda}_{\mathbf{C}}(t_j)\}_{ j =1}^n$ based on $\{\check{\betav}_{b_n}(t_i)\}_{i = 1}^n$.

\State Generate $B$ sets of i.i.d. standard normal random variables $\{u_i^{(r)}\}_{i=1}^{n}$, $r=1,2,\cdots,B$. For each $r$, calculate $\{\mathbf{\Phi}_{n,j}^{o}\}_{j = 1}^n$ as $\{\mathbf{\Phi}_{n,j}^{(r)}\}_{j = 1}^n$ by letting $R_i = u_i^{(r)}$ in Theorem \ref{thm:3}.

\State Calculate $M_r =\max_{i_* \le j \le i^*} |\boldsymbol{\Phi}_{n,j}^{(r)}|_\infty$, $r=1,2,\cdots,B$.

\State For a given $\alpha \in (0,1)$, find the $(1-\alpha)$-th sample quantile of $\{M_r\}_{r=1}^B$, $\hat{q}_{n,1-\alpha}$. 

\State For the Exact Function Test where $H_0: \betav_{\mathbf{C}}(t) = f(t), \quad t \in (0,1)$, reject the null hypothesis if $\max_{i_* \le j \le i^*}|\check{\Lambda}_{\mathbf{C}}(t_j) - \int_{0}^{t_j} f(s)\; ds|_\infty >  \hat{q}_{n,1-\alpha}/\sqrt{n}$. 
\end{algorithmic}
\vspace{4pt}
\hrule
\end{algorithm}    

\subsection{Lack-of-fit Test}\label{sec:LOFT}
The Lack-of-fit Test investigates whether $\Lambda_{\mathbf{C}}(t)$ belongs to a given parametric family; i.e., $\Lambda_{\mathbf{C}}(t) = g(t, \boldsymbol{\theta}), t \in (0,1)$ for a given family of functions $g(t,\cdot)$ with unknown $\boldsymbol{\theta}$. Under $H_0$, oftentimes the unknown $\boldsymbol{\theta}$ can be expressed (solved) in terms of $\Lambda_{\mathbf{C}}(v_i)$ for some given points $v_1, v_2,\cdots,v_k$ $\in[0,1]$. See Section \ref{subsec:pt} for a detailed example of testing whether $\boldsymbol{\beta}_{\mathbf{C}}(t)$ is a polynomial of a given degree. 
In this case, the null hypothesis is written as 
\begin{eqnarray}\label{eq:LOFT}
    \Lambda_{\mathbf{C}}(t) - f(t, \{\Lambda_{\mathbf{C}}(v_i)\}_{i = 1}^k) = \mathbf{0}, \quad t \in (0,1),
\end{eqnarray}
where $k$ is a finite integer. Assumptions on $f(\cdot)$ are discussed in Theorem \ref{thm:test}. Consider 
\begin{eqnarray}
    T_l := \sup_{t\in(0,1)}\sqrt{n}\left|\check{\Lambda}_{\mathbf{C}}(t) - f(t, \{\check{\Lambda}_{\mathbf{C}}(v_i)\}_{i =1}^k)\right|_{\infty},
\end{eqnarray}
then the critical value can be obtained as shown in Algorithm \ref{alg:LOFT}.

\subsubsection{Polynomial Test}\label{subsec:pt}
The polynomial test aims to verify whether the coefficients are $q$-th order polynomials for a fixed $q$. For instance, $q = 0$ allows one to test if  $\betav_{\mathbf{C}}(t)=\boldsymbol{\theta}_0$ for some unknown $\boldsymbol{\theta}_0$.  Therefore under $H_0$, $\Lambda_{\mathbf{C}}(t)=\boldsymbol{\theta}_0t$ and $\boldsymbol{\theta}_0=\Lambda_{\mathbf{C}}(1)$ by setting $v_1=1$. Hence $H_0$ can be written equivalently as $\Lambda_{\mathbf{C}}(t) - t\Lambda_{\mathbf{C}}(1) = \mathbf{0}, t \in (0,1)$. This special case of $q=0$ is closely related to CUSUM tests in structural change detection and has been widely discussed in the statistics literature recently; see for instance \citet{mies2021functional} and the references therein.

In general, testing whether $\betav_{\mathbf{C}}(t)$ is a $q$-th order polynomial is equivalent to testing $H_0: \boldsymbol{\Lambda}_{\mathbf{C}}(t) = \mathbf{a}_1 t + \cdots + \mathbf{a}_{q+1} t^{q+1}, t\in (0,1)$. Choosing $k=q+1$ and $v_i=i/(q+1)$, we solve $\mathbf{a}_i, 1 \le i \le q+1$ via the system of linear equations
\begin{eqnarray*}
    \begin{cases}
      \Lambda_{\mathbf{C}}\left(\frac{1}{q+1}\right) - \frac{1}{q+1}\mathbf{a}_1 - \left(\frac{1}{q+1}\right)^2\mathbf{a}_2 - \cdots -  \left(\frac{1}{q+1}\right)^{q+1}\mathbf{a}_{q+1}= \mathbf{0}\\
        \Lambda_{\mathbf{C}}\left(\frac{2}{q+1}\right) - \frac{2}{q+1}\mathbf{a}_1 - \left(\frac{2}{q+1}\right)^2\mathbf{a}_2 - \cdots -  \left(\frac{2}{q+1}\right)^{q+1}\mathbf{a}_{q+1}= \mathbf{0}\\
      \quad \quad \vdots\\
      \Lambda_{\mathbf{C}}(1) -  \mathbf{a}_1 - \mathbf{a}_2 - \cdots -  \mathbf{a}_q= \mathbf{0}\\
    \end{cases}.      
\end{eqnarray*}
By plugging in the solutions to $\mathbf{a}_i$'s, $H_0$ is of the form \eqref{eq:LOFT}, and the LOFT can be applied. 

Theorem \ref{thm:test} lays a theoretical foundation for the Lack-of-fit Test and we summarise by listing the procedures in Algorithm \ref{alg:LOFT}. 

\begin{theorem}\label{thm:test}
Under the conditions of Theorem \ref{thm:2} and \ref{thm:3}, assume $f(t, \{\mathbf{s}_i\}_{i =1}^k): [0,1]\times \mathbb{R}^{sk} \to \mathbb{R}^{s}$ is continuously differentiable with regard to $\{\mathbf{s}_i\}_{i =1}^k$. Denote the partial derivative of $f(\cdot)$ with regard to $\mathbf{s}_j$ as $\partial_j f(t,\{\mathbf{s}_i\}_{i =1}^k)$, we assume $\partial_j f(t,\{\mathbf{s}_i\}_{i =1}^k), j = 1, 2, \cdots, k$ are Lipschitz continuous for $\{\mathbf{s}_i\}_{i =1}^k$ uniformly over $t \in (0,1)$, and that $\max_{j =1}^k\sup_{t \in (0,1)}|\partial_j f(t,\{\mathbf{s}_i\}_{i =1}^k)| \le C$.
Let $\mathbf{U}(t)$ be a Gaussian process defined in Theorem \ref{thm:3}, $T(t) := {\Lambda}_{\mathbf{C}}(t) - f(t, \{{\Lambda}_{\mathbf{C}}(v_i)\}_{i =1}^k)$, $T_n(t) := \check{\Lambda}_{\mathbf{C}}(t) - f(t, \{\check{\Lambda}_{\mathbf{C}}(v_i)\}_{i =1}^k)$, 
$\tilde{\mathbf{U}}(t) := \mathbf{U}(t) - \sum_{j=1}^k\partial_j f(t, \{\Lambda_{\mathbf{C}}(v_i)\}_{i =1}^k)\mathbf{U}(v_j)$. Let $i_* = \max(\lceil nb_n \rceil, \lceil 2nc_n \rceil)$ and $i^* = n - i_*$, then
\begin{eqnarray}\label{eq:Tn}
    \max_{i_* \le i \le i^*}\left|\sqrt{n}(T_n(t_i) - T(t_i))\right|_\infty \Rightarrow \sup_{t \in (0,1)}\left|\tilde{\mathbf{U}}(t)\right|_\infty.
\end{eqnarray}
Additionally, let $\check{{L}}_n(t) = \tilde{\boldsymbol{\Phi}}^\circ_n (t) - \sum_{j = 1}^k\partial_j f(t, \{\check{\Lambda}_{\mathbf{C}}(v_i)\}_{i =1}^k)\tilde{\boldsymbol{\Phi}}^\circ_n (v_j)$, then on a sequence of events $A_n$ with $\mathbb{P}(A_n) = 1 - o(1)$, we have
\begin{equation}\label{eq:comparison_Ln}
\begin{aligned}
     &\sup _{|x|>d_{n, p}^{\circ}}\left|\mathbb{P}\left(\max_{i_* \le i \le i^*}|\tilde{\mathbf{U}}(t_i)|_\infty \leq x\right)-\mathbb{P}\left(\max_{i_* \le i \le i^*}|\check{{L}}_n(t_i)|_\infty \leq x \mid \{(\mathbf{x}_i,y_i)\}_{i = 1}^n \right)\right|  \to 0, 
\end{aligned}
\end{equation}
where $d_{n, p}^{\circ} \to 0$ as defined in Theorem \ref{thm:3}.
\end{theorem}

\begin{algorithm}
\floatname{algorithm}{\bf Algorithm}
\caption{Lack-of-fit Test}
\vspace{4pt}
\hrule
\vspace{4pt}
\label{alg:LOFT}
\begin{algorithmic}[1]
\State Perform Step 1 - 3 proposed in \textbf{Algorithm} \ref{alg:EFT}. Based on $\{\check{\Lambda}_{\mathbf{C}}(t_j)\}_{ j =1}^n$, obtain the Jackknife bias-corrected estimators $T_n(t) = \check{\Lambda}_{\mathbf{C}}(t) - f(t, \{\check{\Lambda}_{\mathbf{C}}(v_i)\}_{i =1}^k)$, and construct the corresponding bootstrap statistics $\{\check{L}^{(r)}_n(t) = \tilde{\boldsymbol{\Phi}}^{(r)}_n (t) - \sum_{j =1}^k\partial_j f(t, \{\check{\Lambda}_{\mathbf{C}}(v_i)\}_{i =1}^k)\tilde{\boldsymbol{\Phi}}^{(r)}_n (v_j)\}_{r = 1}^B$.

\State Calculate $M_r =\max_{i_* \le j \le i^*} |\check{L}^{(r)}_n(t_j)|_\infty$, $r=1,2,\cdots,B$.

\State For a given $\alpha \in (0,1)$, find the $(1-\alpha)$-th sample quantile of $\{M_r\}_{r=1}^B$, $\hat{q}_{n,1-\alpha}$. 

\State For the Lack-of-fit Test $H_0: \Lambda_{\mathbf{C}}(t) - f(t, \{\Lambda_{\mathbf{C}}(t_i)\}_{i =1}^k) = \mathbf{0}, \quad t \in (0,1)$, reject the null hypothesis if $  \max_{i_* \le j \le i^*}|T_n(t_j)|_\infty >  \hat{q}_{n,1-\alpha}/\sqrt{n}$. 
\end{algorithmic}
\vspace{4pt}
\hrule
\end{algorithm}

It's straightforward to see that the EFT addressed in Section \ref{sec:EFT} is a special case of the LOFT in Section \ref{sec:LOFT}. Therefore, we state the asymptotic rejection rate of the proposed algorithm of LOFT in Proposition \ref{prop: typeIerror_LOFT} and \ref{prop: power_LOFT}, which also applies to EFT. 
\begin{proposition}\label{prop: typeIerror_LOFT}
	Suppose the conditions of Theorem \ref{thm:test} hold, under the null hypothesis, for a given significance level $\alpha \in (0,1)$, the rejection rate of the proposed LOFT satisfies
	\begin{eqnarray*}
		\lim_{n \to \infty}\lim_{B \to \infty}\mathbb{P}\left(\max_{i_* \le j \le i^*}|T_n(t_j)|_\infty >  \hat{q}_{n,1-\alpha}/\sqrt{n}\right) = \alpha.
	\end{eqnarray*}
\end{proposition}	

\begin{proposition}\label{prop: power_LOFT}
Suppose the conditions of Theorem \ref{thm:test} hold, under the alternative hypothesis of LOFT, if $\sup_{t \in (0,1)}|\Lambda_{\mathbf{C}}(t) - f(t, \{\Lambda_{\mathbf{C}}(v_i)\}_{i =1}^k)|_\infty \gg n^{-\frac{1}{2}}$, then
	\begin{eqnarray*}
		\lim_{n \to \infty}\lim_{B \to \infty}\mathbb{P}\left(\max_{i_* \le j \le i^*}|T_n(t_j)|_\infty >  \hat{q}_{n,1-\alpha}/\sqrt{n}\right) = 1.
	\end{eqnarray*}
\end{proposition}

Propositions \ref{prop: typeIerror_LOFT} and \ref{prop: power_LOFT} demonstrate the accuracy and power of our proposed testing framework. In particular, both EFT and LOFT are asymptotically accurate under the null hypothesis, while attaining asymptotic power of $1$ under local alternatives deviating from $H_0$ with a distance much greater than $n^{-1/2}$. These results highlight the asymptotic reliability and robustness of our methodology.

\subsection{Qualitative Test}\label{sec:qualitative}
In this subsection, we are interested in testing qualitative hypotheses on M-regression coefficients in the sense that  the null hypothesis is nonparametric and can be written as 
\begin{eqnarray}\label{eq:null_QT}
    H_0: \Lambda_\mathbf{C}(t) \in \mathcal{N}_0, \quad  t \in (0,1),
\end{eqnarray}
where $\mathcal{N}_0$ is a non-empty subset of $\mathcal{N} := \{g \in \mathcal{C}^3[0,1]: [0, 1] \to \mathbb{R}^s, \quad g(0) = \mathbf{0}\}$. 
Based on the cumulative M-estimator $\check{\Lambda}_{\mathbf{C}}(t)$, consider the optimization problem:
\begin{eqnarray}\label{eq:opt}
    &&\text{minimize} \quad \sup_{t \in (0,1)}|\boldsymbol{\phi}(t) - \check{\Lambda}_{\mathbf{C}}(t)|_\infty , \crcr
&&\text{subject to} \quad \boldsymbol{\phi}(t) \in \mathcal{N}_0.
\end{eqnarray}
Suppose that there exists a projection under the $\mathcal{L}_\infty$ norm on $\mathcal{N}_0$ for any $f \in \mathcal{N}$. Then the solution to \eqref{eq:opt}, denoted by $\tilde{\Lambda}_{\mathbf{C}}(t)$, is the projection of $\check{\Lambda}_{\mathbf{C}}(t)$ onto $\mathcal{N}_0$. Define 
\begin{eqnarray}\label{eq:t_QT}
    T_q:= \sup_{t \in (0,1)} \sqrt{n}|\check{\Lambda}_{\mathbf{C}}(t) - \tilde{\Lambda}_{\mathbf{C}}(t)|_{\infty},
\end{eqnarray}
we observe that $T_q$ becomes a natural indicator of the distance between ${\Lambda}_{\mathbf{C}}(t)$ and $\mathcal{N}_0$, as guaranteed by the uniform consistency of $\check{\Lambda}_{\mathbf{C}}(t)$. Therefore, in principle, $T_q$ should be small under $H_0$ and large under the alternative. The critical value of the test can be easily obtained by the self-convolved bootstrap, similar to the implementation of the EFT in Algorithm \ref{alg:EFT}. Detailed steps are given in Algorithm \ref{alg:qualitative}.

\begin{algorithm}
\floatname{algorithm}{\bf Algorithm}
\caption{Qualitative Test}
\vspace{4pt}
\hrule
\vspace{4pt}
\label{alg:qualitative}
\begin{algorithmic}[1]
\State Perform Step 1 - 5 in \textbf{Algorithm} \ref{alg:EFT} to obtain the critical value of EFT, $\hat{q}_{n,1-\alpha}$.
\State  Formulate a projection problem as in \eqref{eq:opt} and find the solution $\tilde{\Lambda}_{\mathbf{C}}(t)$.
\State  For the Qualitative Test where $H_0: \Lambda_\mathbf{C}(t) \in \mathcal{N}_0, \quad t \in (0,1)$, reject the null hypothesis whenever $\max_{i_* \le j \le i^*}|\check{\Lambda}_{\mathbf{C}}(t_j) - \tilde{\Lambda}_{\mathbf{C}}(t_j)|_\infty >  \hat{q}_{n,1-\alpha}/\sqrt{n}$.
\end{algorithmic}
\vspace{4pt}
\hrule
\end{algorithm}

Proposition \ref{prop: qualitative_typeI} stated below shows the capability of our approach to asymptotically control the type I error under the specified significance level. Meanwhile, it attains an asymptotic power of $1$ when the CRF deviates from the qualitative hypothesis with a $\mathcal{L}_\infty$ distance dominating $1/\sqrt{n}$, as illustrated in Proposition \ref{prop:qualitative_power}.

\begin{proposition}\label{prop: qualitative_typeI}
Suppose conditions of Theorem \ref{thm:2} - \ref{thm:3} hold. Under the null hypothesis of QT, we have
\begin{eqnarray*}
	\lim_{n \to \infty} \lim_{B \to \infty} \mathbb{P}\left(\max_{i_* \le j \le i^*}|\check{\Lambda}_{\mathbf{C}}(t_j) - \tilde{\Lambda}_{\mathbf{C}}(t_j)|_\infty >  \hat{q}_{n,1-\alpha}/\sqrt{n}\right) \le \alpha,
\end{eqnarray*}
where $\tilde{\Lambda}_{\mathbf{C}}(t)$ is the projection of $\check{\Lambda}_{\mathbf{C}}(t)$ onto the feasible region $\mathcal{N}_0$.
\end{proposition}

\begin{proposition}\label{prop:qualitative_power}
Suppose conditions of Theorem \ref{thm:2} - \ref{thm:3} hold. Under the alternative hypothesis of QT, further assume that $\inf_{g \in \mathcal{N}_0}\sup_{t \in (0,1)}|{\Lambda}_{\mathbf{C}}(t)-g(t)|_\infty \gg n^{-\frac{1}{2}}$, we have
	\begin{eqnarray*}
		\lim_{n \to \infty} \lim_{B \to \infty} \mathbb{P}\left(\max_{i_* \le j \le i^*}|\check{\Lambda}_{\mathbf{C}}(t_j) - \tilde{\Lambda}_{\mathbf{C}}(t_j)|_\infty >  \hat{q}_{n,1-\alpha}/\sqrt{n}\right) =1.
	\end{eqnarray*}
\end{proposition}

\subsubsection{Shape Test}\label{sec:shape}
One instance of such Qualitative Tests involves testing the shape of $\betav_{\mathbf{C}}(t)$. For example, an attempt is to check the monotonicity of $\betav_{\mathbf{C}}(t)$, which is equivalent to testing
\begin{eqnarray*}
	H_0: \Lambda_{\mathbf{C}}(t) \text{ is  a convex/concave function.}
\end{eqnarray*}
As a straightforward extension, we may consider the null hypothesis $H_0: \betav_{\mathbf{C}}^{(d)}(t) \ge \mathbf{0}$. Particularly, $d=0$ corresponds to nonnegativity, $d=1$ corresponds to the null hypothesis that $\betav_{\mathbf{C}}(t)$ is a monotonically increasing function, and $d=2$ refers to the convexity of $\betav_{\mathbf{C}}(t)$.
Clearly, such differential-type shape constraints belong to the realm of QT, by letting the feasible region $\mathcal{N}_0 = \{g \in \mathcal{C}^3[0,1]: g(0) = \mathbf{0}, g^{(d+1)}(t) \ge \mathbf{0}\}$. 

In practice, to tackle the optimization problem in the finite sample, the functional optimization problem is reduced to that regarding discrete vectors. Let $s = 1$ for a slight gain of simplicity, we denote the $d$-th differential matrix as $\nabla_d$ and let $\boldsymbol{\phi} = (\phi_0, {\phi}_1, \cdots, {\phi}_n)^\top$, then \eqref{eq:opt} can be formulated as follows:
\begin{eqnarray}\label{eq:shape_minimize}
	&&\text{minimize} \quad  \max_{i_* \le i \le i^*}|{\phi}_{i}-{\check{\Lambda}_\mathbf{C}}(t_i)|_\infty , \crcr
	&&\text{subject to} \quad  \nabla_{d+1}\boldsymbol{\phi} \ge 0, \quad \phi_0 = 0.
\end{eqnarray}
Further, as is discussed in \citet{Knight_2017}, the projection under $\mathcal{L}_{\infty}$ norm in \eqref{eq:shape_minimize} can be expressed as the solution to a linear program, which can be obtained easily. 

\section{Simulation Studies}\label{sec:simu}
This section utilizes Monte Carlo simulations to demonstrate the finite sample performance of the proposed method. All three hypotheses stated in Section \ref{sec:test} are considered. 

Let $a(t) = 1/2 - (t - 1/2)^2$, $b(t) = 1/2 - t/2$, $c(t) = 1/4 + t/2$. For $i = 1, 2, \cdots,n$, define $t_i = i/n$, $e_i = e(t_i)$, where $e(t) = \sum_{j = 0}^\infty a(t)^{j}\zeta_{i-j}/4$, and $x_{i,1} = x_{i,1}(t_i)$, $x_{i,2} = x_{i,2}(t_i)$, where $x_{i,1}(t) = \sum_{j = 0}^\infty b^{j}(t)\epsilon_{i-j}$, $x_{i,2}(t) = \sum_{j = 0}^\infty c^{j}(t)\eta_{i-j}$, and $e_i = (\eta_i + \varepsilon_i)/\sqrt{2}$, $\{\zeta_i\}_{i = -\infty}^\infty$, $\{\eta_i\}_{i = -\infty}^\infty$ and $\{\varepsilon_i\}_{i = -\infty}^\infty$ are i.i.d standard normals. Let $\GG_i = (\eta_{-\infty}, \varepsilon_{-\infty}, \cdots, \eta_i, \varepsilon_i)$, $\FF_i = (\zeta_{-\infty}, \cdots, \zeta_i)$, and $e_{i,\tau}$ be the $\tau$-th quantile of $e_i$. Consider the following models: 
\begin{enumerate}
\item[\textbf{I}] $y_i = \sin(2\pi t_i) + 0.5 x_{i,1} + 2\log(1+2t_i)x_{i,2} + e_i$. Covariates and errors are independent.
\item[\textbf{II}] $y_i = \sin(2\pi t_i) + 0.5 x_{i,1} + 2\log(1+2t_i)x_{i,2} + \sqrt{1 + x_{i,1}^2 + x_{i,2}^2}(e_i - e_{i,\tau})/\sqrt{3}$ when using quantile loss, and $y_i = \sin(2\pi t_i) + 0.5 x_{i,1} + 2\log(1+2t_i)x_{i,2} + \sqrt{1 + x_{i,1}^2 + x_{i,2}^2}
e_i/\sqrt{3}$ when using quadratic loss. In this model, covariates and errors are correlated. 
\end{enumerate}
Case \textbf{I} and \textbf{II} share the same $\betav(t) = (\sin(2\pi t), 0.5, 2\log(1+2t))^{\top}$ but have different errors. By investigating both cases, we can illustrate the applicability of our method under error-covariate independence and dependence. Throughout this section, we use the Epanechnikov kernel. The number of replications is $1000$, and the bootstrap sample size is $B = 1000$. 
\subsection{Bandwidth Selection}\label{sec:bw}
The framework involves smoothing parameters $b_n$ and $c_n$, where $b_n$ is used for estimating $\Lambda
_{\mathbf{C}}(t)$, while $c_n$ is used for constructing the bootstrapped process $\{\tilde{\Phi}_n^{o}(t), t\in [0, 1]\}$. 

To select the appropriate bandwidth $b_n$, we recommend using Leave-One-Out Cross-Validation (LOOCV) to select $\tilde{b}$ from a predetermined grid $\{b_1, b_2, \cdots, b_k\}$. Denote $\hat{\betav}_{b, -i}(t)$ as the estimator of $\betav(t)$ using bandwidth $b$ and the training set $\{(\mathbf{x}, y)_{(-i)}\}$, then we select $\tilde{b}$ via the minimum average error:
\begin{eqnarray}
     \tilde{b} = \argmin_{b \in \{b_1, \cdots, b_k\}} \frac{1}{n}\sum_{i = 1}^n\rho(y_i - \x_i^{\top}\hat{\betav}_{b, -i}(t_i)).
\end{eqnarray}
Ideally $\tilde{b} = O(n^{-1/5})$ by the proof of Theorem \ref{thm:1} and a similar argument from \citet{wu2017nonparametric} regarding quantile regression. Based on \citet{Chen_Hong_2012} where $b_n = \frac{1}{\sqrt{12}}n^{-1/5}$ was chosen as a rule-of-thumb approach, a feasible grid of $b_n$ can be selected around it, say equispaced points within the interval $[0.5, 1.5]\cdot \frac{1}{\sqrt{12}}n^{-1/5}$. The proposed procedure works reasonably well in our simulation studies.

Regarding the bandwidth $c_n$, we can use $c_n = 0.5b_n$ as an easy implementation. For refinements, we recommend the extended Minimum Volatility (MV) method suggested by \citet{zhou2010simultaneous}, which is an extension of the minimal volatility method proposed in \citet{Politis_Romano_Wolf_1999}. Suppose we are to get bootstrap samples of $\betav_{\mathbf{C}}(t)$, where $\mathbf{C}$ is a $s \times p$ matrix. Let the candidates of possible bandwidths be $\{c_1, c_2, \cdots, c_k\}$. For $t \in [2c_h, 1 - 2c_h]$, define $V_h(t) = \sqrt{\frac{c_h}{\mu}}\mathbf{C}(\check{\betav}_{c_h}(t + c_h) + \check{\betav}_{c_h}(t - c_h)-2\check{\betav}_{c_h}(t))$ and $M_h(t) = \sum_{i = \left\lceil 2 n c_h \right \rceil}^{\lceil nt \rceil}V_h(t_i)V_h(t_i)^{\top}$. We further define $M_h(t)$ over interval $[0,1]$ by letting $M_h(t) = M_h(2c_h)$ if $t \in [0, 2c_h)$ and $M_h(t) = M_h(1-2c_h)$ if $t \in (1-2c_h, 1]$. Based on this, we then compute $\{M_1(t_j), M_2(t_j), \cdots, M_k(t_j)\}_{j = 1}^n$, respectively. For a positive integral $r$, say $r = 5$, define the integrated standard error (\textit{ise}) as
\begin{eqnarray*}
    ise\left[\left\{{M}_{h}^{a, b}(t)\right\}_{h=1}^{k-r+1}\right]=\left\{\frac{1}{r-1} \sum_{l=h}^{h + r - 1}\left({M}_{l}^{a, b}(t)-\sum_{l=h}^{h + r - 1} {M}_{l}^{a, b}(t) / r\right)^{2}\right\}^{1 / 2},
\end{eqnarray*}
where ${M}_{l}^{a, b}(t)$ represents the $(a, b)_{th}$ entry of the matrix ${M}_{l}(t)$. Further, define
\begin{eqnarray*}
     ise\left[\left\{{M}_{h}\right\}_{h=1}^{k-r+1}\right]=\frac{1}{n}\sum_{i=1}^n \left(\sum_{a = 1}^s ise\left[\left\{{M}_{h}^{a, a}(t_i)\right\}_{h=1}^{k-r+1}\right]^2\right)^{1/2},
\end{eqnarray*}
then $c_n$ is chosen as $c_{h+\lfloor r/2\rfloor}$ if $h = \argmin_{h\in\{1, \cdots, k - r + 1\}}\Big(ise\left[\left\{{M}_{h}\right\}_{h=1}^{k-r+1}\right]\Big)$.

\subsection{Exact Function Test}\label{sec:simu_EFT}
We generate data under the setting of Case \textbf{I} and Case \textbf{II}, focusing on 2 null hypotheses $H_0: \beta_1(t) = 0.5$ and $H_0: (\beta_1(t), \beta_2(t))^\top = (0.5, 2\log(1+2t))^\top$ for both cases. Sample sizes $N = 300$ and $N = 500$ are considered, and 4 different losses are applied: quadratic loss, quantile losses with $\tau = 0.15, \tau = 0.5$, and $\tau = 0.85$. 

\begin{table}
\centering
\caption{\label{tab:TypeIErrorI} Simulated type I error of EFT in $\%$ with nominal level $\alpha = 5\%, 10\%$ for Case \textbf{I}.} 
\vspace{5pt}
\begin{adjustbox}{width=\columnwidth}
\begin{tabular}{|>{\centering}m{0.06\textwidth}|*{8}{m{0.03\textwidth}m{0.03\textwidth}|}}
\hline 
 &\multicolumn{4}{c|} {\(L_2\)}& \multicolumn{4}{c|} {\(\tau=0.15\)} & \multicolumn{4}{c|} {\(\tau=0.5\)} & \multicolumn{4}{c|} {\(\tau=0.85\)} \\ 
\hline 
& \multicolumn{2}{c|} {\(\beta_{1}(t)\)} & \multicolumn{2}{c|} {\((\beta_1(t), \beta_{2}(t))\)}& \multicolumn{2}{c|} {\(\beta_{1}(t)\)} & \multicolumn{2}{c|} {\((\beta_1(t), \beta_{2}(t))\)} & \multicolumn{2}{c|} {\(\beta_{1}(t)\)} & \multicolumn{2}{c|} {\((\beta_1(t), \beta_{2}(t))\)} & \multicolumn{2}{c|} {\(\beta_{1}(t)\)} & \multicolumn{2}{c|} {\((\beta_1(t), \beta_{2}(t))\)} \\ 
\hline
\(\alpha\) & \multicolumn{2}{c|}{\(5 \%\)\quad\(10 \%\)}  & \multicolumn{2}{c|}{\(5 \%\)\quad\(10 \%\)} & \multicolumn{2}{c|}{\(5 \%\)\quad\(10 \%\)}  & \multicolumn{2}{c|}{\(5 \%\)\quad\(10 \%\)} & \multicolumn{2}{c|}{\(5 \%\)\quad\(10 \%\)}  & \multicolumn{2}{c|}{\(5 \%\)\quad\(10 \%\)} & \multicolumn{2}{c|}{\(5 \%\)\quad\(10 \%\)}  & \multicolumn{2}{c|}{\(5 \%\)\quad\(10 \%\)}\\ 
\hline
\multicolumn{17}{|c|} {\(N = 300\)}\\ 
\hline
\(b/1.15\) &  
\multicolumn{2}{c|}{\(5.9\)\quad\(10.1\)} & 
\multicolumn{2}{c|}{\(4.7\)\quad\(8.8\)} & 
\multicolumn{2}{c|}{\(4.5\)\quad\(8.4\)} &  
\multicolumn{2}{c|}{\(4.5\)\quad\(8.0\)} &  
\multicolumn{2}{c|}{\(5.3\)\quad\(9.8\)} &  
\multicolumn{2}{c|}{\(4.8\)\quad\(9.0\)} & 
\multicolumn{2}{c|}{\(5.5\)\quad\(10.6\)} & 
\multicolumn{2}{c|}{\(5.4\)\quad\(9.3\)} \\
\(b\) &
\multicolumn{2}{c|}{\(5.2\)\quad\(10.5\)} & 
\multicolumn{2}{c|}{\(5.2\)\quad\(10.0\)} & 
\multicolumn{2}{c|}{\(6.2\)\quad\(10.1\)} &  
\multicolumn{2}{c|}{\(5.1\)\quad\(8.7\)} &  
\multicolumn{2}{c|}{\(5.8\)\quad\(9.7\)} &  
\multicolumn{2}{c|}{\(6.0\)\quad\(10.4\)} & 
\multicolumn{2}{c|}{\(6.8\)\quad\(10.7\)} & 
\multicolumn{2}{c|}{\(5.0\)\quad\(9.5\)} \\
\(1.15b\) & 
\multicolumn{2}{c|}{\(5.9\)\quad\(11.0\)} & 
\multicolumn{2}{c|}{\(5.0\)\quad\(9.6\)} & 
\multicolumn{2}{c|}{\(5.8\)\quad\(9.6\)} &  
\multicolumn{2}{c|}{\(4.9\)\quad\(8.8\)} &  
\multicolumn{2}{c|}{\(6.7\)\quad\(10.9\)} &  
\multicolumn{2}{c|}{\(5.4\)\quad\(10.1\)} & 
\multicolumn{2}{c|}{\(6.1\)\quad\(10.1\)} & 
\multicolumn{2}{c|}{\(5.5\)\quad\(8.5\)} \\
\hline
\multicolumn{17}{|c|} {\(N = 500\)}\\ 
\hline
\(b/1.15\) &
\multicolumn{2}{c|}{\(4.9\)\quad\(9.0\)} & 
\multicolumn{2}{c|}{\(4.1\)\quad\(9.6\)} & 
\multicolumn{2}{c|}{\(6.3\)\quad\(11.2\)} &  
\multicolumn{2}{c|}{\(4.8\)\quad\(9.2\)} &  
\multicolumn{2}{c|}{\(6.0\)\quad\(9.7\)} &  
\multicolumn{2}{c|}{\(5.3\)\quad\(8.7\)} & 
\multicolumn{2}{c|}{\(5.3\)\quad\(10.6\)} & 
\multicolumn{2}{c|}{\(4.7\)\quad\(8.6\)} \\
\(b\) & 
\multicolumn{2}{c|}{\(5.1\)\quad\(10.2\)} & 
\multicolumn{2}{c|}{\(5.6\)\quad\(10.1\)} & 
\multicolumn{2}{c|}{\(6.3\)\quad\(10.8\)} &  
\multicolumn{2}{c|}{\(4.7\)\quad\(9.0\)} &  
\multicolumn{2}{c|}{\(5.5\)\quad\(8.9\)} &  
\multicolumn{2}{c|}{\(5.3\)\quad\(8.9\)} & 
\multicolumn{2}{c|}{\(5.3\)\quad\(9.2\)} & 
\multicolumn{2}{c|}{\(4.9\)\quad\(8.6\)} \\
\(1.15b\) & 
\multicolumn{2}{c|}{\(5.4\)\quad\(10.2\)} & 
\multicolumn{2}{c|}{\(5.7\)\quad\(9.9\)} & 
\multicolumn{2}{c|}{\(6.3\)\quad\(9.9\)} &  
\multicolumn{2}{c|}{\(4.9\)\quad\(9.4\)} &  
\multicolumn{2}{c|}{\(5.3\)\quad\(9.3\)} &  
\multicolumn{2}{c|}{\(4.9\)\quad\(8.8\)} & 
\multicolumn{2}{c|}{\(5.8\)\quad\(9.6\)} & 
\multicolumn{2}{c|}{\(4.6\)\quad\(8.6\)} \\
\hline
\end{tabular}
\end{adjustbox}
\end{table}    

\begin{table}
\centering
\caption{\label{tab:TypeIErrorII}  Simulated type I error of EFT in $\%$ with nominal level $\alpha = 5\%, 10\%$ for Case \textbf{II}.}
\vspace{5pt}
\begin{adjustbox}{width=\columnwidth}
\begin{tabular}{|>{\centering}m{0.06\textwidth}|*{8}{m{0.03\textwidth}m{0.03\textwidth}|}}
\hline 
 &\multicolumn{4}{c|} {\(L_2\)}& \multicolumn{4}{c|} {\(\tau=0.15\)} & \multicolumn{4}{c|} {\(\tau=0.5\)} & \multicolumn{4}{c|} {\(\tau=0.85\)} \\ 
\hline 
& \multicolumn{2}{c|} {\(\beta_{1}(t)\)} & \multicolumn{2}{c|} {\((\beta_1(t), \beta_{2}(t))\)}& \multicolumn{2}{c|} {\(\beta_{1}(t)\)} & \multicolumn{2}{c|} {\((\beta_1(t), \beta_{2}(t))\)} & \multicolumn{2}{c|} {\(\beta_{1}(t)\)} & \multicolumn{2}{c|} {\((\beta_1(t), \beta_{2}(t))\)} & \multicolumn{2}{c|} {\(\beta_{1}(t)\)} & \multicolumn{2}{c|} {\((\beta_1(t), \beta_{2}(t))\)} \\ 
\hline
\(\alpha\) & \multicolumn{2}{c|}{\(5 \%\)\quad\(10 \%\)}  & \multicolumn{2}{c|}{\(5 \%\)\quad\(10 \%\)} & \multicolumn{2}{c|}{\(5 \%\)\quad\(10 \%\)}  & \multicolumn{2}{c|}{\(5 \%\)\quad\(10 \%\)} & \multicolumn{2}{c|}{\(5 \%\)\quad\(10 \%\)}  & \multicolumn{2}{c|}{\(5 \%\)\quad\(10 \%\)} & \multicolumn{2}{c|}{\(5 \%\)\quad\(10 \%\)}  & \multicolumn{2}{c|}{\(5 \%\)\quad\(10 \%\)}\\ 
\hline
\multicolumn{17}{|c|} {\(N = 300\)}\\ 
\hline
\(b/1.15\) &  
\multicolumn{2}{c|}{\(5.2\)\quad\(10.9\)} & \multicolumn{2}{c|}{\(5.2\)\quad \(9.7\)} &
\multicolumn{2}{c|}{\(4.7\)\quad \(9.8\)} & 
\multicolumn{2}{c|}{\(4.7\)\quad \(8.9\)} & 
\multicolumn{2}{c|}{\(6.2\)\quad\(10.2\)} & 
\multicolumn{2}{c|}{\(4.8\)\quad \(9.0\)} &
\multicolumn{2}{c|}{\(5.6\)\quad\(10.2\)}&
\multicolumn{2}{c|}{\(5.3\)\quad \(9.1\)}  \\
\(b\) &
\multicolumn{2}{c|}{\(5.5\)\quad\(11.3\)} & \multicolumn{2}{c|}{\(5.1\)\quad \(10.2\)} &
\multicolumn{2}{c|}{\(6.1\)\quad \(10.5\)} & 
\multicolumn{2}{c|}{\(4.6\)\quad \(9.1\)} & 
\multicolumn{2}{c|}{\(6.1\)\quad\(9.2\)} & 
\multicolumn{2}{c|}{\(5.6\)\quad \(9.8\)} &
\multicolumn{2}{c|}{\(6.7\)\quad\(10.7\)}&
\multicolumn{2}{c|}{\(5.0\)\quad \(9.5\)}  \\
\(1.15b\) & 
\multicolumn{2}{c|}{\(6.3\)\quad\(11.1\)} & 
\multicolumn{2}{c|}{\(6.0\)\quad\(10.9\)} & 
\multicolumn{2}{c|}{\(6.2\)\quad\(9.7\)} &  
\multicolumn{2}{c|}{\(4.5\)\quad\(8.0\)} &  
\multicolumn{2}{c|}{\(5.4\)\quad\(10.7\)} &  
\multicolumn{2}{c|}{\(5.2\)\quad\(8.9\)} & 
\multicolumn{2}{c|}{\(6.9\)\quad\(11.8\)} & 
\multicolumn{2}{c|}{\(5.6\)\quad\(9.5\)} \\
\hline
\multicolumn{17}{|c|} {\(N = 500\)}\\ 
\hline
\(b/1.15\) &
\multicolumn{2}{c|}{\(5.4\)\quad\(9.9\)} & 
\multicolumn{2}{c|}{\(5.7\)\quad\(9.8\)} & 
\multicolumn{2}{c|}{\(6.1\)\quad\(11.0\)} &  
\multicolumn{2}{c|}{\(4.5\)\quad\(9.1\)} &  
\multicolumn{2}{c|}{\(6.6\)\quad\(9.8\)} &  
\multicolumn{2}{c|}{\(5.5\)\quad\(10.1\)} & 
\multicolumn{2}{c|}{\(6.2\)\quad\(10.2\)} & 
\multicolumn{2}{c|}{\(4.1\)\quad\(9.2\)} \\
\(b\) & 
\multicolumn{2}{c|}{\(6.1\)\quad\(10.9\)} & 
\multicolumn{2}{c|}{\(6.2\)\quad\(10.0\)} & 
\multicolumn{2}{c|}{\(6.2\)\quad\(9.5\)} &  
\multicolumn{2}{c|}{\(4.4\)\quad\(8.3\)} &  
\multicolumn{2}{c|}{\(5.6\)\quad\(10.5\)} &  
\multicolumn{2}{c|}{\(4.9\)\quad\(9.9\)} & 
\multicolumn{2}{c|}{\(5.9\)\quad\(9.0\)} & 
\multicolumn{2}{c|}{\(4.7\)\quad\(8.6\)} \\
\(1.15b\) & 
\multicolumn{2}{c|}{\(5.9\)\quad\(10.9\)} & 
\multicolumn{2}{c|}{\(6.4\)\quad\(10.8\)} & 
\multicolumn{2}{c|}{\(5.5\)\quad\(9.7\)} &  
\multicolumn{2}{c|}{\(4.9\)\quad\(8.1\)} &  
\multicolumn{2}{c|}{\(5.6\)\quad\(10.7\)} &  
\multicolumn{2}{c|}{\(4.4\)\quad\(9.7\)} & 
\multicolumn{2}{c|}{\(6.1\)\quad\(9.7\)} & 
\multicolumn{2}{c|}{\(4.6\)\quad\(9.1\)} \\
\hline
\end{tabular}
\end{adjustbox}
\end{table}

Table \ref{tab:TypeIErrorI} and \ref{tab:TypeIErrorII} demonstrate the type I error of EFT with nominal levels $\alpha = 5\%, 10\%$. Using the bandwidth $b$ selected via the procedures described in Section \ref{sec:bw}, a reasonable type I error can be observed for both cases. We also present results using $b/1.15$ and $1.15b$ for sensitivity analysis. The results show a good approximation of the nominal level, as long as the bandwidths are not very far from the ones chosen by the proposed procedures.

\begin{figure}
\centering
\begin{minipage}[b]{0.48\textwidth}
\centering
\includegraphics[width=\textwidth]{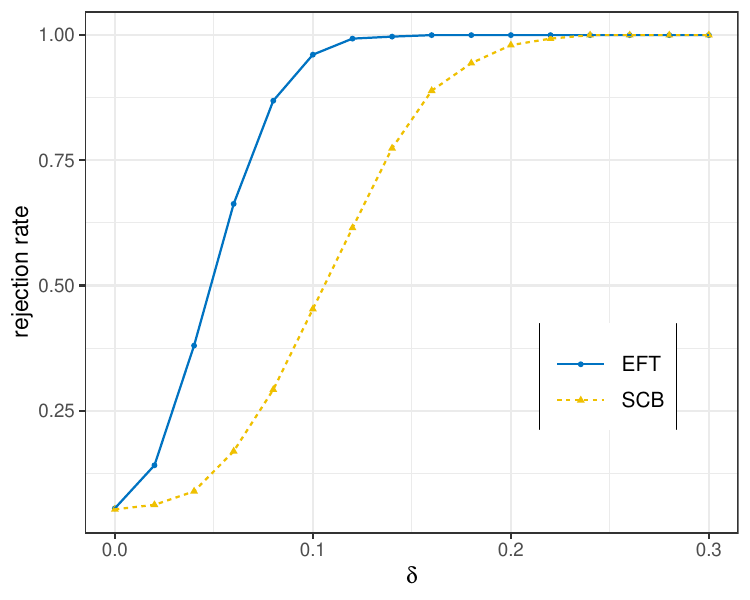}
\caption{\small Power of EFT and SCB: Case \textbf{I}}
\label{fig:exactI}
\end{minipage}
\hfill
\begin{minipage}[b]{0.48\textwidth}
\centering
\includegraphics[width=\textwidth]{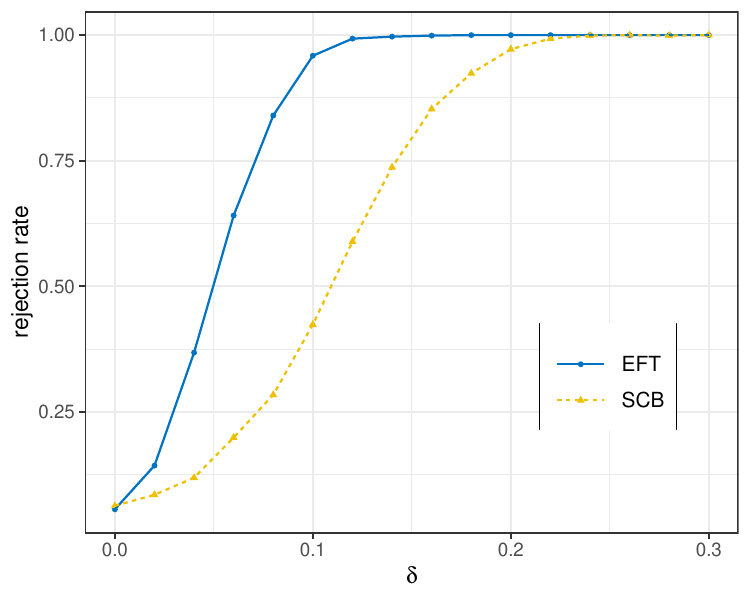}
\caption{\small Power of EFT and SCB: Case \textbf{II}}
\label{fig:exactII}
\end{minipage}
\label{fig:powerexact}
\end{figure}

To demonstrate the power of EFT, we generate data under the mechanism $\beta_1(t) = 0.5 + \delta$, while $\beta_0(t)$ and $\beta_2(t)$ remain unchanged from the original setting. Let $\delta$ be a positive constant, then the power of the test $H_0: \beta_1(t) = 0.5$ should increase with $\delta$. For comparison, we also replicate the SCB test proposed by \citet{wu2017nonparametric} under the same setting. For both Case \textbf{I} and Case \textbf{II}, we present the power when using quantile loss with $\tau = 0.5$ and sample size $N = 500$. 
The nominal level used here is $5\%$. Figure \ref{fig:exactI} and Figure \ref{fig:exactII} show that, in both cases, the power of both methods goes to $1$ as $\delta$ increases from $0$ to $0.3$. As expected, the power of EFT converges faster to $1$ than the other method, which is guaranteed by the $\sqrt{n}$ convergence rate theoretically.

\subsection{Lack-of-fit Test}\label{sec:simu_LOFT}
\begin{table}
\centering
\caption{\label{tab:LOFT_constancy}  Simulated type I error of constancy test of $\beta_1(t)$ in Case \textbf{I} and \textbf{II}.}
\vspace{5pt}
\begin{adjustbox}{width=\columnwidth}
\begin{tabular}{|>{\centering}m{0.06\textwidth}|*{8}{m{0.03\textwidth}m{0.03\textwidth}|}}
\hline 
 &\multicolumn{4}{c|} {\(L_2\)}& \multicolumn{4}{c|} {\(\tau=0.15\)} & \multicolumn{4}{c|} {\(\tau=0.5\)} & \multicolumn{4}{c|} {\(\tau=0.85\)} \\ 
\hline 
& \multicolumn{2}{c|} {\(\text{Case I}\)} & \multicolumn{2}{c|} {\(\text{Case II}\)}& \multicolumn{2}{c|} {\(\text{Case I}\)} & \multicolumn{2}{c|} {\(\text{Case II}\)} & \multicolumn{2}{c|} {\(\text{Case I}\)} & \multicolumn{2}{c|} {\(\text{Case II}\)} & \multicolumn{2}{c|} {\(\text{Case I}\)} & \multicolumn{2}{c|} {\(\text{Case II}\)} \\ 
\hline
\(\alpha\) & \multicolumn{2}{c|}{\(5 \%\)\quad\(10 \%\)}  & \multicolumn{2}{c|}{\(5 \%\)\quad\(10 \%\)} & \multicolumn{2}{c|}{\(5 \%\)\quad\(10 \%\)}  & \multicolumn{2}{c|}{\(5 \%\)\quad\(10 \%\)} & \multicolumn{2}{c|}{\(5 \%\)\quad\(10 \%\)}  & \multicolumn{2}{c|}{\(5 \%\)\quad\(10 \%\)} & \multicolumn{2}{c|}{\(5 \%\)\quad\(10 \%\)}  & \multicolumn{2}{c|}{\(5 \%\)\quad\(10 \%\)}\\ 
\hline
\multicolumn{17}{|c|} {\(N = 300\)}\\ 
\hline
\(b/1.15\) &
\multicolumn{2}{c|}{\(5.1\)\quad\(8.6\)} & 
\multicolumn{2}{c|}{\(5.0\)\quad\(10.0\)} & 
\multicolumn{2}{c|}{\(5.4\)\quad\(8.9\)} &  
\multicolumn{2}{c|}{\(5.6\)\quad\(10.0\)} &  
\multicolumn{2}{c|}{\(5.0\)\quad\(8.7\)} &  
\multicolumn{2}{c|}{\(5.0\)\quad\(8.8\)} & 
\multicolumn{2}{c|}{\(5.1\)\quad\(8.9\)} & 
\multicolumn{2}{c|}{\(5.6\)\quad\(10.7\)} \\
\(b\) & 
\multicolumn{2}{c|}{\(5.0\)\quad\(9.1\)} & 
\multicolumn{2}{c|}{\(6.1\)\quad\(10.2\)} & 
\multicolumn{2}{c|}{\(6.0\)\quad\(9.1\)} &  
\multicolumn{2}{c|}{\(5.7\)\quad\(10.2\)} &  
\multicolumn{2}{c|}{\(5.3\)\quad\(9.4\)} &  
\multicolumn{2}{c|}{\(5.6\)\quad\(9.5\)} & 
\multicolumn{2}{c|}{\(5.4\)\quad\(9.0\)} & 
\multicolumn{2}{c|}{\(5.6\)\quad\(9.9\)} \\
\(1.15b\) & 
\multicolumn{2}{c|}{\(4.4\)\quad\(8.0\)} & 
\multicolumn{2}{c|}{\(6.0\)\quad\(9.4\)} & 
\multicolumn{2}{c|}{\(4.7\)\quad\(8.0\)} &  
\multicolumn{2}{c|}{\(5.8\)\quad\(9.6\)} &  
\multicolumn{2}{c|}{\(5.2\)\quad\(9.1\)} &  
\multicolumn{2}{c|}{\(5.9\)\quad\(10.2\)} & 
\multicolumn{2}{c|}{\(5.3\)\quad\(9.8\)} & 
\multicolumn{2}{c|}{\(6.5\)\quad\(10.2\)} \\
\hline
\multicolumn{17}{|c|} {\(N = 500\)}\\ 
\hline
\(b/1.15\) & 
\multicolumn{2}{c|}{\(5.9\)\quad\(9.2\)} & 
\multicolumn{2}{c|}{\(4.7\)\quad\(9.6\)} & 
\multicolumn{2}{c|}{\(6.1\)\quad\(10.3\)} &  
\multicolumn{2}{c|}{\(5.5\)\quad\(10.6\)} &  
\multicolumn{2}{c|}{\(5.0\)\quad\(9.1\)} &  
\multicolumn{2}{c|}{\(5.9\)\quad\(9.6\)} & 
\multicolumn{2}{c|}{\(6.7\)\quad\(10.8\)} & 
\multicolumn{2}{c|}{\(5.9\)\quad\(9.5\)} \\
\(b\) & 
\multicolumn{2}{c|}{\(5.2\)\quad\(9.1\)} & 
\multicolumn{2}{c|}{\(6.0\)\quad\(11.2\)} & 
\multicolumn{2}{c|}{\(6.1\)\quad\(9.7\)} &  
\multicolumn{2}{c|}{\(6.2\)\quad\(10.8\)} &  
\multicolumn{2}{c|}{\(5.4\)\quad\(8.8\)} &  
\multicolumn{2}{c|}{\(6.1\)\quad\(9.8\)} & 
\multicolumn{2}{c|}{\(5.9\)\quad\(9.5\)} & 
\multicolumn{2}{c|}{\(6.0\)\quad\(9.8\)} \\
\(1.15b\) & 
\multicolumn{2}{c|}{\(5.6\)\quad\(9.9\)} & 
\multicolumn{2}{c|}{\(5.1\)\quad\(10.0\)} & 
\multicolumn{2}{c|}{\(5.4\)\quad\(9.0\)} &  
\multicolumn{2}{c|}{\(5.9\)\quad\(10.6\)} &  
\multicolumn{2}{c|}{\(5.5\)\quad\(9.1\)} &  
\multicolumn{2}{c|}{\(5.1\)\quad\(9.6\)} & 
\multicolumn{2}{c|}{\(6.6\)\quad\(10.3\)} & 
\multicolumn{2}{c|}{\(6.1\)\quad\(9.6\)} \\
\hline
\end{tabular}
\end{adjustbox}
\end{table}

\begin{table}
\centering
\caption{\label{tab:LOFT_linearity}  Simulated type I error of linearity test of $\beta_1(t)$ in Case \textbf{I} and \textbf{II}.}
\vspace{5pt}
\begin{adjustbox}{width=\columnwidth}
\begin{tabular}{|>{\centering}m{0.06\textwidth}|*{8}{m{0.03\textwidth}m{0.03\textwidth}|}}
\hline 
 &\multicolumn{4}{c|} {\(L_2\)}& \multicolumn{4}{c|} {\(\tau=0.15\)} & \multicolumn{4}{c|} {\(\tau=0.5\)} & \multicolumn{4}{c|} {\(\tau=0.85\)} \\ 
\hline 
& \multicolumn{2}{c|} {\(\text{Case I}\)} & \multicolumn{2}{c|} {\(\text{Case II}\)}& \multicolumn{2}{c|} {\(\text{Case I}\)} & \multicolumn{2}{c|} {\(\text{Case II}\)} & \multicolumn{2}{c|} {\(\text{Case I}\)} & \multicolumn{2}{c|} {\(\text{Case II}\)} & \multicolumn{2}{c|} {\(\text{Case I}\)} & \multicolumn{2}{c|} {\(\text{Case II}\)} \\ 
\hline
\(\alpha\) & \multicolumn{2}{c|}{\(5 \%\)\quad\(10 \%\)}  & \multicolumn{2}{c|}{\(5 \%\)\quad\(10 \%\)} & \multicolumn{2}{c|}{\(5 \%\)\quad\(10 \%\)}  & \multicolumn{2}{c|}{\(5 \%\)\quad\(10 \%\)} & \multicolumn{2}{c|}{\(5 \%\)\quad\(10 \%\)}  & \multicolumn{2}{c|}{\(5 \%\)\quad\(10 \%\)} & \multicolumn{2}{c|}{\(5 \%\)\quad\(10 \%\)}  & \multicolumn{2}{c|}{\(5 \%\)\quad\(10 \%\)}\\ 
\hline
\multicolumn{17}{|c|} {\(N = 300\)}\\ 
\hline
\(b/1.15\) &
\multicolumn{2}{c|}{\(4.6\)\quad\(8.5\)} & 
\multicolumn{2}{c|}{\(5.6\)\quad\(9.5\)} & 
\multicolumn{2}{c|}{\(5.4\)\quad\(9.8\)} &  
\multicolumn{2}{c|}{\(5.5\)\quad\(9.5\)} &  
\multicolumn{2}{c|}{\(5.7\)\quad\(10.8\)} &  
\multicolumn{2}{c|}{\(5.9\)\quad\(9.7\)} & 
\multicolumn{2}{c|}{\(6.0\)\quad\(9.1\)} & 
\multicolumn{2}{c|}{\(6.0\)\quad\(10.5\)} \\
\(b\) & 
\multicolumn{2}{c|}{\(4.5\)\quad\(8.3\)} & 
\multicolumn{2}{c|}{\(6.5\)\quad\(9.8\)} & 
\multicolumn{2}{c|}{\(6.3\)\quad\(8.9\)} &  
\multicolumn{2}{c|}{\(5.0\)\quad\(9.1\)} &  
\multicolumn{2}{c|}{\(5.6\)\quad\(9.4\)} &  
\multicolumn{2}{c|}{\(5.7\)\quad\(9.7\)} & 
\multicolumn{2}{c|}{\(6.1\)\quad\(10.2\)} & 
\multicolumn{2}{c|}{\(5.4\)\quad\(9.7\)} \\
\(1.15b\) & 
\multicolumn{2}{c|}{\(5.6\)\quad\(8.8\)} & 
\multicolumn{2}{c|}{\(6.6\)\quad\(9.6\)} & 
\multicolumn{2}{c|}{\(5.8\)\quad\(9.5\)} &  
\multicolumn{2}{c|}{\(5.6\)\quad\(8.7\)} &  
\multicolumn{2}{c|}{\(5.0\)\quad\(9.3\)} &  
\multicolumn{2}{c|}{\(5.5\)\quad\(10.3\)} & 
\multicolumn{2}{c|}{\(5.8\)\quad\(9.0\)} & 
\multicolumn{2}{c|}{\(5.4\)\quad\(9.9\)} \\
\hline
\multicolumn{17}{|c|} {\(N = 500\)}\\ 
\hline
\(b/1.15\) & 
\multicolumn{2}{c|}{\(5.3\)\quad\(9.0\)} & 
\multicolumn{2}{c|}{\(6.3\)\quad\(10.0\)} & 
\multicolumn{2}{c|}{\(5.0\)\quad\(8.9\)} &  
\multicolumn{2}{c|}{\(5.5\)\quad\(9.1\)} &  
\multicolumn{2}{c|}{\(5.6\)\quad\(9.0\)} &  
\multicolumn{2}{c|}{\(5.3\)\quad\(9.3\)} & 
\multicolumn{2}{c|}{\(5.9\)\quad\(10.2\)} & 
\multicolumn{2}{c|}{\(5.6\)\quad\(9.6\)} \\
\(b\) & 
\multicolumn{2}{c|}{\(5.9\)\quad\(8.9\)} & 
\multicolumn{2}{c|}{\(6.4\)\quad\(10.5\)} & 
\multicolumn{2}{c|}{\(5.7\)\quad\(9.1\)} &  
\multicolumn{2}{c|}{\(5.6\)\quad\(9.1\)} &  
\multicolumn{2}{c|}{\(6.8\)\quad\(9.9\)} &  
\multicolumn{2}{c|}{\(6.0\)\quad\(9.5\)} & 
\multicolumn{2}{c|}{\(5.5\)\quad\(10.4\)} & 
\multicolumn{2}{c|}{\(6.0\)\quad\(10.2\)} \\
\(1.15b\) & 
\multicolumn{2}{c|}{\(5.2\)\quad\(9.0\)} & 
\multicolumn{2}{c|}{\(5.3\)\quad\(9.9\)} & 
\multicolumn{2}{c|}{\(6.2\)\quad\(10.3\)} &  
\multicolumn{2}{c|}{\(6.0\)\quad\(9.9\)} &  
\multicolumn{2}{c|}{\(5.7\)\quad\(9.0\)} &  
\multicolumn{2}{c|}{\(6.8\)\quad\(10.6\)} & 
\multicolumn{2}{c|}{\(5.8\)\quad\(10.6\)} & 
\multicolumn{2}{c|}{\(5.1\)\quad\(8.9\)} \\
\hline
\end{tabular}
\end{adjustbox}
\end{table}
For illustration of the Lack-of-fit Test, we focus on $\beta_1(t)$ in Case \textbf{I} and \textbf{II}. Under both settings, $\beta_1(t) = 0.5$ satisfies constancy and linearity, thus the two types of Lack-of-fit Tests can be conducted to demonstrate the accuracy of our method. Following the procedures in Section \ref{sec:LOFT}, we obtain the simulated type I errors for the 4 losses and sample size $N = 300, 500$. The optimal bandwidths are the same as those selected in Section \ref{sec:simu_EFT} in that we are considering the same settings. Table \ref{tab:LOFT_constancy} and Table \ref{tab:LOFT_linearity} show that the accuracy is quite satisfactory and is not very sensitive to the selection of the bandwidths. 

\begin{figure}
\centering
\begin{minipage}[b]{0.48\textwidth}
\centering
\includegraphics[width=\textwidth]{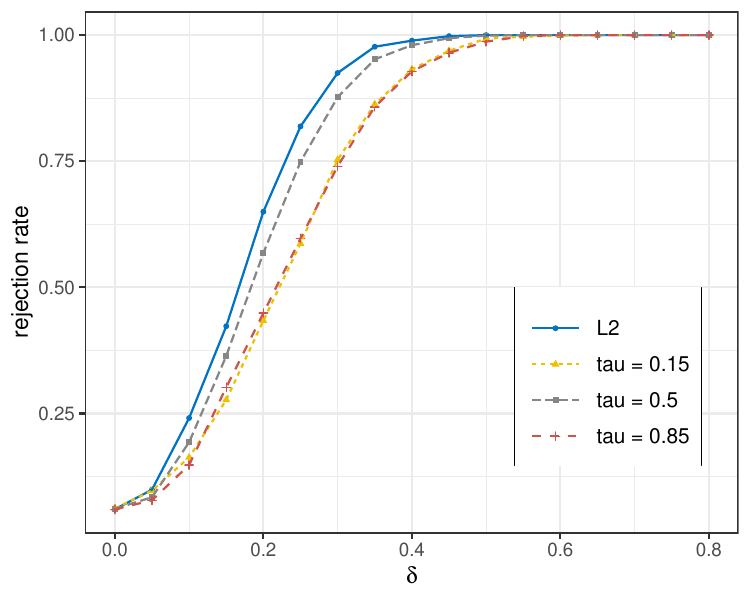}
\caption{\small Power of LOFT: Constancy}
\label{fig:constancy}
\end{minipage}
\hfill
\begin{minipage}[b]{0.48\textwidth}
\centering
\includegraphics[width=\textwidth]{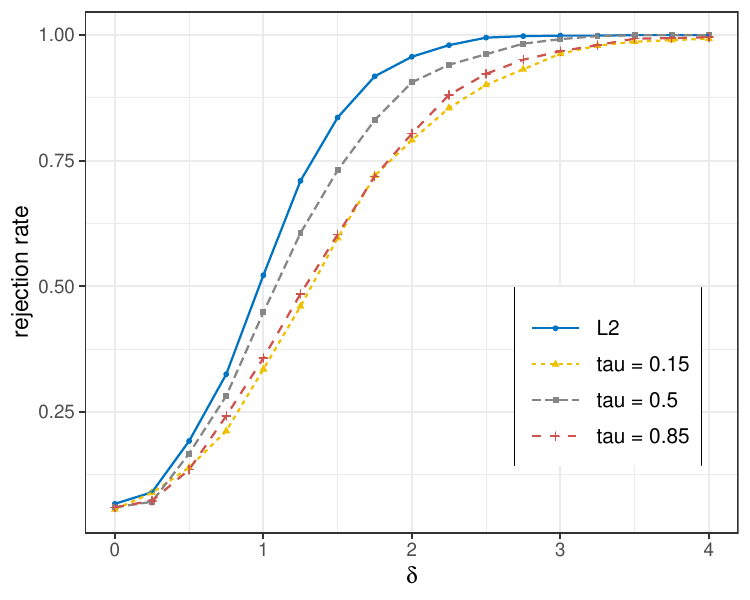}
\caption{\small Power of LOFT: Linearity}
\label{fig:linearity}
\end{minipage}
\end{figure}

Under the setting of Case \textbf{II}, we also conducted the power analysis under 2 scenarios: (1) $\beta_1(t) = 0.5 + \delta t$; (2) $\beta_1(t) = 0.5 + \delta t^2$, where the first scenario allows us to investigate constancy while the other enables the testing of linearity. The sample size is chosen to be $N = 500$ and the nominal level is $5\%$. For each test, power curves regarding the 4 different losses are simulated and compared in Figure \ref{fig:constancy} and Figure \ref{fig:linearity}. The quadratic loss is the most powerful and the quantile loss with $\tau = 0.5$ comes second. As for the extreme quantiles $\tau = 0.15$ and $\tau = 0.85$, the convergence is relatively slower. 

We then compare our method with \citet{Chen_Hong_2012} to test the constancy of the coefficients. Since their method assumes stationary covariates and errors, we consider the following stationary model for a more fair comparison:
\begin{enumerate}
\item[\textbf{III}] $y_i = \delta \cdot \sin(2\pi t_i) + 0.5 x_{i,1} + \delta \cdot 2\log(1+2t_i)x_{i,2} + e_i$, where $x_{i,1} = 0.5x_{i-1,1} + \varepsilon_i$, $x_{i,2} = 0.5x_{i-1,2} + \eta_i$, and $\{\varepsilon_i\}, \{\eta_i\}, \{e_i\}$ are i.i.d standard normal random variables independent of each other. 
\end{enumerate}
\begin{figure}[htbp]
    \centering
    \includegraphics[width = 0.48\textwidth]{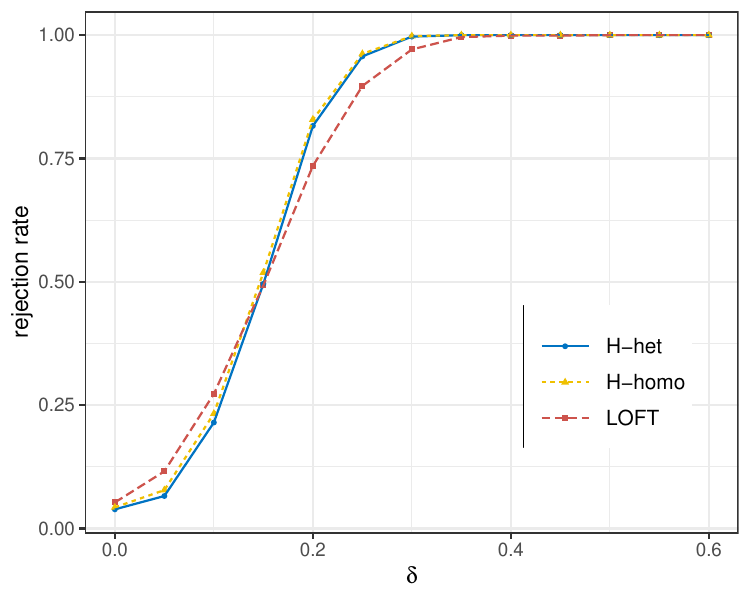}
    \caption{\small Power comparison of LOFT}
    \label{fig:ChenHong}
\end{figure}

Note that when $\delta = 0$, the model corresponds to constant coefficients $\betav(t) = (0, 0.5, 0)^{\top}$, and for the joint constancy test of $\betav(t)$, we expect to see an increasing rejection rate as $\delta \ge 0$ increases. We generate the data under $N = 500$ and use $L_2$ loss to implement our method. The nominal level is chosen as $\alpha = 5\%$. Figure \ref{fig:ChenHong} indicates that our method has comparable accuracy and power with \citet{Chen_Hong_2012}'s method, where `H-homo' and `H-het' represent the homoscedastic and heteroscedasticity-robust version of the generalized Hausman test proposed in \citet{Chen_Hong_2012}. Considering that their method is designed for the constancy test of $L_2$-regression models and applies to stationary variables, our Lack-of-fit Test enjoys more flexibility and broader applicability, with a faster convergence rate theoretically. 

\subsection{Qualitative Test}
For the Qualitative Test, exemplarily, we mainly focus on the hypothesis of monotonicity. Under the setting of Case \textbf{II} with $N = 500$, we generate the data using the following mechanism: $\beta_1(t) = 0.5 -\delta \cdot 10 \exp(-(t - 0.5)^2)$ for $\delta \ge 0$, and we expect to see an increasing power curve as $\delta$ increases. 
 
\begin{figure}
     \centering
     \begin{minipage}[b]{0.48\textwidth}
        \centering
        \includegraphics[width=\textwidth]{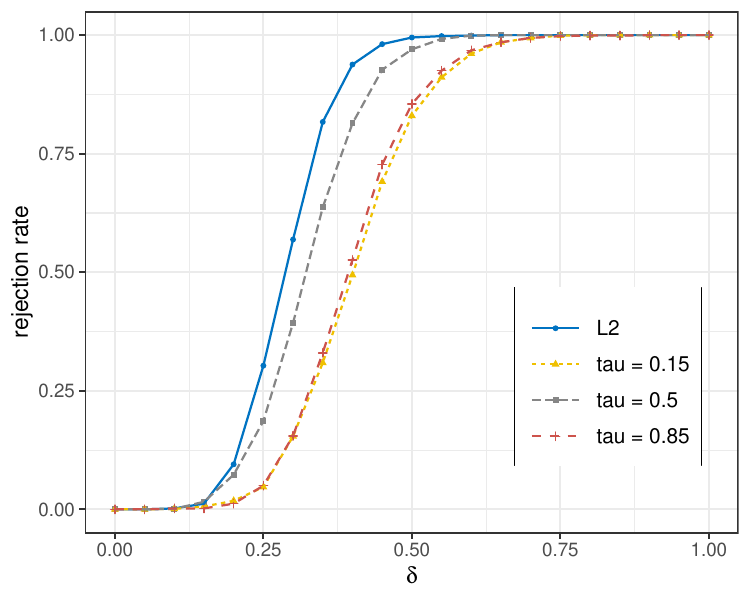}
        \caption{\small Power of QT under different losses}
        \label{fig:Shape4Losses}
     \end{minipage}
     \hfill
     \begin{minipage}[b]{0.48\textwidth}
         \centering
         \includegraphics[width=\textwidth]{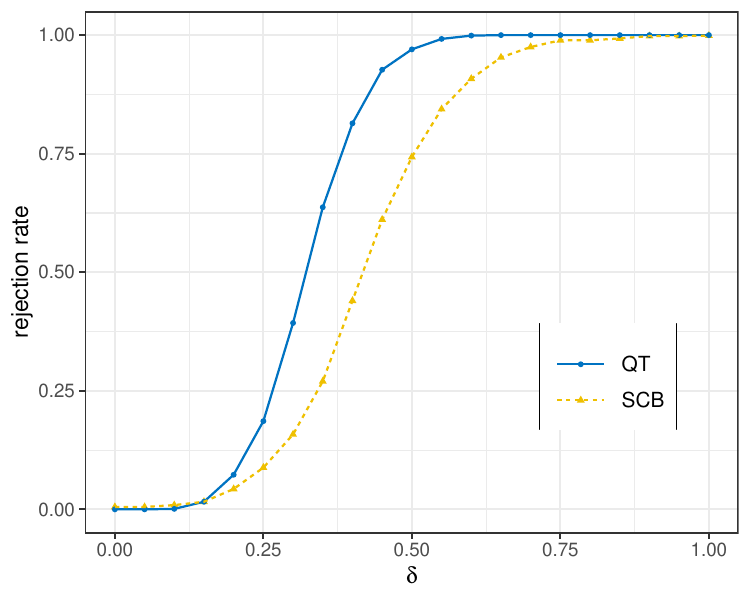}
         \caption{\small Power comparison of QT and SCB}
         \label{fig:ShapeComparison}
     \end{minipage}
\end{figure}

Applying a similar projection idea to the SCB test in \citet{wu2017nonparametric}, we also manage to obtain the power curve regarding the monotonicity test for quantile loss with $\tau = 0.5$ for their method. Figure \ref{fig:Shape4Losses} provides the results with nominal level $\alpha = 5\%$, demonstrating the performance of our method regarding 4 losses and the comparison with the SCB test. The power curves of quadratic loss and quantile loss with $\tau = 0.5$ converge faster than the extreme quantiles, as we expect. Figure \ref{fig:ShapeComparison} also shows that, under the same quantile loss with $\tau = 0.5$, our qualitative test is more powerful than the SCB test with projection. This is due to the fact that our CRF-based test can detect local alternatives of the order $O(1/\sqrt{n})$ while the SCB in \citet{wu2017nonparametric} are sensitive to those of the order $O(\sqrt{\log n}/\sqrt{nb_n})$ with bandwidth $b_n\rightarrow 0$.   

\section{Real Data Illustrations}\label{sec:data}
\subsection{Global Temperature Anomalies}
In this section, we delve into monthly global temperature anomalies from 1882.1 to 2005.12 (available at \href{https://crudata.uea.ac.uk/cru/data/temperature}{HadCRUT5} dataset), investigating the anthropogenic warming trend and the time-varying relationship between the anomalies and potential factors. The initial candidate factors include lags of multivariate ENSO index (\href{https://psl.noaa.gov/enso/mei.ext/table.ext.html}{MEI}), total solar irradiance (\href{https://solarisheppa.geomar.de/solarisheppa/sites/default/files/data/CMIP5/TSI_WLS_mon_1882_2008.txt}{TSI}), aerosol optical depth (\href{https://data.giss.nasa.gov/modelforce/strataer/tau.line_2012.12.txt}{AOD}) and Atlantic multidecadal oscillation (\href{https://psl.noaa.gov/data/timeseries/AMO}{AMO}). These factors are typical regressors in existing studies, where multivariate linear regression is commonly employed to filter out the fluctuations caused by the factors and to reveal the underlying anthropogenic warming \citep{zhou2013deducing}. The authors believe that 
the warming trend has been remarkably steady since the mid-twentieth century. \citet{foster2011global} also applied a similar approach to the global temperature anomalies from 1979 to 2010, where the deduced warming rate was also found to be steady over the investigated time interval. All previous analyses assume a stationary structure of the errors and constant coefficients over time, in addition to which a linear trend of anthropogenic warming is also hypothesized. Nevertheless, by utilizing the change-point detection method in \citep{zhou2013heteroscedasticity}, nonstationarity of the residuals has been detected after fitting the multivariate linear model, suggesting the necessity to use a more general model that allows for the nonstationarity of the time series. Furthermore, no rigorous proof has been provided in the existing literature to draw convincing conclusions about the shape of the warming trend. 

We hereby apply our method to the aforementioned time series. Following the steps stated in Section \ref{sec:test}, a full model consisting of all the factors is first constructed and analyzed, then a backward variable selection is implemented based on our EFT. Most covariates are therefore removed due to insignificance or multicollinearity. The final model includes the intercept, MEI, and the 6-month lag of MEI under $L_2$ loss, quantile loss with $\tau = 0.15$ and $\tau = 0.5$, while the $\tau = 0.85$ quantile regression model is composed of the intercept, 2-month and 6-month lags of MEI, and 5-month lag of AOD. All models involve the intercept and corresponding lags of MEI, enabling us to conduct tests regarding the anthropogenic warming trend and the coefficient of the ENSO effect. The outcome agrees with the fact that ENSO has conventionally been recognized as a leading contributor to global temperature fluctuations \citep{Trenberth_Caron_Stepaniak_Worley_2002,foster2011global}. The 6-month lag of MEI chosen in each model also aligns with the belief that El Ni\~no warms up the global temperature with a lag of $\sim$6 months \citep{Trenberth_Caron_Stepaniak_Worley_2002}. As a distinctive predictor, the 5-month lag of AOD is selected under the $\tau = 0.85$ quantile, emphasizing the influence of volcano activities on extreme high temperatures. The model selection result coincides with vast literature in climatology, where ENSO and volcano events are identified as significant sources of variance in global temperature, compared to solar activities and other factors \citep{Lean_Rind_2008, foster2011global}. On the other hand, our conclusion is drawn under general non-stationary assumptions which could be more reliable.


\renewcommand{\arraystretch}{1.2}
\begin{table}
\centering
\caption{Tests of monotonicity (increasing), linearity and constancy.\label{tab:realdatatest}}
\vspace{5pt}
\begin{adjustbox}{width=\columnwidth}
\begin{tabular}{|>{\centering}m{0.12\textwidth}|*{4}{>{\centering\arraybackslash}m{0.14\textwidth}>{\centering\arraybackslash}m{0.11\textwidth}>{\centering\arraybackslash}m{0.06\textwidth}|}}
\hline
 &\multicolumn{3}{c|}{\(\text{Intercept}, \text{Monotonicity}\)}& \multicolumn{3}{c|} {\(\text{Intercept}, \text{Linearity}\)} & \multicolumn{3}{c|} {\(\text{MEI or MEI (lag 2)}, \text{Constancy}\)}& \multicolumn{3}{c|} {\(\text{MEI (lag 6)}, \text{Constancy}\)} \\ 
\hline
 & \(\text{Test Statistic}\) & \({95\%}\text{ C.V.}\)& \(\text{p-val}\) & \(\text{Test Statistic}\)&\({95\%} \text{ C.V.}\)& \(\text{p-val}\)& \(\text{Test Statistic}\)& \({95\%}\text{ C.V.}\)&\(\text{p-val}\) & \(\text{Test Statistic}\) & \({95\%}\text{ C.V.}\)& \(\text{p-val}\)
 \\
\hline
\(L_2\) & 0.132 &1.355 &1& 0.853&0.69 &0.007& 0.217&0.702&0.986& 0.221 & 0.499 & 0.842
\\
\(\tau = 0.15\) &0.157 &1.02 &1&0.886 & 0.647&0.002& 0.401&0.711&0.613&0.4 &0.528&0.244
\\
\(\tau = 0.5\) &0.143  &1.464 &1&0.877 & 0.752&0.013& 0.333&0.729 &0.804&0.216&0.643 & 0.978
 \\
 \(\tau = 0.85\) & 0.293 &1.495&0.999 &0.954 &0.733&0.001 & 0.322& 0.527&0.468&0.431&0.496&0.121
 \\
\hline
\end{tabular}
\end{adjustbox}
\end{table}

\begin{figure}
\centering
\begin{minipage}[b]{0.48\textwidth}
\centering
\includegraphics[width=\textwidth]{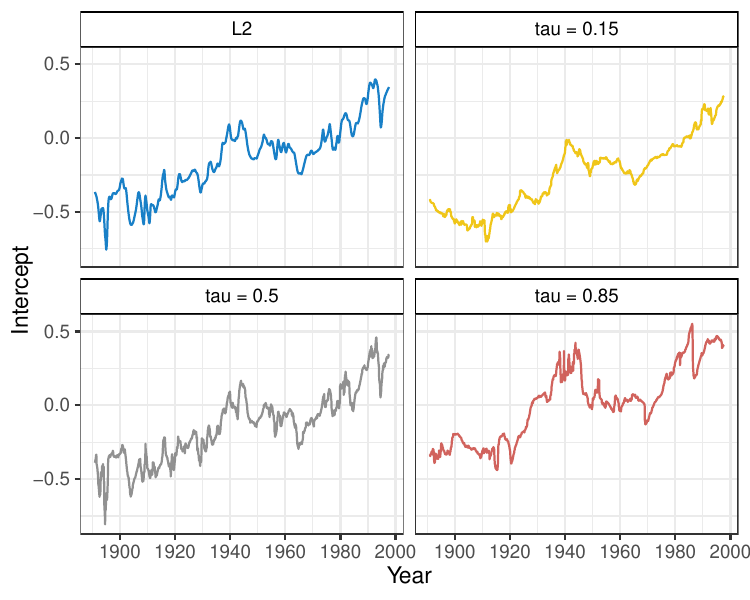}
\caption{\small Estimated warming trend}
\label{fig:intercept}
\end{minipage}
\hfill
\begin{minipage}[b]{0.48\textwidth}
\centering
\includegraphics[width=\textwidth]{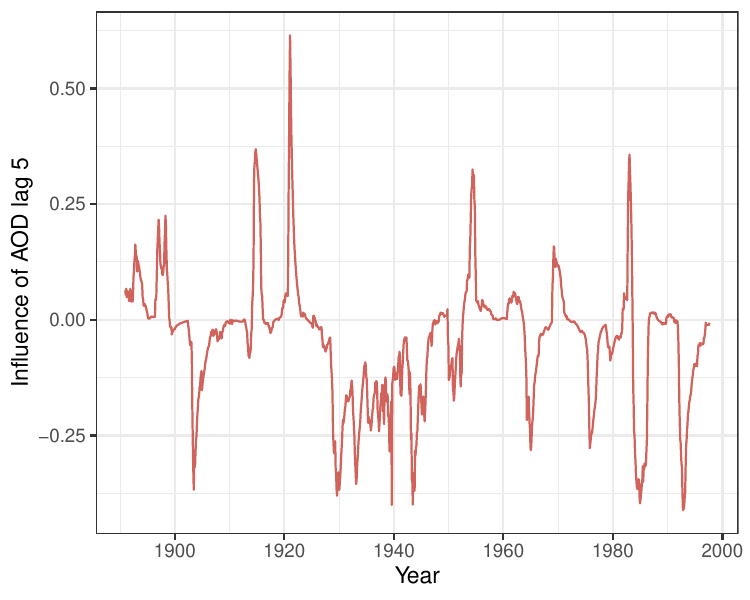}
\caption{\small Influence of the 5-month lag of AOD}
\label{fig:influence_aod}
\end{minipage}
\end{figure}    

We are interested in the hypotheses on the shape of the anthropogenic trend and the coefficients regarding MEI and AOD. It is commonly believed that the anthropogenic trend is increasing, and we can further check whether it is steady over time, which is equivalent to testing whether the intercept is a linear function. As for the coefficients of MEI and AOD, we examine whether they are time-invariant, which amounts to testing the constancy. The test results are summarized in Table \ref{tab:realdatatest}, which demonstrates similar results under different losses. For the shape of the intercept, the positive monotonicity is not rejected, while the linearity is rejected at $5\%$ significance level. This strengthens the belief of discernible human influences on global warming and also suggests that the warming rate is not steady from 1882 to 2005. The non-uniform rate is consistent with the conclusion from \citet{change2007climate}, where global warming is observed to accelerate until 2005 after a cooling period in the 1960s and 1970s. This can also be illustrated by Figure \ref{fig:intercept}, with a relatively flat trend from 1960 to 1970, and a steep increase afterward. While a similar pattern can be detected by solely analyzing the temperature anomalies, our approach rigorously assesses the anthropogenic fluctuations by separating them from the natural ones, making the conclusion more reliable and statistically verifiable. As for the coefficients of MEI and its lags, we fail to reject the null hypothesis of constancy at $5\%$ significance level. This gives us the insight that the overall ENSO effect on global temperature is likely to be steady from 1882 to 2005. 

To represent the temperature change induced by volcano activities, we demonstrate the influence of the 5-month lag of AOD in Figure \ref{fig:influence_aod}, which is computed by multiplying the estimated coefficient and the factor itself. Unlike the traditional multivariate linear regression where only the overall negative effect of AOD can be statistically detected, Figure \ref{fig:influence_aod} reveals the effect of AOD as a combination of heating and long-term cooling. This finding complies with the mechanism of volcano activities, where volcano eruptions can cool the surface due to an increased aerosol loading in the stratosphere that scatters solar radiation back to space. Simultaneously, these eruptions can lead to regional heating, particularly evident during winters in the Northern Hemisphere \citep{Robock_2000, Christiansen_2008}. Such unstable and complicated impact is also confirmed using our Lack-of-fit Test, where the $p$-values of the test of constancy and linearity turn out to be 0.029 and 0.002, respectively.
\subsection{Microsoft Stock Return}
As another application, we incorporate the time-varying M-regression framework with the Fama-French 5-factor (FF5) model to study the monthly return of Microsoft stock from 1986.5 to 2023.10. Initially proposed by \citet{FAMA20151}, the FF5 model aims to capture the size, value, profitability, and investment patterns in average stock returns. There have been subsequent discussions on the time-stability of the regression coefficients of FF5. \citet{racicot2019} accounted for the time-varying nature of the model using a GMM approach, presenting evidence against a static market systematic risk parameter $\beta$. Similarly, the time-varying parameters were examined by \cite{HORVATH2021101848}, especially during selected events such as the dot-com bubble around 2000, the 2008 financial crisis, and the COVID-19 outbreak. \citet{Noda_2022} further argued that the parameters change over time, exhibiting distinct patterns of change across different countries and regions. 

In the meantime, a quantile version of the FF5 model has been discussed in order to analyze the extremes of the stock returns \citep{Allen2011}. Emerging as an alternative to least squares estimation, quantile regression not only looks beyond the conditional mean, but also alleviates some of the statistical problems which plague $L_2$ model including errors-in-variables, omitted variables bias, sensitivity to outliers, and non-normal error distributions \citep{Barnes_Hughes_2002}. 

While the aforementioned alternatives have been primarily studied in a parallel way in the literature, both directions can be accommodated under our framework. Denoting the monthly return of Microsoft stock as $R$, the model is written as
\begin{eqnarray}
    R_i - {R_F}_i = \alpha_i + \beta_{1,i} ({R_M}_i - {R_F}_i) + \beta_{2,i}SMB_i + \beta_{3,i}HML_i + \beta_{4,i}RMW_i + \beta_{5,i}CMA_i + e_i,
\end{eqnarray}
where $R_F$ represents the risk-free rate, $R_M$ is the monthly return on the value-weight (VW) market portfolio, and $SMB, HML, RMW, CMA$ are size factor, value factor, profitability factor, and investment factor, respectively. The data are accessible from \href{https://mba.tuck.dartmouth.edu/pages/faculty/ken.french/data_library.html#Research}{Professor French's website}, and we refer to \citet{FAMA20151} for a detailed explanation of the variables. 

We estimate the regression coefficients under 4 losses: $L_2$ loss, quantile losses with $\tau = 0.15$, $0.5$, and $0.85$. Using the test proposed in \citet{zhou2013heteroscedasticity}, second-order stationarity of the residuals is rejected under all losses with $p$-value $0.0085, 0.0145, 0.0175$ and $0$, suggesting the existence of time non-stationarity in the underlying data. Additionally, following the procedures of LOFT as presented in Section \ref{sec:LOFT}, a joint constancy test is applied to the regression coefficients, which yields a $p$-value of $0.002$, $0.003$, $0.002$, $0.049$ under the 4 losses, respectively. These results confirm the necessity of utilizing a time-varying coefficient model, and the original FF5 model that features static parameters might be misspecified. 

\begin{figure}
\centering
\begin{minipage}[b]{0.48\textwidth}
\centering
\includegraphics[width=\textwidth]{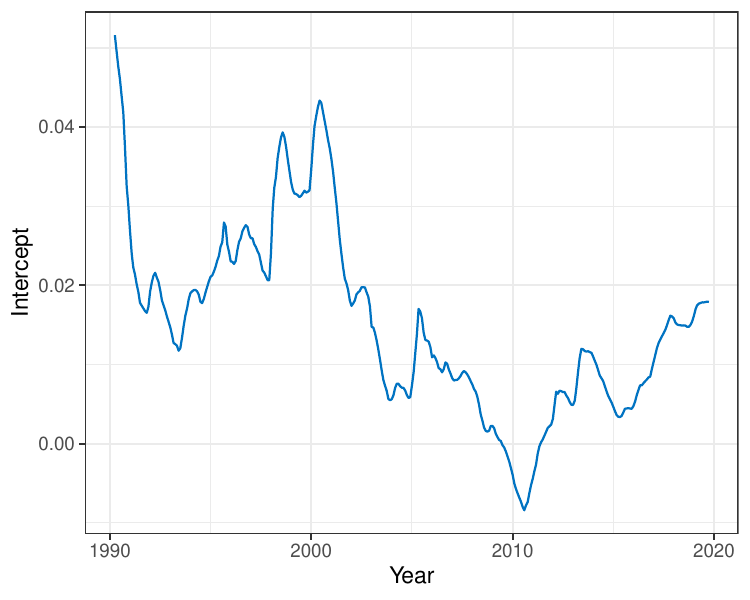}
\caption{\small Estimated intercept $\alpha(t)$}
\label{fig:alpha}
\end{minipage}
\hfill
\begin{minipage}[b]{0.48\textwidth}
\centering
\includegraphics[width=\textwidth]{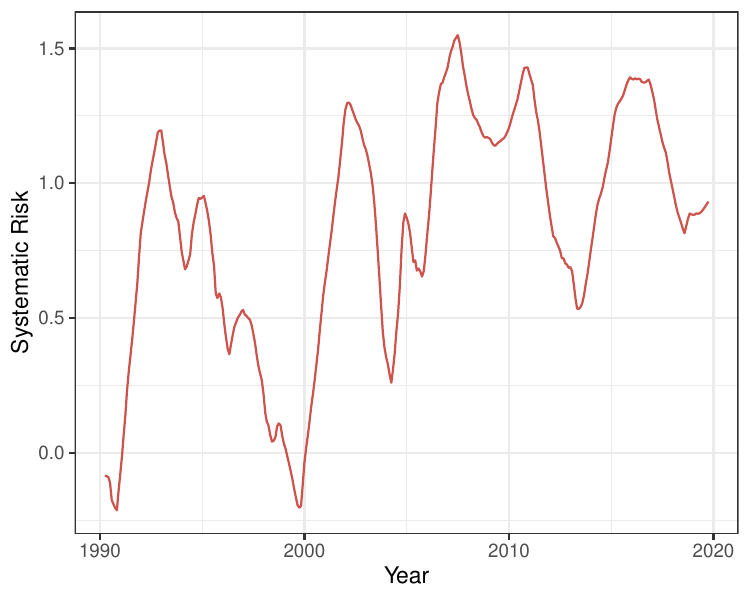}
\caption{\small Estimated systematic risk $\beta_1(t)$}
\label{fig:beta}
\end{minipage}
\end{figure}    
We further focus on the behavior of the estimators under $L_2$ loss, as the intercept $\alpha(t)$ and the market coefficient $\beta_{1}(t)$ have been widely recognized as important indicators for investment. \citet{Jesnsen} proposed the intercept as a measure of performance, where a positive value suggests that the stock tends to outperform the average market. Meanwhile, $\beta_{1}(t)$ is regarded as the measure of systematic risk \citep{sharpe}. 
As is indicated in Figure \ref{fig:alpha}, the estimated intercept is greater than $0$ most of the time, which aligns with the result of non-negativity test with a $p$-value equal to $1$. When it comes to the systematic market risk $\beta_1(t)$, fluctuations and extreme values have been observed from Figure \ref{fig:beta}, especially during the dot-cum bubble around 2000. Additionally, $\beta_1(t)$ is smaller than $1$ most of the time, indicating a relatively lower risk of Microsoft stock compared to the average market.

\begin{table}
\centering
\caption{Tests of significance of the predictors ($p$-values).}
\vspace{5pt}
\begin{adjustbox}{width=0.85\columnwidth}
\begin{tabular}{|>{\centering}m{0.15\textwidth}|*{6}{>{\centering\arraybackslash}m{0.13\textwidth}}|}
\hline
 & {\(\text{Intercept}\)}& {\(R_M - R_F\)} & {\(\text{SMB}\)}& {\(\text{HML}\)} & {\(\text{RMW}\)} & {\(\text{CMA}\)}\\ 
\hline
\(L_2\) & 0 &0 &0& 0&0.482 &0.028
\\
\(\tau = 0.15\) &0 &0.029 &0&0 & 0.976&0
\\
\(\tau = 0.5\) &0  & 0&0&0 & 0.368&0.375
 \\
 \(\tau = 0.85\) & 0 &0&0.68 &0.002 &0.075&0.428 
 \\
\hline
\end{tabular}
\end{adjustbox}
\label{tab:pval_mcsf}
\end{table}

While it remains unclear how to interpret the quantile coefficients in an economic context due to scarce literature on the quantile FF5 model, we gain some insights into the predictors by conducting significance tests on the regression coefficients. Presented in Table \ref{tab:pval_mcsf} are $p$-values of the significance tests. The intercept and the market factor serve as significant predictors under all losses. Similarly, the value factor HML continues to be significant as well. The size factor SMB plays its role under $L_2$ loss and $L_1$ loss with $\tau = 0.15$ and $\tau = 0.5$, but becomes insignificant under the higher quantile $\tau = 0.85$. This suggests that the higher quantile of Microsoft stock return is no longer attributable to the size factor. An opposite pattern can be observed for the profitability factor RMW, where it remains insignificant while starting to make a difference under higher quantile $\tau = 0.85$. Finally, the investment factor CMA might contribute to the stock return under the $L_2$ loss and lower quantile $\tau = 0.15$, but is considered insignificant otherwise with $\tau = 0.5$ or $\tau = 0.85$. 

\section{Assumptions and Auxiliary Theoretical Results}\label{sec:aux}
Let $z_i(t) = y_i - \x_i^\top\betav(t) - \x_i^\top\betav^{\prime}(t)(t_i-t)$, and define
\begin{eqnarray}\label{eq:local_linear_trans}
\tilde{\thetav}_{b_n}(t) := \left(\begin{array}{c}\hat{\thetav}_{b_{n}}(t) \\ \hat{\thetav}_{b_{n}}^{\prime}(t)\end{array}\right)=\argmin_{\cv_0, \cv_1\in\R^p}\sum_{i=1}^n\rho(z_i(t)-\x_i^\top\cv_0 - \x_i^\top\cv_1(t_i-t)/b_n) K_{b_n}(t_i-t),
\end{eqnarray}
where $K_{b_n}(\cdot):=K(\cdot/b_n)$. Compare this with equation \eqref{eq:local_linear}, simple calculations yield that
\begin{eqnarray}\label{eq:estimate_diff}
\hat{\thetav}_{b_{n}}(t) = \hat{\betav}_{b_n}(t) - \betav(t), \quad \hat{\thetav}_{b_{n}}^{\prime}(t) = b_n(\hat{\betav}_{b_{n}}^{\prime}(t)-{\betav}^{\prime}(t)).
\end{eqnarray}
Let $\thetav = (\thetav_0^\top, \thetav_1^{\top})^\top = (\theta_{01}, \theta_{02},\cdots \theta_{0p}, \theta_{11}, \theta_{12}, \cdots, \theta_{1p})^\top$ be a $(2p\times1)$ vector, $\mathbf{w}_{i n}(t)=\left(\mathbf{x}_{i}^{\top}, \mathbf{x}_{i}^{\top}(i / n-t) / b_{n}\right)^{\top}\allowbreak =\left(w_{i n, 1}(t), \ldots, w_{i n, 2 p}(t)\right)^{\top}$. Write $\eta_i(t, \thetav) = \psi(z_i(t) - \bw_{in}^{\top}(t)\thetav)K_{b_n}(t_i-t)$, $S_n(t, \thetav)=\sum_{i=1}^n\eta_i(t,\thetav)\bw_{in}(t)$. We assume $\sup_{t \in (0,1)}S_n(t, \tilde{\thetav}_{b_n}(t)) = O_{\mathbb{P}}(\log n)$. 
\begin{remark}\label{remark: mag_of_Sn}
The assumption $\sup_{t \in (0,1)}S_n(t, \tilde{\thetav}_{b_n}(t)) = O_{\mathbb{P}}(\log n)$ controls the magnitude of $S_n(t, \tilde{\thetav}_{b_n}(t)$, which is used to establish the Bahadur representation. If $\psi(\cdot)$ is continuous, this assumption is automatically satisfied as $\sup_{t \in (0,1)}S_n(t, \tilde{\thetav}_{b_n}(t)) = 0$. While in the case where $\psi(\cdot)$ is discontinuous, the solution may not always exist. 
A key example of such discontinuous $\psi(\cdot)$ arises in quantile regression \cite[Remark 4]{wu2007m} and the proof of it for quantile regression is provided in Lemma \ref{lem:4} of the supplementary.
\end{remark}

\subsection{Model Assumptions}\label{subsec:ass}
For $q \ge 0, j = 0,1,\cdots,r$, define for $t\in (w_j , w_{j+1}]$,
\begin{eqnarray*}
	 \Xi_{j}^{(q)}\left(t, x \mid \mathcal{F}_{k-1}, \mathcal{G}_{k}\right) &=&\dfrac{\partial^{q}}{\partial x^{q}} \mathbb{E}\left\{\psi\left(G_{j}\left(t, \mathcal{F}_{k}, \mathcal{G}_{k}\right)+x\right) \mid \mathcal{F}_{k-1}, \mathcal{G}_{k}\right\}, \crcr 
	 \bar{\kappa}_{j}\left(t, x \mid \GG_{k}\right) &=&\dfrac{\partial}{\partial x} \mathbb{E}\left(\psi\left(G_{j}\left(t, \mathcal{F}_{k}, \mathcal{G}_{k}\right)+x\right) \mid \GG_{k}\right). 
\end{eqnarray*}
\begin{itemize}
\item[(A1)] For $0\le j\le r$, assume that
\begin{eqnarray}\label{eq:A1_1}
	\sup_{t\in (w_j, w_{j+1}), x,y,\in \R}|\Xi_{j}(t,x\mid \FF_{-1},\GG_0) - \Xi_{j}(t,y\mid \FF_{-1},\GG_0)|\crcr
	\le C_{j,1}|x-y| + C_{j,2}|x^2-y^2| + C_{j,3}|x-y|^2,
\end{eqnarray}
where $C_{j,1}, C_{j,2}, C_{j,3}$ are $(\FF_{-1}, \GG_0)$ measurable r.v. with finite fourth moments. We also require that for $0 \le j \le r$, $\Xi_{j}^{(1)}(t,0 \mid \FF_{-1},\GG_0)$ is stochastically Lipschitz continuous for $t \in (w_j, w_{j+1}]$, that is, $\exists M < \infty$, s.t. $\forall t_1, t_2 \in (w_j, w_{j+1}]$, $0 \le j \le r$,
\begin{eqnarray}\label{eq:A1_2}
	||\Xi_{j}^{(1)}(t_1,0\mid \FF_{-1},\GG_0)-\Xi_{j}^{(1)}(t_2,0\mid \FF_{-1},\GG_0)|| \le M|t_1 -t_2|.
\end{eqnarray}

\item[(A2)] 
Assume for the covariate process, $\exists t_x > 0$, s.t. $\max _{1 \leq i \leq n} \E(\exp(t_x |\bx_i|)) \le \infty$. In addition, $\max_{1 \le j \le r}\sup_{t \in (w_j, w_{j+1}]}||\HH_j(t, \GG_i) - \HH_j(t, \GG_i^*)||_1 = O(\chi^{|i|})$ for some constant $\chi \in (0,1)$. For all $t_1, t_2 \in (w_j, w_{j+1}], 0 \le j \le r$, $||\HH_j(t_1, \GG_0) - \HH_j(t_2, \GG_0)|| \le M|t_1 - t_2|$ for some constant $M < \infty$.  Recall the definition of $\GG_i^*$ in Section 2.1.

\item[(A3)] For $\epsilon \in \R$, $0 \le q \le 2p+1$  and any integer $s$, define
\begin{eqnarray*}
	\delta_s^{(q)}(k-1, \epsilon) :=\max_{0 \le j\le r}\sup_{t\in (w_j, w_{j+1}], |x| \le \epsilon}|| \Xi_{j}^{(q)}\left(t, x \mid \mathcal{F}_{k-1}, \mathcal{G}_{k}\right)  -  \Xi_{j}^{(q)}\left(t, x \mid  \mathcal{F}_{k-1}^*, \mathcal{G}_{k}\right) ||_s.
\end{eqnarray*}
We assume that $\exists \epsilon_0 \in \R$, s.t. $\delta_1^{(q)}(k-1, \epsilon_0) = O(\chi^{|k-1|})$ for some constant $0<\chi < 1$. 
\item[(A4)] 
 Define $\bar{\Xi}^{(q)}(x \mid \GG_i) = \frac{\partial^q}{\partial x^q}\E[\psi(e_i + x) \mid \GG_i]$, we require for all $i = 1,2,\cdots,n$ and any $p$-dimensional vector $\mathbf{h}$ that 
\begin{eqnarray*}
	\bar{\Xi}(0 \mid \GG_i) = \E[\psi(e_i) \mid \GG_i] = 0, a.s. \crcr
	\bar{\Xi}(\bx^{\top}_i \mathbf{h} \mid \GG_i) = \bar{\Xi}^{(1)}(0 \mid \GG_i)\bx^{\top}_i \mathbf{h} + O_{\mathbb{P}}(|\bx_i^2||\mathbf{h}|^2),
\end{eqnarray*}
and $\bar{\Xi}^{(1)}(x \mid \GG_i) > 0$ a.s. for $|x| \le \epsilon$ for some $\epsilon > 0$. 
Let $\bar{\lambda}_n(t)$ be the smallest eigenvalue of $\mathbb{E}\{\sum_{i=1}^n \bar{\Xi}^{(1)}(0 \mid \GG_i)\}\bx_i \bx^{\top}_i K_{b_n}(t_i - t)\}$, assume that $\liminf_{n\to \infty} \inf_{t\in(0,1)} \frac{\bar{\lambda}_n(t)}{nb_n} > 0$.
Let $\lambda(t)$ be the smallest eigenvalue of $\Sigma(t):= \E\{\bar{\kappa}_{j}\left(t, 0 \mid \GG_{i}\right)\HH_{j}(t, \GG_i)\HH_{j}^{\top}(t, \GG_i)\}$ if $t \in (w_j, w_{j+1}]$. We require that i) $\max_{0 \le j \le r}\sup_{t \in (w_j ,w_{j+1}]}||\bar{\kappa}_{j}\left(t, 0 \mid \GG_{i}\right)-\bar{\kappa}_{j}\left(t, 0 \mid \GG_{i}^*\right)|| = O(\chi^{|i|})$ for some constant $\chi \in (0,1)$, and $\max_{0 \le j \le r}\sup_{t \in (w_j ,w_{j+1}]}|{\bar{\kappa}}_{j}\left(t, 0 \mid \GG_{0}\right)| \le M$ a.s. for some constant $M < \infty$; ii) $\inf_{t \in (0,1)} \lambda(t) \ge \eta > 0$ for some positive constant $\eta$; iii) $\max_{0 \le j \le r}\sup_{t \in (w_j ,w_{j+1}]}||\dot{\bar{\kappa}}_{j}\left(t, 0 \mid \GG_{i}\right) - \dot{\bar{\kappa}}_{j}\left(t, 0 \mid \GG_{i}^*\right)|| = O(\chi^{|i|})$ for some constant $\chi \in (0,1)$, and $\max_{0 \le j \le r}\sup_{t \in (w_j ,w_{j+1}]}|\dot{\bar{\kappa}}_{j}\left(t, 0 \mid \GG_{0}\right)| \le M$ a.s. for some constant $M < \infty$, where $\dot{\bar{\kappa}}_{j}\left(t, 0 \mid \GG_{i}\right):=\frac{\partial}{\partial t}\bar{\kappa}_{j}\left(t, 0 \mid \GG_{i}\right)$.
\item[(A5)] Assume $\delta_{\W}(k,4)=O(\chi^{k})$ for some $\chi \in (0,1)$, where $\mathbf{W}(t,\FF_i, \GG_i) = \HH_j(t,\GG_i)\psi(G_j(t, \FF_i, \GG_i))$ if $t \in (w_j, w_{j+1}]$, $j = 0,\cdots,r$. Also assume for the error process that $\sup_{i}||\psi(G_j(t, \FF_i, \GG_i))||_4 < \infty$ for all $j = 0 , \cdots, r$ and $t \in (w_j, w_{j+1}]$. Further assume for all $t_1, t_2 \in (w_j, w_{j+1}], 0 \le j \le r$, $||\psi(G_j(t_1, \FF_0, \GG_0))-\psi(G_j(t_2, \FF_0, \GG_0))|| \le M|t_1 - t_2|$ for some constant $M < \infty$. 

\item[(A6)] Define the long-run covariance function of $\{\mathbf{W}_i := \mathbf{W}(t_i,\FF_i, \GG_i)=\bx_i \psi(e_i)\}$
\begin{eqnarray*}
	\Psi(t)=\sum_{i=-\infty}^\infty\mbox{ Cov}[\mathbf{W}(t,\FF_0, \GG_0),\mathbf{W}(t,\FF_i, \GG_i)].
\end{eqnarray*}
Let $\Psi(0)=\lim_{t\downarrow 0}\Psi(t)$. Assume that the minimum eigenvalue of $\Psi(t)$ is bounded away from 0 on $[0,1]$.
\item[(A7)] Define $\Sigma_m = \sum_{i=1}^m \Sigma^{-1}(t_i)\Psi(t_i)\Sigma^{-1}(t_i)$. Assume that, for sufficiently large $n$, there exists $L < \infty$, such that $|\Sigma_{\lfloor s_1 n\rfloor}-\Sigma_{\lfloor s_2 n\rfloor}| \le L (\lfloor s_2 n\rfloor-\lfloor s_1 n\rfloor)$ holds for $0 < s_1 < s_2 < 1$.
\end{itemize}

Here are some insights on the above regularity conditions. Assumption (A1) guarantees the smoothness of the conditional loss functions. \eqref{eq:A1_1} holds if $\sup_{t \in (0,1)}|\psi^{(1)}(t)| < \infty$, which is satisfied by least square regression, and can be achieved by quantile regression under some mild constraints. A sufficient condition for \eqref{eq:A1_1} is provided in \cite{wu2018gradient} by virtue of the robustness of loss functions. (A2) requires that the covariate process is stochastic Lipschitz continuous with geometrically decaying dependence measures. {The moment requirement of $\bx_i$ in (A2) can also be significantly relaxed when $\psi(\cdot)$ is bounded or light-tailed.} (A3) controls the dependence measures of the derivatives of the loss functions' conditional expectations, and is easy to verify for a large class of non-stationary processes, see \cite{wu2018gradient} for further details. (A4) plays a key role in the consistency of $\check{\betav}(t)$ by endowing smoothness on $\bar{\Xi}(\cdot \mid \GG_i)$ to some extent. These conditions altogether enable Bahadur representation of the estimators. Condition (A6) means the time-varying long-run covariance matrices of $\{\mathbf{W}_i\}_{i = 1}^n$ are non-degenerate on $[0,1]$, which is mild in most applications. Condition (A7) assumes Lipschitz continuity of $\Sigma_m$. Observe that $\Sigma_m$ is the integration of $\Sigma^{-1}(t_i)\Psi(t_i)\Sigma^{-1}(t_i)$ and hence the assumption is mild even though both $\Sigma(t)$ and $\Psi(t)$ may experience jumps. (A7) is utilized to guarantee the tightness of the estimators, see Lemma \ref{lem:7} in the supplementary material for more details.

\subsection{Uniform Bahadur Representation}\label{subsec:bahadur}
This section establishes uniform Bahadur representations for the local linear M-estimators and the estimated CRF, by which the limiting distribution of the estimators can be derived in conjunction with some Gaussian approximation results.

Define a $2p \times 2p$ matrix $\Sigma_1(t) = \resizebox{4\baselineskip}{!}{$\begin{pmatrix}
	\Sigma(t) & 0\\ 0 & \mu_2 \Sigma(t)
\end{pmatrix}$}$, where $\Sigma(t)$ is defined in (A4) and $\mu_k := \int |x|^k K(x)\, dx$ for some positive integer $k$. 
\begin{theorem}\label{thm:1}
Suppose (A1)-(A4), $\frac{nb_n ^4}{\log^8 n} \to \infty$, and $n^c b_n \to 0$ for some positive constant $c$. Let $\mathfrak{T}_{n}=\left[b_{n}, 1-b_{n}\right]$, and $\mathfrak{T}_{n}^\prime=\mathfrak{T}_{n} \setminus \cup_{j = 1}^r [w_j - b_n, w_j + b_n]$, then 
\begin{eqnarray}\label{eq:bahadur_original}
\sup_{t \in \mathfrak{T}_{n}^\prime}\left|\Sigma_{1}(t) \tilde{\thetav}_{b_{n}}(t)-\frac{\sum_{i=1}^{n} \psi\left(z_{i}(t)\right) \mathbf{w}_{i n}(t) K_{b_{n}}(i / n-t)}{n b_{n}}\right| = O_{\mathbb{P}}\left(\frac{\pi_n}{\sqrt{nb_n}}\right), 
\end{eqnarray}
where $\pi_n = b_n \log^6 n + (nb_n)^{-\frac{1}{4}}\log ^3 n + \sqrt{nb_n}b_n^3\log^3 n$.
\end{theorem}

The Bahadur representation in \eqref{eq:bahadur_original} involves a bias term involving $\betav^{\prime\prime}(t)$. Recall the Jackknife bias-corrected M-estimator $\check{\Lambda}(t)$ defined in \eqref{eq:jack} and \eqref{eq:hat_crf}. We extend the above Bahadur representation result to the estimated CRF as follows. 

\begin{corol}\label{col:1}
Suppose the conditions of Theorem \ref{thm:1} hold, then for any $s\times p$, $s \le p$ full rank matrix $\mathbf{C}$, $\pi_n = b_n \log^6 n + (nb_n)^{-\frac{1}{4}}\log ^3 n + \sqrt{nb_n}b_n^3\log^3 n$, 
\begin{eqnarray*}\label{eq:partialsum_bahadur}
    \sup_{t \in \mathfrak{T}_{n}}\left|\sqrt{n}\mathbf{C} \left(\check{\Lambda}(t) - {\Lambda}(t)\right)-\frac{1}{\sqrt{n}}\sum_{i=1}^{\lfloor nt\rfloor} \mathbf{C} \Sigma^{-1}\left(t_{i}\right) \psi\left(e_{i}\right) \mathbf{x}_{i}\right|=O_{\mathbb{P}}\left(\frac{\pi_n}{\sqrt{b_n}}\right).
\end{eqnarray*}
\end{corol}

\bigskip
\begin{center}
{\large\bf SUPPLEMENTARY MATERIAL}
\end{center}

\begin{description}
\item[Theoretical proof:] This file contains detailed proof of all theorems. (.pdf file)
\end{description}

\setlength{\bibsep}{2pt}
\bibliographystyle{chicago}
{\refersize \bibliography{Mregression_abbre}}

\makeatletter\@input{tmp_for_main.tex}\makeatother
\end{document}


\def\spacingset#1{\renewcommand{\baselinestretch}%
{#1}\small\normalsize} \spacingset{1}

\if0\blind
{
  \title{\bf Supplementary material for ``Self-convolved Bootstrap for M-regression under Complex Temporal Dynamics"}
  \author{Miaoshiqi Liu\hspace{.2cm}\\
    Department of Statistical Sciences, University of Toronto\\
    and \\
    Zhou Zhou \thanks{Email: \texttt{zhou.zhou@utoronto.ca}}\hspace{.2cm}\\
    Department of Statistical Sciences, University of Toronto}
  \maketitle
} \fi

\if1\blind
{
  \bigskip
  \bigskip
  \bigskip
  \begin{center}
    {\LARGE\bf Supplementary material for ``Self-convolved Bootstrap for M-regression under Complex Temporal Dynamics"}
\end{center}
  \medskip
} \fi

\spacingset{1.45} 
This supplementary material presents additional theoretical results for ``Self-convolved Bootstrap for M-regression under Complex Temporal Dynamics". A brief outline of the proof can be found in Section \ref{sec:outline}, while the comprehensive and detailed proofs of all the theoretical results are deferred to Section \ref{sec:proof}.

\section{Outline of Proofs}\label{sec:outline}
In this section, we provide a sketch of the proofs of Theorem \ref{thm:1} and \ref{thm:3}. The former targets at Bahadur representation and greatly facilitates the analysis on $\check{\betav}(t)$ and $\check{{\Lambda}}(t)$, while the latter quantifies how the estimators can be approximated by the self-convolved bootstrapped. These two theorems, together with Gaussian approximation, trace the blueprint of the methodology we propose. Theorem \ref{thm:2} follows from Corollary \ref{col:1} and Lemma \ref{lem:7}. Theorem \ref{thm:test} is a straightforward result from Theorem \ref{thm:2} and \ref{thm:3}. Detailed proofs are presented in Section \ref{sec:proof} of this supplementary material.

\begin{proof}[Proof of Theorem \ref{thm:1}]
Write 
\begin{eqnarray*}
	\tilde{K}_{n}(t, \thetav):=S_{n}(t, \thetav)-\mathbb{E}\left(S_{n}(t, \thetav) \mid \mathcal{G}_{n}\right)=M_{n}(t, \thetav)+N_{n}(t, \thetav),
\end{eqnarray*}
then Lemma \ref{lem:1} - \ref{lem:3} controls the magnitude of the M-estimator $\tilde{\thetav}_{b_n}(t)$ and derives the upper bound of $\sup_{t\in (0,1)}\left|\tilde{K}_{n}(t, \thetav) - \tilde{K}_{n}(t, 0)\right|$ when $\thetav$ is small. Additionally, under the constraint $\sup_{t\in (0,1)}S_n(t, \tilde{\thetav}_{b_n}(t) = O_{\mathbb{P}}(\log n)$, uniformly on $\mathfrak{T}_n:=[b_n, 1-b_n]$, $S_n(t,0)$ can be well approximated by $\E(S_n(t, 0) \mid \GG_n) - \E(S_n(t, \tilde{\thetav}_{b_n}(t))\mid \GG_n)$. Consider $A_n = \{\sup_{t \in (0,1)}|\tilde{\thetav}_{b_n}(t)| \le \frac{\log^4 n}{\sqrt{n b_n}}+b_n^2 \log n\}$, then $\lim_{n \to \infty}\mathbb{P}(A_n) = 1$ by Lemma \ref{lem:3}. Therefore, on $A_n$, by (A4) and similar argument in Lemma \ref{lem:3}, 
\begin{align*}
& \sup_{t \in \mathfrak{T}_n}\left|[\E(S_n(t, 0) \mid \GG_n) - \E(S_n(t, \tilde{\thetav}_{b_n}(t))\mid \GG_n)] -\sum_{i = 1}^n \bar{\kappa}_{\zeta_i}(t_i, 0 \mid \GG_i)\bw_{in}(t)\bw_{in}^{\top}(t)\tilde{\thetav}_{b_n}(t)K_{b_n}(t_i - t)\right| \\
& O_{\mathbb{P}}(\log^{11}n + nb_n^5 \log^5 n).
\end{align*}
Furthermore, write $\bar{\Lambda}_n(t) = \frac{1}{nb_n}\sum_{i =1}^n \bar{\kappa}_j(t,0 \mid \GG_i)\bw_{in}(t)\bw_{in}^{\top}(t)K_{b_n}(t_i - t)$ when $t \in (w_j, w_{j+1}]$, by (A2)(A4), one can show that 
\begin{eqnarray*}
	\sup_{t \in \mathfrak{T}_n^\prime}|\bar{\Lambda}_n(t) - \Sigma_1(t)| =  O_{\mathbb{P}}(b_n).
\end{eqnarray*}
Then the theorem follows from elementary calculations.
\end{proof}
\begin{proof}[Proof of Theorem \ref{thm:3}]
    Let $\mu$ defined as in Theorem \ref{thm:3}, denote
\begin{eqnarray*}
	\mathbf{S}_i &=& \Sigma^{-1}(t_i + b_n)\sum_{k=1}^n \mathbf{x_k}\psi(e_k) K_{b_n}({t_k - t_i - b_n}) + \Sigma^{-1}(t_i - b_n)\sum_{k=1}^n \mathbf{x_k}\psi(e_k) K_{b_n}({t_k - t_i + b_n})\cr
	&-& 2 \Sigma^{-1}(t_i)\sum_{k=1}^n \mathbf{x_k}\psi(e_k) K_{b_n}({t_k - t_i}).
\end{eqnarray*}
For appropriately chosen $b_n$, the uniform Bahadur representation shows that
\begin{eqnarray*}
	\max_{\lceil 2nb_n \rceil \le i \le n - \lceil 2nb_n \rceil }\left|\sqrt{\frac{b_n}{\mu}}(\check{\betav}(t_i + b_n) + \check{\betav}(t_i - b_n)-2\check{\betav}(t_i))-\frac{1}{\sqrt{\mu}}\frac{1}{n\sqrt{b_n}}\mathbf{S}_i\right| = O_{\mathbb{P}}\left(\frac{\pi_n}{\sqrt{n}}+b_n^{\frac{5}{2}}\right),
\end{eqnarray*}
which enables us to connect the bootstrapped sample $ \boldsymbol{\Phi}_{n,j}^{o}$ to $\mathbf{S}_i$. Define $\Lambda_{r, s} = \frac{1}{n^2 b_n}\sum_{i=r}^s \mathbf{S}_i\mathbf{S}_i^{\top}$, then conditional on $\{(\mathbf{x}_i,y_i)\}_{i = 1}^n$, the conditional covariance of $\boldsymbol{\Phi}_{n,j}^{o}$ can be well expressed by $\Lambda_{r, s}$.
Combining Lemma \ref{lem:8} to Lemma \ref{lem:11}, we have
\begin{eqnarray*}
	\left\| \max _{\lceil 2n b_n\rceil \leq r \leq s \leq n-\lceil 2n b_n\rceil}\left|\frac{1}{\mu} \Lambda_{r, s} -\int_{r / n}^{s / n} \Sigma^{-1}(u)\Psi(u)\Sigma^{-1}(u)\right|\right\| = O(\sqrt{b_n}),
\end{eqnarray*}
which indicates that the covariance of the bootstrapped sample is uniformly close to that of a Gaussian process. \\
The result follows from deriving the magnitude of $\Delta$ and Theorem C.1 of \cite{zhou2020frequency_sup}.
\end{proof}

\section{Proofs}\label{sec:proof}
In this section, the symbol $C$ represents a generic finite constant that may vary from place to place. 
For $q \ge 0, j = 0,1,\cdots,r$, define for $t\in (w_j , w_{j+1}]$,
\begin{eqnarray*}
	 F_{j}^{(q)}\left(t, x \mid \mathcal{F}_{k-1}, \mathcal{G}_{k}\right) &=&\frac{\partial^{q}}{\partial x^{q}} \mathbb{E}\left\{\mathbf{1}\left(G_{j}\left(t, \mathcal{F}_{k}, \mathcal{G}_{k}\right) \leq x\right) \mid \mathcal{F}_{k-1}, \mathcal{G}_{k}\right\}, \crcr
	  F_{j}^{(q)}\left(t, x \mid \mathbf{x}_{k}\right) &=&\frac{\partial^{q}}{\partial x^{q}} \mathbb{E}\left\{\mathbf{1}\left(G_{j}\left(t, \mathcal{F}_{k}, \mathcal{G}_{k}\right) \leq x\right) \mid \mathbf{x}_{k}\right\}, \crcr 
	  f_{j}\left(t, x \mid \mathcal{F}_{k-1}, \mathcal{G}_{k}\right) &=&F_{j}^{(1)}\left(t, x \mid \mathcal{F}_{k-1}, \mathcal{G}_{k}\right), \quad f_{j}\left(t, x \mid \mathbf{x}_{k}\right)=F_{j}^{(1)}\left(t, x \mid \mathbf{x}_{k}\right) .
\end{eqnarray*}
For $\delta >0$ and $z_i(t) = y_i - \x_i^\top\betav(t) - \x_i^\top\betav^{\prime}(t)(t_i-t)$, let
\begin{equation}
    \begin{aligned}
    \tau_n(\delta) := \mathbb{E}\{\sup_{t\in(0,1)}\sum_{i=1}^n |\mathbf{x}_i|^2 \left[\psi(z_i(t) + |\mathbf{x}_i|\delta) - \psi(z_i(t) - |\mathbf{x}_i|\delta)\right]^2K^2_{b_n}(t_i - t)\},
    \end{aligned}
\end{equation}
and for the rest of this paper, assume $\tau_n(\delta) = O(nb_n\delta)$ for $\delta \to 0$. Without the assumption, via the same techniques, all the results can still be established but with more factors of logarithms involved.

\begin{remark}\label{rm: magnitude of tau_n}
The assumption $\tau_n(\delta) = O(nb_n\delta)$ for $\delta \to 0$ can be satisfied by most commonly-used loss functions. Based on the detailed discussion in Remark 2.1 of \cite{wu2018gradient_sup}, for quantile regression, if for $i = 1, 2, \cdots, n$, the conditional densities satisfy
\begin{equation*}
    \max_{1\le j \le r}\sup_{t\in(w_j, w_{j+1}]}|f_j(t,x \mid \x_i)| \le M_0 < \infty, \quad a.s.,
\end{equation*}
then we have $\tau_n(\delta) = O(nb_n\delta)$. Similarly, for the Huber, $L_2$ and expectile regressions, it can be proved that $\tau_n(\delta) = O(nb_n\delta^2) \le O(nb_n\delta)$ when $\delta \to 0$. For robust $L_q$ regression where $1 < q < 2$, 
\begin{equation}
    \tau_n(\delta)=\left\{
\begin{array}{lcl}
O(nb_n\delta^{2q-1}) & & {1 < q < \frac{3}{2}}\\
O(nb_n \delta^2 \log(\delta^{-1})) & & {q = \frac{3}{2}}\\
O(nb_n\delta^2) & & {\frac{2}{3} < q < 2}
\end{array} \right. ,
\end{equation}
then for $\delta$ sufficiently small, we also have $\tau_n(\delta) = O(nb_n \delta)$. 
\end{remark}
Write 
\begin{eqnarray}\label{eq:Kn}
	\tilde{K}_{n}(t, \thetav):=S_{n}(t, \thetav)-\mathbb{E}\left(S_{n}(t, \thetav) \mid \mathcal{G}_{n}\right)=M_{n}(t, \thetav)+N_{n}(t, \thetav),
\end{eqnarray}
\begin{eqnarray}\label{eq:Mn}
	M_n(t, \thetav) = \sum_{i=1}^n [\eta_i(t, \thetav)\bw_{in}(t)-\E(\eta_i(t, \thetav) \mid \FF_{i-1},\GG_n)\bw_{in}(t)],
\end{eqnarray}
\begin{eqnarray}\label{eq:Nn}
	N_n(t, \thetav) = \sum_{i=1}^n[\E(\eta_i(t, \thetav) \mid \FF_{i-1},\GG_n)\bw_{in}(t)-\E(\eta_i(t, \thetav) \mid \GG_n)\bw_{in}(t)].
\end{eqnarray}
Define $l_n(t) = \max(\lfloor n t-nb_n\rfloor, 1)$, $s_n(t) = \min(\lfloor nt + nb_n \rfloor, n)$, and $\NN_n(t) = \{i \in \mathbb{Z}: l_n(t) \le i \le s_n(t)\} $, $\NN_n(t-) = \NN_n(t) \cap (0, nt]$, and $\NN_n(t+) = \NN_n(t) \cap (nt, n]$.

\begin{lemma}\label{lem:1}
Assume (A1), (A2), (A3) and (A4), $\gamma_n \to 0$, $b_n \to 0$, $nb_n^2 \to \infty$, then $$\sup_{t\in(0,1), |\thetav| \le \gamma_n}|M_n(t, \thetav) - M_n(t,0)|=O_\p(\sqrt{nb_n\gamma_n}\log n+n^{-3}).$$
\end{lemma}
\begin{proof}
Let $w_{in,l}(t)$ be the $l$-th component of $\bw_{in}(t)$, write
\begin{eqnarray*}
	M_{nl}(t, \thetav) = \sum_{i=1}^n [\eta_i(t,\thetav) - \E(\eta_i(t, \thetav)\mid \FF_{i-1},\GG_{n})]w_{in,l}(t ), 1\le l \le p,\crcr
	M_{nl1}(t, \thetav) = \sum_{i\in \NN_n(t+)}^n [\eta_i(t,\thetav) - \E(\eta_i(t, \thetav)\mid\FF_{i-1},\GG_{n})]w_{in,l}(t), p+1 \le l \le 2p, \crcr
	M_{nl2}(t, \thetav) = \sum_{i\in \NN_n(t-)}^n [\eta_i(t,\thetav) - \E(\eta_i(t, \thetav)\mid \FF_{i-1},\GG_{n})]w_{in,l}(t), p+1 \le l \le 2p,
\end{eqnarray*}
then it suffices to show Lemma 1 with $M_n(t,\thetav)$ therein replaced by $M_{nl}(t, \thetav)$ for $1 \le l \le p$ and $M_{nl1}(t, \thetav)$, $M_{nl2}(t, \thetav)$ for $p+1 \le l \le 2p$. We consider the case of $M_{nl1}(t, \thetav), l = p+1$ for presentation clarity. Let $g_n$ be a real sequence which goes to infinity arbitrarily slow with $g_n \ge 3$ for all $n$, and $u_n = \sqrt{nb_n\gamma_n}g_n/ \log(g_n)$, $\phi_n = \sqrt{nb_n\gamma_n}g_n \log(n), l = p+1$,
\begin{eqnarray*}
	A_n(t) = \max_{i\in \NN_n(t+)}\sup_{|\thetav|\le \gamma_n}|(\eta_i(t,\thetav)-\eta_i(t,0))w_{in,l}(t)|,
\end{eqnarray*}
\begin{eqnarray*}
	U_n(t) = \sum_{i\in \NN_n(t+)} \E\{(\psi(z_i(t)+|\bw_{in}(t)||\gamma_n|)- \psi(z_i(t)-|\bw_{in}(t)||\gamma_n|))^2 \mid \FF_{i-1}, \GG_{n}\}w_{in,l}(t)^2K_{b_n}^2(t_i-t).
\end{eqnarray*}
By monotonicity of $\psi(\cdot)$, we have
\begin{eqnarray}
	&\sup_{|\thetav|\le \gamma_n}|(\eta_i(t,\thetav)-\eta_i(t,0))w_{in,l}(t)| \cr
	&\le |w_{in,l}(t)|K_{b_n}(t_i-t)[\psi(z_i(t)+|\bw_{in}(t)|\gamma_n)- \psi(z_i(t)-|\bw_{in}(t)|\gamma_n)].
\end{eqnarray}
Observe that
\begin{equation}
    \begin{aligned}
    \mathbb{E}[\sup_{t\in(0,1)}A_n^2(t)] & \le \mathbb{E}\{\sup_{t\in(0,1)}\sum_{i=1}^n |\mathbf{x}_i|^2 \left[\psi(z_i(t) + |\mathbf{x}_i|\gamma_n) - \psi(z_i(t) - |\mathbf{x}_i|\gamma_n)\right]^2K^2_{b_n}(t_i - t)\} = \tau_n(\gamma_n),
    \end{aligned}
\end{equation}
consequently,
\begin{eqnarray}\label{eq:An}
	\mathbb{P}[\sup_{t\in(0,1)}A_n(t)\ge u_n]\le u_n^{-2}\E(\sup_{t \in (0,1)}A_n(t)^2) = O\left(\left(\log g_{n} / g_{n}\right)^{2}\right)=o(1).
\end{eqnarray}
On the other hand, we can also show that 
$\mathbb{E}[\sup_{t\in(0,1)}U_n(t)] \le \tau_n(\gamma_n) = O(nb_n\gamma_n)$, thus
\begin{eqnarray}\label{eq:Un}
	\mathbb{P}[\sup _{t \in(0,1)} U_{n}(t) \geq u_{n}^{2}]=O\left(\left(\log g_{n} / g_{n}\right)^{2}\right)=o(1).
\end{eqnarray}
Let $\Pi_p = \{-1,+1\}^p$, for $i \in \N$, define $D_{\bx}(i) = (2\times I(x_{i,1}\ge 0)-1,\cdots,2\times I(x_{i,p}\ge 0)-1) \in \Pi_p$, and $M_{nl1,\mathbf{d}}(t,\thetav) = \sum_{i \in \NN_n(t+)}[\eta_i(t,\thetav) - \E(\eta_i(t,\thetav)\mid \FF_{i-1},\GG_n)]w_{in,l}I(D_{\bx}(i) =\bd)$. By the fact that $M_{nl1}(t,\thetav) = \sum_{\bd \in \Pi_p} M_{nl1,\mathbf{d}}(t,\thetav) $, it suffices to prove the lemma with $M_{nl1,\mathbf{d}}(t,\thetav) $ for all $\bd \in \Pi_p$. Write
\begin{eqnarray*}
	\zeta_{i,\bd,l}(t,\thetav) = (\eta_i(t,\thetav)-\eta_i(t,0))w_{in,l}I(D_{\bx}(i) =\bd),\crcr
	B_n(t,\thetav) = \sum_{i \in \NN_n(t+)} \E\{\zeta_{i,\bd,l}(t,\thetav)  I(|\zeta_{i,\bd,l}(t,\thetav)|\ge u_n)\mid \FF_{i-1},\GG_n\},
\end{eqnarray*}
hence for large $n$, by our choice of $u_n, \phi_n$, since $u_n = o(\phi_n)$,
\begin{eqnarray}\label{eq:Bn}
	\mathbb{P}[\sup_{t\in(0,1)}|B_n(t,\thetav)|\ge \phi_n, \sup_{t\in(0,1)}U_n(t)\le u_n^2] \le \mathbb{P}[\sup_{t\in(0,1)}U_n(t)\ge \phi_n u_n, \sup_{t\in(0,1)}U_n(t)\le u_n^2] = 0. 
	\crcr
\end{eqnarray}
Let $q = n^{10}$ and $\mathfrak{G}_{q}=\left\{\left(k_{1} / q, \ldots, k_{2 p} / q\right): k_{i} \in \mathbb{Z},\left|k_{i}\right| \leq n^{10}\right\} \cap\left[-\gamma_{n}, \gamma_{n}\right]^{2 p}$, $\mathfrak{H}_{q}=\{k_{p+1}/q:0\le k_{p+1} \le n^{10}, k_{p+1}\in \Z\}$, so there are at most $(2n^{10}+1)^{2p}\times (n^{10}+1)$ points in $\fG_q \times \fH_q$. Combining \eqref{eq:An}\eqref{eq:Un}\eqref{eq:Bn}, the similar argument of \citet{wu2007m_sup} and \citet{freedman1975tail}'s exponential inequality for martingale differences, we have that
\begin{eqnarray}\label{eq:mart_diff}
\mathbb{P}\{\sup _{\thetav \in \mathfrak{G}_{q}, t \in \mathfrak{H}_{q}}\left|M_{n l 1, \mathbf{d}}(t, \thetav)-M_{n l 1, \mathbf{d}}(t, 0)\right| \geq 2 \phi_{n}, \sup _{t \in(0,1)} A_{n}(t) \leq u_{n}, \sup _{t \in(0,1)} U_{n}(t) \leq u_{n}^{2}\}\crcr
\leq\left(2 n^{10}+1\right)^{2 p} \times\left(n^{10}+1\right) O\left(\exp \left\{-\phi_{n}^{2} /\left(4 u_{n} \phi_{n}+2 u_{n}^{2}\right)\right\}\right).
\end{eqnarray}
Consequently, by \eqref{eq:An}\eqref{eq:Un}\eqref{eq:mart_diff} and the definition of $u_n, \phi_n$, we have
\begin{eqnarray}
	\lim_{n\to \infty} \mathbb{P}\left\{\sup _{\thetav \in \mathfrak{G}_{q}, t \in \mathfrak{H}_{q}}\left|M_{n l 1, \mathbf{d}}(t, \thetav)-M_{n l 1, \mathbf{d}}(t, 0)\right| \geq 2 \phi_{n}\right\} = 0.
\end{eqnarray}
For any $a\in \R$, define $\langle a\rangle_{q,-1}=\lceil a\rceil_{q}=\lceil a q\rceil / q$, $\langle a\rangle_{q, 1}=\lfloor a\rfloor_{q}=\lfloor a q\rfloor / q$. For presentation clarity we consider $\bd =(-1,-1,1,1,\cdots,1)$, write $\langle \thetav \rangle_{q,\bd} = (\lceil \theta_{01} \rceil_q, \lceil \theta_{02} \rceil_q, \lfloor \theta_{03} \rfloor_q,\cdots, \lfloor \theta_{0p} \rfloor_q, \allowbreak \lceil \theta_{11} \rceil_q, \lceil \theta_{12} \rceil_q, \lfloor \theta_{13} \rfloor_q, \allowbreak \cdots, \lfloor \theta_{1p} \rfloor_q)$ and $\langle \thetav \rangle_{q,-\bd} = (\lfloor \theta_{01} \rfloor_q, \lfloor \theta_{02} \rfloor_q, \lceil \theta_{03} \rceil_q,\cdots, \lceil \theta_{0p} \rceil_q, \lfloor \theta_{11} \rfloor_q, \lfloor \theta_{12} \rfloor_q,\allowbreak \lceil \theta_{13} \rceil_q,\cdots, \lceil \theta_{1p} \rceil_q)$. By the monotonicity of $\psi(\cdot)$, for $i \in \NN_n(t+), t\in \fH_q, \thetav \in \fG_q, l = p+1$,
\begin{eqnarray}
	\zeta_{i, \bd, l}\left(t,\langle\thetav\rangle_{q, -\bd}\right) \leq \zeta_{i, \bd, l}(t, \thetav) \leq \zeta_{i, \bd, l}\left(t,\langle\thetav\rangle_{q,\bd}\right)
\end{eqnarray}
Note that by (A1) and the magnitude assumption of $\x_i$ in (A2),
\begin{eqnarray}\label{eq:zeta}
	&&\sup_{|\thetav| \le \gamma_n} \sum_{i \in \NN_n(t+)} |\E[\zeta_{i, \bd, l}(t, \thetav)-\zeta_{i, \bd, l}\left(t,\langle\thetav\rangle_{q, \bd}\right)\mid \FF_{i-1},\GG_n]| \cr
	&\le & \sup_{|\thetav| \le \gamma_n}\sum_{i \in \NN_n(t+)}|\E[\psi(z_i(t) - \bw_{in}^{\prime}(t)\thetav) - \psi(z_i(t)-\bw_{in}^{\prime}(t)\langle\thetav\rangle_{q, \bd})\mid \FF_{i-1},\GG_n ]||\bw_{in}t)|K_{b_n}(t_i - t) \crcr
	&\le &\sup_{|\thetav| \le \gamma_n}\sum_{i \in \NN_n(t+)}[C_{\zeta_i,1}\frac{|\bw_{in}(t)|}{q} + C_{\zeta_i,2}\frac{|\bw_{in}(t)|}{q}|2\x_i^\top[\betav(t_i) -\betav(t) - \betav^{\prime}(t)(t_i-t)]-\bw_{in}^{\prime}(t)(\thetav+\langle\thetav\rangle_{q, \bd})| \crcr
	&+&C_{\zeta_i,3}\left(\frac{|\bw_{in}(t)|}{q}\right)^2]|\bw_{in}t)|K_{b_n}(t_i - t) \crcr
	& \le & C\sum_{i \in \NN_n(t+)} D_i /q,
\end{eqnarray}
where $D_i \in \mathcal{L}_2$ is a $(\FF_{i-1}, \GG_i)$ measurable random variable. Similar inequality to \eqref{eq:zeta} holds for $\langle\thetav\rangle_{q, -\bd}$. By similar arguments in Lemma 5 of \citet{zhou2009local} and the fact that $g_n \to \infty$ arbitrarily slowly, we can show that 
\begin{eqnarray}\label{eq:tr1}
	\sup _{|\thetav| \le \gamma_n, t \in \mathfrak{H}_{q}}\left|M_{n l 1, \mathbf{d}}(t, \thetav)-M_{n l 1, \mathbf{d}}(t, 0)\right|  = O_{\mathbb{P}}(\sqrt{nb_n\gamma_n}\log n + n^{-4}).
\end{eqnarray}
Let $t_k = k/q, k = 0,1,\cdots,q$. By similar chaining argument, we can get 
\begin{eqnarray}\label{eq:tr2}
	\max _{0 \leq k \leq q^{-1}} \sup _{0<t-t_{k}<q^{-1},|\thetav| \leq \gamma_{n}}\left|M_{n l 1, \mathbf{d}}(t, \thetav)-M_{n l 1, \mathbf{d}}\left(t_{k}, \thetav\right)\right|=o_{\mathbb{P}}\left(n^{-3}\right).
\end{eqnarray}
Then by triangle inequality and \eqref{eq:tr1}\eqref{eq:tr2}, the lemma holds. 
\end{proof}

\begin{lemma}\label{lem:2}
Under the conditions of Lemma \ref{lem:1}, we have 
\begin{eqnarray*}
	\sup_{t\in(0,1), |\thetav| \le \gamma_n}|N_n(t,\thetav) - N_n(t,0)| = O_{\mathbb{P}}(\sqrt{nb_n}\gamma_n\log^{2p+\frac{7}{2}}n).
\end{eqnarray*}
\end{lemma}
\begin{proof}
Let $J=\{\alpha_1, \alpha_2, \cdots, \alpha_q\} \subseteq \{1,2,\cdots,2p\}$ be a nonempty set and $1 \le \alpha_1 < \alpha_2 \cdots < \alpha_{q} \le 2p$. For a $2p$-dimensional vector $\bu = (u_1,u_2,\cdots,u_{2p})$, let $\bu_{J} = (u_1I(1 \in J), \cdots,u_{2p}I(2p \in J))$, then write
\begin{eqnarray*}
	\int_{0}^{\thetav_{I}}\frac{\partial^{q} N_{n}\left(t, \mathbf{u}_{J}\right)}{\partial \mathbf{u}_{J}} d \mathbf{u}_{J} = \int_{0}^{\theta_{\alpha_1}}\cdots \int_{0}^{\theta_{\alpha_q}}\frac{\partial^{q} N_{n}\left(t, \mathbf{u}_{J}\right)}{\partial u_{\alpha_1}\partial u_{\alpha_2}\cdots \partial u_{\alpha_q}} d  u_{\alpha_1}\cdots  d u_{\alpha_q}.
\end{eqnarray*}
Let $\mathbf{W}_{in,J}(t) = \bw_{in}(t)w_{in,\alpha_1}(t)\cdots w_{in,\alpha_q}(t)$, we have that for $1\le q \le 2p$,
\begin{eqnarray}\label{eq:partial_derivative}
	\frac{\partial^{q} N_{n}\left(t, \mathbf{u}_{J}\right)}{\partial \mathbf{u}_{J}} &=& \sum_{i=1}^n \sum_{k = 1}^{\infty}\PP_{i-k,n}[\E^{(q)}(\psi(z_i(t) - \bw_{in}^{\top}(t)\bu_J)\mid\FF_{i-1},\GG_n)]\mathbf{W}_{in,J}(t)K_{b_n}(t_i-t), \qquad
\end{eqnarray}
where $\PP_{i-k,n}(\cdot) = \mathbb{E}(\cdot \mid \FF_{i-k},\GG_n) - \mathbb{E}(\cdot \mid \FF_{i-k-1},\GG_n)$.
By triangle inequality, we have
\begin{eqnarray}\label{eq:Nn_control}
    && \sup_{t\in(0,1), |\thetav| \le \gamma_n}|N_n(t,\thetav) - N_n(t,0)| \crcr
    && \le  \sum_{J \subseteq \{1,\cdots, 2p\}}\int_{-\gamma_n}^{\gamma_n}\cdots \int_{\gamma_n}^{\gamma_n}\sup_{t\in(0,1)}\left|\frac{\partial^{q} N_{n}\left(t, \mathbf{u}_{J}\right)}{\partial u_{\alpha_1}\partial u_{\alpha_2}\cdots \partial u_{\alpha_q}}\right| d  u_{\alpha_1}\cdots  d u_{\alpha_q}. 
\end{eqnarray}
By Burkholder inequality, for $s > 1$,
\begin{eqnarray}\label{eq:burkholder}
	||\sum_{i=1}^n \PP_{i-k,n}[\E^{(q)}(\psi(z_i(t) - \bw_{in}^{\top}(t)\bu_J)\mid\FF_{i-1},\GG_n)]\mathbf{W}_{in,J}(t)K_{b_n}(t_i-t)||^2_s\crcr
	\le s\sum_{i=1}^n ||\PP_{i-k,n}[\E^{(q)}(\psi(z_i(t) - \bw_{in}^{\top}(t)\bu_J)\mid\FF_{i-1},\GG_n)]\mathbf{W}_{in,J}(t)K_{b_n}(t_i-t)||^2_s.
\end{eqnarray}
Since $z_i(t) = e_i + \x_i^\top[\betav(t_i) -\betav(t) - \betav^{\prime}(t)(t_i-t)]$, define $\Delta(i,t, \bu_J) = \x_i^\top[\betav(t_i) -\betav(t) - \betav^{\prime}(t)(t_i-t)]-\bw_{in}^{\top}(t)\bu_J$, we have $\psi(z_i(t) - \bw_{in}^{\top}(t)\bu_J)=\psi(e_i + \Delta(i,t, \bu_J))$. 
Let $|\bu_{J}| \le \gamma_n$, then $|\bw_{in}^{\top}(t)\bu_J| \le C\gamma_n \to 0$, also since $\betav(\cdot) \in {\cal C}^2[0,1]$, we have
\begin{eqnarray*}
	|\x_i^\top[\betav(t_i) -\betav(t) - \betav^{\prime}(t)(t_i-t)]| \le C b_n^2 \to 0,
\end{eqnarray*}
which implies that $|\Delta(i,t, \bu_J)| \le \epsilon_0$ for large n. 
Note that $\E[\psi(e_i + \Delta(i,t, \bu_J)) \mid \FF_{i-1}, \GG_{i}] = \E[\psi(e_i + \Delta(i,t, \bu_J)) \mid \FF_{i-1}, \GG_{n}]$, by using the similar argument of Lemma 1 in \citet{wu2007m_sup} and Holder's inequality, under condition (A3), we shall have
\begin{eqnarray}\label{eq:proj_s}
	&&||\PP_{i-k,n}[\E^{(q)}(\psi(z_i(t) - \bw_{in}^{\top}(t)\bu_J)\mid\FF_{i-1},\GG_n)]\mathbf{W}_{in,J}(t)||_s \crcr
	&& \le \sup_{t \in (0,1)}||\mathbf{W}_{in,J}(t)||_{2s}	||\PP_{i-k,n}[\E^{(q)}(\psi(z_i(t) - \bw_{in}^{\top}(t)\bu_J)\mid\FF_{i-1},\GG_n)]||_{2s} \crcr
	&& \le \sup_{t \in (0,1)}||\mathbf{W}_{in,J}(t)||_{2s}\delta_{2s}(k-1, \epsilon_0).
\end{eqnarray}
By Cauchy-Schwarz inequality, Proposition 6 in \citet{wu2017nonparametric_sup} and (A2), we know that $\sup_{t \in (0,1)}\max_{1 \le i \le n}||\mathbf{W}_{in,J}(t)||_{2s} \le C s^{2p+1}$. Also by Condition (A3) and similar argument of Proposition 5 of \citet{wu2017nonparametric_sup}, we have $\delta_{2s}(k-1, \epsilon_0)=O(\chi^{\frac{k-1}{2s}})$. Consequently, by \eqref{eq:partial_derivative}\eqref{eq:burkholder}\eqref{eq:proj_s} and the triangle inequality, for some large constant $C$,\begin{eqnarray}\label{eq:snorm}
	\sup _{\left|\mathbf{u}_{J}\right| \leq \gamma_{n}}\left\|\frac{\partial^{q} N_{n}\left(t, \mathbf{u}_{J}\right)}{\partial \mathbf{u}_{J}}\right\|_{s} \leq C \sqrt{n b_{n}} s^{2 p+3 / 2} \frac{1}{1-\chi^{\frac{1}{2 s}}}=O\left(\sqrt{n b_{n}} s^{2 p+5 / 2}\right).
\end{eqnarray}
Now consider $\frac{\partial}{\partial t}\frac{\partial^{q} N_{n}\left(t, \mathbf{u}_{J}\right)}{\partial \mathbf{u}_{J}}$. Based on \eqref{eq:partial_derivative}, elementary calculations suggests that
\begin{eqnarray}
	\frac{\partial}{\partial t}\frac{\partial^{q} N_{n}\left(t, \mathbf{u}_{J}\right)}{\partial \mathbf{u}_{J}} &= &\sum_{i=1}^n \sum_{k = 1}^{\infty}\PP_{i-k,n}\frac{\partial}{\partial t}\{[\E^{(q)}(\psi(z_i(t) - \bw_{in}^{\top}(t)\bu_J)\mid\FF_{i-1},\GG_n)]\mathbf{W}_{in,J}(t)K_{b_n}(t_i-t)\}\crcr
	&=& \sum_{i=1}^n \sum_{k = 1}^{\infty}\PP_{i-k,n}\{[\E^{(q)}(\psi(z_i(t) - \bw_{in}^{\top}(t)\bu_J)\mid\FF_{i-1},\GG_n)]\mathbf{R}_1(i,n,J,t) \cr
	&+& [\E^{(q+1)}(\psi(z_i(t) - \bw_{in}^{\top}(t)\bu_J)\mid\FF_{i-1},\GG_n)]\mathbf{R}_2(i,n,J,t)\},
\end{eqnarray}
where $\mathbf{R}_1(i,n,J,t) = \frac{\partial}{\partial t}[\mathbf{W}_{in,J}(t)K_{b_n}(t_i-t)]$, $\mathbf{R}_2(i,n,J,t) = \frac{\partial}{\partial t}[\Delta(i,t, \bu_J)]\times \mathbf{W}_{in,J}(t)K_{b_n}(t_i-t)$. Then similar to \eqref{eq:snorm}, elementary calculation shows that
\begin{eqnarray}
	\sup _{\left|\mathbf{u}_{J}\right| \leq  \gamma_{n}}\left\|\frac{\partial}{\partial t}\frac{\partial^{q} N_{n}\left(t, \mathbf{u}_{J}\right)}{\partial \mathbf{u}_{J}}\right\|_{s} =O\left(\frac{\sqrt{n b_{n}} s^{2 p+7 / 2}}{b_n}\right).
\end{eqnarray}
According to Proposition 1 of \citet{wu2017nonparametric_sup}, we have
\begin{eqnarray}
	\sup _{\left|\mathbf{u}_{J}\right| \leq \gamma_{n}}\left\|\sup _{t \in(0,1)}\left|\frac{\partial^{q} N_{n}\left(t, \mathbf{u}_{J}\right)}{\partial \mathbf{u}_{J}}\right|\right\|_{s}=O\left(\sqrt{n b_{n}} b_{n}^{-1 / s} s^{2 p+7 / 2}\right).
\end{eqnarray}
Combining with \eqref{eq:Nn_control}, the lemma follows by taking $s = \lfloor\log \frac{1}{b_n} \rfloor$.
\end{proof}

\begin{lemma}\label{lem:3}
Suppose (A1)-(A4), $b_n \to 0$, $nb_n^4/ \log ^7n \to \infty$, then we have  
\begin{eqnarray*}
	\sup_{t \in (0,1)}|\tilde{\thetav}_{b_n}(t)| \le_{\mathbb{P}}\left(n b_{n}\right)^{-1 / 2} \log ^{4} n+b_{n}^{2} \log n.
\end{eqnarray*}
\end{lemma}
\begin{proof}
	Let $g_n = \left(n b_{n}\right)^{-1 / 2} \log ^{4} n+b_{n}^{2} \log n$. Write $\Theta_n(t,\thetav) = \sum_{i = 1}^n [\rho(z_i(t) - \bw^{\top}_{in}(t)\thetav) - \rho(z_i(t))]K_{b_n}(t_i - t)$, and 
\begin{eqnarray}
	A_{n}(t, \thetav)=-\sum_{i=1}^{n} \int_{0}^{1} \mathbb{E}\left[\psi\left(z_{i}(t)-\mathbf{w}_{i n}^{\top}(t) \thetav s\right) \mid \mathcal{G}_{n}\right] \mathbf{w}_{i n}^{\top}(t) \thetav K_{b_{n}}(t_i-t) \;d s.
\end{eqnarray}
By the fact that $\rho(z_i(t)) - \rho(z_i(t) - \bw^{\top}_{in}(t)\thetav) = \int _{0}^1 \psi(z_i(t) - \bw^{\top}_{in}(t)\thetav s)\bw^{\top}_{in}(t)\thetav \;ds$, we have
\begin{eqnarray}
	\sup_{t\in(0,1), |\thetav| \le g_n }|\Theta_n(t,\thetav)-A_{n}(t, \thetav)| = \sup_{t\in(0,1), |\thetav| \le g_n}|\int_{0}^1 \thetav^{\top}\tilde{K}_n(t, \thetav s)\;ds|,
\end{eqnarray}
where $\tilde{K}_n(t, \thetav)$ is defined in \eqref{eq:Kn}. Via the similar argument of Lemma \ref{lem:1} and Lemma \ref{lem:2}, $\allowbreak \sup_{t \in (0,1)}|\tilde{K}_n(t, 0)| \allowbreak = O_{\mathbb{P}}(\sqrt{nb_n}\log^{\frac{7}{2}}n)$. Consequently, we get
\begin{eqnarray*}
	\sup_{t\in(0,1), |\thetav| \le g_n}|\tilde{K}_n(t, \thetav)| \le \sup_{t\in(0,1), |\thetav| \le g_n}|\tilde{K}_n(t, \thetav)- \tilde{K}_n(t, 0)| + \sup_{t \in (0,1)}|\tilde{K}_n(t, 0)|  = O_{\mathbb{P}}(\sqrt{nb_n}\log^{\frac{7}{2}}n),
\end{eqnarray*}
which implies that
\begin{eqnarray*}
	\sup_{t\in(0,1), |\thetav| \le g_n}|\Theta_n(t,\thetav)-A_{n}(t, \thetav)| =O_{\mathbb{P}}(\sqrt{nb_n}g_n\log^{\frac{7}{2}}n).
\end{eqnarray*}
On the other hand, since $z_i(t) = e_i + \x_i^\top[\betav(t_i) -\betav(t) - \betav^{\prime}(t)(t_i-t)]$, $\mathbf{w}_{i n}^{\top}(t) \thetav s = \bx^{\top}_i(\thetav_0s + \thetav_1\frac{t_i - t}{b_n}s)$, we have $z_i(t) -\mathbf{w}_{i n}^{\top}(t) \thetav s = e_i + \bx^{\top}_i[\betav(t_i) -\betav(t) - \betav^{\prime}(t)(t_i-t)-\thetav_0s - \thetav_1\frac{t_i - t}{b_n}s]:= e_i + \bx^{\top}_i\Lambda_n(t,i,\thetav,s)$. Then by (A4), let $|\thetav| = g_n$, 
\begin{eqnarray}
	&& A_n(t, \thetav) = \int_{0}^1-\sum_{i=1}^n {\bar{\Xi}(\bx^{\top}_i\Lambda_n(t,i,\thetav,s)\mid \GG_i)}\mathbf{w}_{i n}^{\top}(t) \thetav K_{b_{n}}(t_i-t) \;d s \crcr
	 && =  \int_{0}^1\sum_{i=1}^n[\bar{\Xi}^{(1)}(0\mid \GG_i)\bx^{\top}_i(-\Lambda_n(t,i,\thetav,s))+ O_{\mathbb{P}}(|\bx_i^2||\Lambda_n(t,i,\thetav,s)|^2)]\mathbf{w}_{i n}^{\top}(t) \thetav K_{b_{n}}(t_i-t) \;d s. \qquad
	\end{eqnarray}
Elementary calculations shows that $\betav(t_i) -\betav(t) - \betav^{\prime}(t)(t_i-t) = O((t_i - t)^2)$. Note that by our choice of $g_n$, $g_n/b_n^2 \to \infty$ as $n \to \infty$, which implies that $|\Lambda_n(t,i,\thetav,s)| = O(g_n)$ when $|\thetav| = g_n$. By (A4) and similar argument of Lemma 2.1 in \cite{wu2018gradient_sup}, there exists some constant $\iota >0$ such that 
\begin{eqnarray}
	\inf_{t \in (0,1), |\thetav|=g_n}A_n(t, \thetav) \ge_{\mathbb{P}} \iota nb_n g_n^2.
\end{eqnarray}
Therefore, 
\begin{eqnarray}
	\inf_{t \in (0,1), |\thetav|=g_n}\Theta_n(t, \thetav) &\ge& \inf_{t \in (0,1), |\thetav|=g_n}A_n(t, \thetav)-\sup_{|\thetav| \le g_n, t\in(0,1)}|\Theta_n(t,\thetav)-A_{n}(t, \thetav)|\crcr
	&\ge_{\mathbb{P}}& \iota nb_n g_n^2-\sqrt{nb_n}g_n\log^{\frac{7}{2}}n
\end{eqnarray}
Hence $\inf_{|\thetav| = g_n, t\in(0,1)}\Theta_n(t, \thetav) \ge_{\mathbb{P}} \epsilon \log^8n$ for some small positive $\epsilon$. By convextity of $\Theta_n(t, \cdot)$,
\begin{eqnarray}
	\left\{\inf _{t \in(0,1),|\thetav|=g_{n}} \Theta_{n}(t, \thetav)\right\}=\inf _{t \in(0,1)}\left\{\inf _{|\thetav|=g_{n}} \Theta_{n}(t, \thetav)\right\} \crcr =\inf _{t \in(0,1)}\left\{\inf _{|\thetav| \geq g_{n}} \Theta_{n}(t, \thetav)\right\}=\left\{\inf _{t \in(0,1),|\thetav| \geq g_{n}} \Theta_{n}(t, \thetav)\right\}.
\end{eqnarray}
Observe that $\tilde{\thetav}_{b_n} (t)$ is the minimizer of $\Theta_n(t, \thetav)$, we obtain $\sup_{t \in (0,1)}|\tilde{\thetav}_{b_n}(t)| \le_{\mathbb{P}} g_n = \left(n b_{n}\right)^{-1 / 2} \log ^{4} n+b_{n}^{2} \log n$.
\end{proof}

\begin{lemma}\label{lem:4}
Assume $\max\limits_{0 \le j \le r}\sup\limits_{t \in (w_j, w_{j+1}], x \in \R}|f_j(t,x \mid \FF_{-1}, \GG_0)| < \infty$ and (A2). Then for quantile regression with $\rho_{\tau}(x)=\tau x^{+}+(1-\tau)(-x)^{+}$, we have $\sup_{t \in (0,1)}S_n(t, \tilde{\thetav}_{b_n}(t))=O_{\mathbb{P}}(\log n)$.
\begin{proof}
	The lemma can be shown by similar techniques to Lemma 8 of \cite{zhou2009local} and Lemma 4 of \cite{wu2017nonparametric_sup}. Based on their arguments, following Lemma A.2 of \cite{ruppert1980trimmed}, we just need to show $\sup_{t\in (0,1)}\sum_{i=1}^n |K_{b_n}(t_i - t)\bw_{in}(t)|I(y_i = \bx_i^{\top}\hat{\betav}(t)+\bx_i^{\top}\hat{{\betav}^\prime}(t)(t_i - t)) = O_{\mathbb{P}}(\log n)$. Note that
	\begin{eqnarray}\label{ieq:Sn bound}
		&&\sup_{t\in (0,1)}\sum_{i=1}^n |K_{b_n}(t_i - t)\bw_{in}(t)|I(y_i = \bx_i^{\top}\hat{\betav}(t)+\bx_i^{\top}\hat{{\betav}^\prime}(t)(t_i - t)) \cr
		&\le & C(\sup_{t\in (0,1)}|\bw_{in}(t)|)(\sup_{t \in (0,1)}\sum_{i=1}^n I(y_i = \bx_i^{\top}\hat{\betav}(t)+\bx_i^{\top}\hat{{\betav}^\prime}(t)(t_i - t)) ),
	\end{eqnarray}
where $\sup_{t\in (0,1)}|\bw_{in}(t)| = O_\mathbb{P}(\log n)$ by (A2) and Proposition 6 in \citet{wu2017nonparametric_sup}. By \cite{babu1989strong}, 
\begin{eqnarray}\label{ieq:ind_bound}
	\sup_{t \in (0,1)}\sum_{i=1}^n I(y_i = \bx_i^{\top}\hat{\betav}(t)+\bx_i^{\top}\hat{{\betav}^\prime}(t)(t_i - t)) \le \sup_{a_1, a_2 \in \R}\sum_{i=1}^n I(y_i = a_1 + a_2 \frac{i}{n}) =O_{\mathbb{P}}(1).
\end{eqnarray}
Therefore the lemma follows from \eqref{ieq:Sn bound} and \eqref{ieq:ind_bound}. 
\end{proof}
\end{lemma}

\subsubsection*{Proof of Theorem \ref{thm:1}} 
\begin{proof}
Write $\bar{\Lambda}_n(t) = \frac{1}{nb_n}\sum_{i =1}^n \bar{\kappa}_j(t,0 \mid \GG_i)\bw_{in}(t)\bw_{in}^{\top}(t)K_{b_n}(t_i - t)$ when $t \in (w_j, w_{j+1}]$, then by elementary calculation, (A2) and (A4), one can show that
\begin{eqnarray}\label{eq:exp_sigma}
	\sup_{t \in \mathfrak{T}_n^\prime}|\E(\bar{\Lambda}_n(t)) - \Sigma_1(t)| = O(b_n).
\end{eqnarray}
Similar to the arguments in Proposition 2 of \citet{wu2017nonparametric_sup}, by (A2)(A4), we have for $v \ge 1$ and some $\tilde{\chi} \in (0,1)$,
\begin{eqnarray}
	||\bar{\Lambda}_n(t) - \E(\bar{\Lambda}_n(t))||_v = O_{\mathbb{P}}\left(\frac{v^2 \sqrt{v}}{\sqrt{n b_n}(1 - \tilde{\chi}^{\frac{1}{2v}})}\right)
\end{eqnarray}
and 
\begin{eqnarray}
	\left\|\frac{\partial}{\partial t}(\bar{\Lambda}_n(t) - \E(\bar{\Lambda}_n(t)))\right\|_v = O_{\mathbb{P}}\left(\frac{v^2 \sqrt{v}}{\sqrt{n b_n}b_n(1 - \tilde{\chi}^{\frac{1}{2v}})}\right).
\end{eqnarray}
Take $v = \lfloor \log \frac{1}{b_n} \rfloor$, then by Proposition 1 of \citet{wu2017nonparametric_sup} and \eqref{eq:exp_sigma}, we get 
\begin{eqnarray}\label{proposition3}
	\sup_{t \in \mathfrak{T}_n^\prime}|\bar{\Lambda}_n(t) - \Sigma_1(t)| =  O_{\mathbb{P}}(b_n).
\end{eqnarray}
Combining Lemma \ref{lem:1}, Lemma \ref{lem:2}, Lemma \ref{lem:3} and the constraint $\sup_{t\in (0,1)}S_n(t, \tilde{\thetav}_{b_n}(t) = O_{\mathbb{P}}(\log n)$, we have
\begin{equation}\label{eq:thm1_1}
\begin{aligned}
      &\sup_{t \in \mathfrak{T}_n}\left|S_n(t, 0) -\E(S_n(t, 0) \mid \GG_n) + \E(S_n(t, \tilde{\thetav}_{b_n}(t))\mid \GG_n)\right| \\
      &= O_{\mathbb{P}}((n b_n)^{1/4}\log ^3 n + n^{1/2}b_n^{3/2}\log ^{3/2}n). 
\end{aligned}
\end{equation}
Consider $A_n = \{\sup_{t \in (0,1)}|\tilde{\thetav}_{b_n}(t)| \le \frac{\log^4 n}{\sqrt{n b_n}}+b_n^2 \log n\}$, then $\lim_{n \to \infty}\mathbb{P}(A_n) = 1$ by Lemma \ref{lem:3}. \\
Observe that $\bar{\kappa}_{\zeta_i}(t_i, x \mid \GG_i) = \bar{\Xi}^{(1)}(x \mid \GG_i)$, then by (A4) and similar argument in Lemma \ref{lem:3}, on the set $A_n$, we have
\begin{equation}\label{eq:thm1_modified}
\begin{aligned}
	&\sup_{t \in \mathfrak{T}_n}\left|[\E(S_n(t, 0) \mid \GG_n) - \E(S_n(t, \tilde{\thetav}_{b_n}(t))\mid \GG_n)]-\sum_{i = 1}^n \bar{\kappa}_{\zeta_i}(t_i, 0 \mid \GG_i)\bw_{in}(t)\bw_{in}^{\top}(t)\tilde{\thetav}_{b_n}(t)K_{b_n}(t_i - t)\right| \\
 	& =  O_{\mathbb{P}}(\log^{11}n + nb_n^5 \log^5 n).
\end{aligned}
\end{equation}
Note that $\bar{\kappa}_j(t,x \mid \GG_i) = \E(\Xi_{j}^{(1)}(t,x\mid \FF_{i-1}, \GG_i) \mid \GG_i)$, thus by stochastic lipschitz continuity of $\Xi_{j}^{(1)}(t,0\mid \FF_{-1}, \GG_0)$ in (A1), we also know that $\bar{\kappa}_j(t,0 \mid \GG_i)$ is stochastically lipschitz continuous on $(w_j, w_{j+1}]$ for $j = 0,1,\cdots,r$. Let $\zeta(t) = j$ if $t \in (w_j ,w_{j+1}]$, on the set $A_n$, we have
\begin{equation}\label{eq:thm1_4}
\begin{aligned}
  	&\sup_{t \in \mathfrak{T}_n}\left|\sum_{i = 1}^n\left[ \bar{\kappa}_{\zeta_i}(t_i, 0 \mid \GG_i)-\bar{\kappa}_{\zeta(t)}(t, 0 \mid \GG_i)\right]\bw_{in}(t)\bw_{in}^{\top}(t)\tilde{\thetav}_{b_n}(t)K_{b_n}(t_i - t)\right| \\
	&= O_{\mathbb{P}}((nb_n^2+r)[(nb_n)^{-1/2}\log^6 n+b_n^2\log^3 n])\\
	&= O_{\mathbb{P}}(\sqrt{nb_n}b_n\log^6 n+nb_n^4\log^3 n).  
\end{aligned}
\end{equation}
By \eqref{proposition3}\eqref{eq:thm1_1}\eqref{eq:thm1_modified}\eqref{eq:thm1_4} and Proposition A.1 of \citet{wu2018gradient_sup}, the theorem follows.
\end{proof}

\begin{lemma}\label{lem:5}
	Suppose the conditions of Theorem 1 hold. Then
	\begin{eqnarray}
		\sup_{t \in \mathfrak{T}_n^\prime}\left|\sqrt{n b_n}(\hat{\betav}_{b_n}(t)-\betav(t) -\frac{\mu_2 b_n^2}{2}\ddot{\betav}(t))-\Sigma^{-1}(t)\frac{\sum_{i = 1}^n \psi(e_i)\bx_i K_{b_n}(t_i - t)}{\sqrt{nb_n}}\right| = O_{\mathbb{P}}\left(\pi_n\right),
	\end{eqnarray}
	where $\pi_n = b_n \log^6 n + (nb_n)^{-\frac{1}{4}}\log ^3 n + \sqrt{nb_n}b_n^3\log^3 n$.
\end{lemma}
\begin{proof}
Write $\eta(i,t) = (\psi(z_i(t)) - \psi(e_i))K_{b_n}(t_i - t)$. Define $\mathbf{M}_n(t) = \sum_{i = 1}^n [\eta(i,t)-\E(\eta(i,t) \mid \FF_{i-1},\GG_n)]\bw_{in}(t)$, $\mathbf{N}_n(t) = \sum_{i = 1}^n [\E(\eta(i,t) \mid \FF_{i-1},\GG_n)-\E(\eta(i,t) \mid \GG_n)]\bw_{in}(t)$, by similar argument of Lemma \ref{lem:1} and Lemma \ref{lem:2}, we know that
\begin{eqnarray}
	\sup_{t \in \mathfrak{T}_n}|\mathbf{M}_n(t)| = O_{\mathbb{P}}(\sqrt{nb_n} b_n \log n),\quad \sup_{t \in \mathfrak{T}_n}|\mathbf{N}_n(t)| = O_{\mathbb{P}}\left(\sqrt{nb_n} b_n^2 \log ^{2p+\frac{7}{2}}n\right).
\end{eqnarray}
Then by elementary calculations and (A4), we have
\begin{eqnarray}
	\sup_{t \in \mathfrak{T}_n}\frac{|\sum_{i=1}^n \left[(\psi(z_i(t)) - \psi(e_i))\bw_{in}(t)-\E\{\psi(z_i(t))\bw_{in}(t)\mid \GG_n\}\right]K_{b_n}(t_i - t)|}{nb_n} = O_{\mathbb{P}}\left(\frac{b_n \log n}{\sqrt{nb_n}}\right). 
\end{eqnarray}
Let $\ddot{B}(t) = (\ddot{\betav}(t)^{\top}, \mathbf{0}_{p\times 1}^{\top})^{\top}$ be a $2p \times 1$ vector, $\tilde{\Sigma}(t) = \begin{pmatrix}
	\mu_2 \Sigma(t) & 0 \\ 0 & \mu_4 b_n ^2 \Sigma(t)
\end{pmatrix}$ be a $2p \times 2p$ matrix. According to (A4), elementary calculation shows that 
\begin{eqnarray}
	\sup _{t \in \mathfrak{T}_{n}} \frac{\left|\sum_{i=1}^{n}\left[\mathbb{E}\left\{\psi\left(z_{i}(t)\right) \mid \mathcal{G}_{i}\right\}-\bar{\kappa}_{\zeta_i}(t_i, 0 \mid \GG_i) \mathbf{x}_{i}^{\top} \ddot{\betav}(t)\frac{(t_i-t)^{2}}{2}\right] \mathbf{w}_{i n}(t) K_{b_{n}}(t_i-t)\right|}{nb_n} = O_{\mathbb{P}}(b_n^3 \log ^2n).
\end{eqnarray}
Also note that
\begin{eqnarray}
	&&\frac{1}{nb_n}\sum_{i=1}^n \bar{\kappa}_{\zeta_i}(t_i, 0 \mid \GG_i) \mathbf{x}_{i}^{\top} \ddot{\betav}(t)(t_i-t)^{2}\mathbf{w}_{i n}(t) K_{b_{n}}(t_i-t)\crcr
	&& = \left\{\frac{1}{nb_n}\sum_{i=1}^n \bar{\kappa}_{\zeta_i}(t_i, 0 \mid \GG_i) (t_i-t)^{2}\mathbf{w}_{i n}(t)\bw_{in}^{\top}(t)K_{b_{n}}(t_i-t)\right\}\ddot{B}(t) := \tilde{\Lambda}_n(t)\ddot{B}(t).
\end{eqnarray}
Similar to the argument of \eqref{proposition3}, we can show that
\begin{eqnarray}
	\sup_{t \in \mathfrak{T}_n^\prime}|\tilde{\Lambda}_n(t) - b_n^2\tilde{\Sigma}(t)| = O_{\mathbb{P}}(b_n^3),
\end{eqnarray}
which further implies
\begin{eqnarray}
	\sup_{t \in \mathfrak{T}_n^\prime}\left|\frac{1}{nb_n}\sum_{i=1}^{n}\mathbb{E}\left\{\psi\left(z_{i}(t)\right) \mid \mathcal{G}_{i}\right\}\bw_{in}(t)K_{b_n}(t_i - t) - \frac{b_n^2\tilde{\Sigma}(t)\ddot{B}(t)}{2}\right| = O_{\mathbb{P}}(b_n^3\log^2 n).
\end{eqnarray}
Combining the equations above with Theorem 1, the result follows.
\end{proof}

\begin{lemma}\label{lem:6}
Suppose the conditions of Theorem 1 hold. Then
\begin{eqnarray}
    \max_{\lceil nb_n\rceil \le j \le n-\lceil nb_n \rceil}\left|\sum_{i =\lceil nb_n\rceil }^j \Sigma^{-1}(t_i)\frac{\sum_{k = 1}^n \psi(e_k)\bx_k K_{b_n}(t_i - t_k)}{{nb_n}} - \sum_{i =1}^j \Sigma^{-1}(t_i)\psi(e_i)\bx_i\right| = O_{\mathbb{P}}(\sqrt{nb_n}).
\end{eqnarray}
\end{lemma}
\begin{proof}
\begin{eqnarray*}
    &&\sum_{i = \lceil nb_n\rceil }^j \Sigma^{-1}(t_i)\frac{\sum_{k = 1}^n \psi(e_k)\bx_k K_{b_n}(t_i - t_k)}{{nb_n}}\crcr
    &=& \sum_{i = \lceil nb_n\rceil }^j \sum_{k=1}^{n} \Sigma^{-1}(t_k)\frac{\mathbf{x}_{k} \psi(e_{k})K_{b_{n}}\left(t_{k}-t_i\right) }{nb_n}+ \sum_{i = \lceil nb_n\rceil }^j  \sum_{k=1}^{n}[\Sigma^{-1}(t_i)-\Sigma^{-1}(t_k)]\frac{\mathbf{x}_{k} \psi(e_{k})K_{b_{n}}\left(t_{k}-t_i\right) }{nb_n} \cr
    &:=&  I(t_j)+I\!I(t_j).
\end{eqnarray*}
\begin{eqnarray*}
I\!I(t_j)&=& \sum_{t_i \in \cup_{m = 1}^{r}[w_m-b_n,w_m + b_n]}^j  \sum_{k=1}^{n}[\Sigma^{-1}(t_i)-\Sigma^{-1}(t_k)]\frac{\mathbf{x}_{k} \psi(e_{k})K_{b_{n}}\left(t_{k}-t_i\right) }{nb_n} \cr &+& \sum_{t_i \notin \cup_{m = 1}^{r}[w_m-b_n,w_m + b_n]}^j  \sum_{k=1}^{n}[\Sigma^{-1}(t_i)-\Sigma^{-1}(t_k)]\frac{\mathbf{x}_{k} \psi(e_{k})K_{b_{n}}\left(t_{k}-t_i\right) }{nb_n} := I\!I\!I(t_j)+I\!V(t_j).
\end{eqnarray*}
Observe that $\mathbb{E}\left([\Sigma^{-1}(t_i)-\Sigma^{-1}(t_k)]\frac{\mathbf{x}_{k} \psi(e_{k})K_{b_{n}}\left(t_{k}-t_i\right) }{nb_n}\right) = 0$, then under condition (A5), by Lemma 6(ii) of \cite{zhou2013heteroscedasticity_sup}, we have
\begin{eqnarray*}
&||\sum_{k=1}^{n}[\Sigma^{-1}(t_i)-\Sigma^{-1}(t_k)]\frac{\mathbf{x}_{k} \psi(e_{k})K_{b_{n}}\left(t_{k}-t_i\right) }{nb_n}||_{4} \cr &= ||\sum_{k=i_{*}}^{i^*} [\Sigma^{-1}(t_i)-\Sigma^{-1}(t_k)]\frac{\mathbf{x}_{k} \psi(e_{k})K_{b_{n}}\left(t_{k}-t_i\right) }{nb_n}||_4 = O\left(\frac{1}{\sqrt{nb_n}}\right).
\end{eqnarray*}
Therefore, 
\begin{eqnarray}\label{eq: third}
\max_{1\le j\le n}| I\!I\!I(t_j)| &\le & \sum_{t_i \in \cup_{m = 1}^{r}[w_m-b_n,w_m + b_n]}^n \left|\sum_{k=1}^{n}[\Sigma^{-1}(t_i)-\Sigma^{-1}(t_k)]\frac{\mathbf{x}_{k} \psi(e_{k})K_{b_{n}}\left(t_{k}-t_i\right) }{nb_n}\right|\cr
&\le & 2rnb_n\cdot O_{\mathbb{P}}\left(\frac{1}{\sqrt{nb_n}}\right) = O_{\p}(\sqrt{nb_n}).
\end{eqnarray}
From the piecewise stochastic Lipschitz  continuity of $\bar{\kappa}_j(t,0 \mid \GG_i)$ and $\HH_j(t, \GG_i)$, we know that $\Sigma(t)$ is also piecewise stochastic Lipschitz continuous. Then for $t_i \notin \cup_{m = 1}^{r}[w_m-b_n,w_m + b_n]$ and $t_i - t_k \in [-b_n, b_n]$, we have $|\Sigma(t_i)-\Sigma(t_k)| = O_{\mathbb{P}}(b_n)$. In this case, by Lemma 6(ii) of \cite{zhou2013heteroscedasticity_sup}, we have
\begin{eqnarray}\label{eq: fourth}
&&\max_{1\le j\le n}\left| I\!V(t_j)\right| \cr
&= & \max_{1\le j\le n}\left|\sum_{t_i \notin \cup_{m = 1}^{r}[w_m-b_n,w_m + b_n]}^j  \sum_{k=1}^{n}[\Sigma^{-1}(t_i)-\Sigma^{-1}(t_k)]\frac{\mathbf{x}_{k} \psi(e_{k})K_{b_{n}}\left(t_{k}-t_i\right) }{nb_n}\right|\cr
&\le & \sum_{t_i \notin \cup_{m = 1}^{r}[w_m-b_n,w_m + b_n]}^n b_n\cdot \max_{1 \le i \le n}\left|\Sigma^{-1}(t_i)\right|\cdot \left|\sum_{k=1}^n \frac{[\Sigma(t_k) -\Sigma(t_i)]}{b_n}\Sigma^{-1}(t_k)\frac{\mathbf{x}_{k} \psi(e_{k})K_{b_{n}}\left(t_{k}-t_i\right) }{nb_n}\right| \crcr
&=& nb_n\cdot O_{\mathbb{P}}\left(\frac{1}{\sqrt{nb_n}}\right) = O_{\p}(\sqrt{nb_n}).
\end{eqnarray}
Next consider $I(t_j)$, we approximate it under $2$ different situations: $\lceil nb_n \rceil  \le j  \le  \lceil nb_n \rceil+2\lfloor nb_n\rfloor$, and $j >\lceil nb_n \rceil+2\lfloor nb_n\rfloor$.
\begin{eqnarray*}
	I(t_j) &=&  \sum_{k=1}^{n} \Sigma^{-1}(t_k) \mathbf{x}_{k} \psi(e_{k})\sum_{i = \lceil nb_n \rceil}^j \dfrac{1}{nb_n}K_{b_{n}}\left(t_{k}-t_i\right)\cr
	&=&\left[\sum_{k=1}^{\lceil nb_n \rceil-1}+ \sum_{k=\lceil nb_n \rceil}^{j}+\sum_{k=j+1}^{n\wedge(j+\lfloor nb_n \rfloor)}\right]\Sigma^{-1}(t_k) \mathbf{x}_{k} \psi(e_{k})\sum_{i = \lceil nb_n \rceil}^j \dfrac{1}{nb_n}K_{b_{n}}\left(t_{k}-t_i\right),
\end{eqnarray*}
therefore, 
\begin{eqnarray}\label{eq: indicator}
	&&\left|I(t_j) - \sum_{k = \lceil nb_n \rceil}^j \Sigma^{-1}(t_k)\psi(e_k)\bx_k\right| \cr
	&\le &  \left|\sum_{k=1}^{\lceil nb_n \rceil-1}\Sigma^{-1}(t_k) \mathbf{x}_{k} \psi(e_{k})\sum_{i = \lceil nb_n \rceil}^j \dfrac{1}{nb_n}K_{b_{n}}\left(t_{k}-t_i\right)\right| \crcr
	&+& \left|\sum_{k=\lceil nb_n \rceil}^{j}\Sigma^{-1}(t_k) \mathbf{x}_{k} \psi(e_{k})\left[\left(\sum_{i = \lceil nb_n \rceil}^j \dfrac{1}{nb_n}K_{b_{n}}\left(t_{k}-t_i\right)\right)-1\right]\right|\crcr
	&+& \left|\sum_{k=j+1}^{j+\lfloor nb_n \rfloor}\Sigma^{-1}(t_k) \mathbf{x}_{k} \psi(e_{k})\sum_{i = \lceil nb_n \rceil}^j \dfrac{1}{nb_n}K_{b_{n}}\left(t_{k}-t_i\right)\right|.
\end{eqnarray}
Under the first case, $\lceil nb_n \rceil  \le j  \le  \lceil nb_n \rceil+2\lfloor nb_n\rfloor$. Since $\int_{-\infty}^{+\infty} K(u) d u=1$, then the Riemann sum satisfies $ \dfrac{1}{nb_n}\sum_{i = k-\lfloor nb_n \rfloor}^{k+\lfloor nb_n \rfloor}K_{b_{n}}\left(t_{k}-t_i\right)=1+O\left(\frac{1}{n}\right)$, which implies that the aforementioned $ \frac{1}{nb_n}\sum_{i = \lceil nb_n \rceil}^jK_{b_{n}}\left(t_{k}-t_i\right)$ always ranges within $[0,1]$ for sufficiently large $n$. Under condition (A2) and (A5), by Lemma 6(ii) of \cite{zhou2013heteroscedasticity_sup}, each part of \eqref{eq: indicator} can be uniformly controlled by $O_{\p}(\sqrt{n b_n})$.\\
Under the second case where $ j > \lceil nb_n \rceil+2\lfloor nb_n\rfloor$,
the second term of (\ref{eq: indicator}) can be decomposed and controlled, where
\begin{eqnarray}\label{eq: riemann}
	&&\left|\sum_{k=\lceil nb_n \rceil}^{j}\Sigma^{-1}(t_k) \mathbf{x}_{k} \psi(e_{k})\left[\left(\sum_{i = \lceil nb_n \rceil}^j \dfrac{1}{nb_n}K_{b_{n}}\left(t_{k}-t_i\right)\right)-1\right]\right| \cr &\le & \left|\sum_{k=\lceil nb_n \rceil}^{\lceil nb_n \rceil+\lfloor nb_n \rfloor}\Sigma^{-1}(t_k) \mathbf{x}_{k} \psi(e_{k})\left[\left(\sum_{i = \lceil nb_n \rceil+\lfloor nb_n \rfloor}^j \dfrac{1}{nb_n}K_{b_{n}}\left(t_{k}-t_i\right)\right)-1\right]\right|\crcr
	&+& \left|\sum_{k=\lceil nb_n \rceil+\lfloor nb_n \rfloor+1}^{j-\lfloor nb_n \rfloor}\Sigma^{-1}(t_k) \mathbf{x}_{k} \psi(e_{k})\left[\left(\sum_{i = \lceil nb_n \rceil}^j \dfrac{1}{nb_n}K_{b_{n}}\left(t_{k}-t_i\right)\right)-1\right]\right|\crcr
	&+& \left|\sum_{k=j-\lfloor nb_n \rfloor+1}^{j}\Sigma^{-1}(t_k) \mathbf{x}_{k} \psi(e_{k})\left[\left(\sum_{i = \lceil nb_n \rceil}^j \dfrac{1}{nb_n}K_{b_{n}}\left(t_{k}-t_i\right)\right)-1\right]\right|.
\end{eqnarray}
Since $\dfrac{1}{nb_n}\sum_{i = k-\lfloor nb_n \rfloor}^{k+\lfloor nb_n \rfloor}K_{b_{n}}\left(t_{k}-t_i\right)-1=O(\frac{1}{n})$, then by Lemma 6(ii) of \cite{zhou2013heteroscedasticity_sup}, we have 
\begin{eqnarray*}
	 &&\max_{ \lceil nb_n \rceil+2\lfloor nb_n\rfloor <j \le n} \left|\sum_{k=\lceil nb_n \rceil+\lfloor nb_n \rfloor+1}^{j-\lfloor nb_n \rfloor}\Sigma^{-1}(t_k) \mathbf{x}_{k} \psi(e_{k})\left[\left(\sum_{i = k-\lfloor nb_n \rfloor}^{k+\lfloor nb_n \rfloor} \dfrac{1}{nb_n}K_{b_{n}}\left(t_{k}-t_i\right)\right)-1\right]\right|\cr
	 &=& O\left(\frac{1}{n}\right)\cdot  O_{\mathbb{P}}(\sqrt{n}) = O_{\p}\left(\dfrac{1}{\sqrt{n}}\right),
\end{eqnarray*}
while for the rest parts, according to the similar arguments of the first case, they are also uniformly controlled by $O_{\p}(\sqrt{n b_n})$.\\
Therefore, by combining the 2 cases we get 
\begin{eqnarray}
	\max_{\lceil nb_n \rceil \le j \le n -\lceil nb_n \rceil}\left|I(t_j) - \sum_{k = \lceil nb_n \rceil}^j \Sigma^{-1}(t_k)\psi(e_k)\bx_k\right| = O_{\p}(\sqrt{n b_n}),
\end{eqnarray}
which, together with \eqref{eq: third} and \eqref{eq: fourth}, shows that
\begin{eqnarray*}
	\max_{\lceil nb_n\rceil \le j \le n-\lceil nb_n \rceil}\left|\sum_{i =\lceil nb_n\rceil }^j \Sigma^{-1}(t_i)\frac{\sum_{k = 1}^n \psi(e_k)\bx_k K_{b_n}(t_i - t_k)}{{nb_n}} - \sum_{i =\lceil nb_n\rceil}^j \Sigma^{-1}(t_i)\psi(e_i)\bx_i \right| = O_{\mathbb{P}}(\sqrt{nb_n}).
\end{eqnarray*}
Note that $\left|\sum_{i=1}^{\lceil nb_n\rceil-1}\Sigma^{-1}(t_i)\psi(e_i)\bx_i\right|=O_{\mathbb{P}}(\sqrt{nb_n})$ by Lemma 6(ii) of \cite{zhou2013heteroscedasticity_sup}, the lemma follows. 
\end{proof}

\subsubsection*{Proof of Corollary \ref{col:1}} 
\begin{proof}
Denote $\mathbb{A}:=\{i: t_i \in \cup_{k=1}^r[w_k - b_n, w_k + b_n]\}$, combining Lemma \ref{lem:5} and Lemma \ref{lem:6}, we have
\begin{eqnarray}\label{eq: thm2_proof1}
	\max_{\substack{\lceil nb_n \rceil \le j \le n- \lceil nb_n \rceil, \\ j \notin \mathbb{A}}}\left|\sum_{i = \lceil nb_n \rceil}^j\mathbf{C}(\hat{\betav}_{b_n}(t_i)-\betav(t_i) -\frac{\mu_2 b_n^2}{2}\ddot{\betav}(t_i))-\sum_{i =1}^j \mathbf{C} \Sigma^{-1}(t_i)\psi(e_i)\bx_i \right| = O_{\mathbb{P}}\left(\pi_n\sqrt{\frac{n}{b_n}}\right).
\end{eqnarray}
As for bias correction, we shall use the Jackknife bias-corrected estimator $\check{\betav}_{b_n}(t)$ and omit the subscript $b_n$ when no confusion is caused. According to \cite{zhou2010simultaneous_sup}, the bias of  $\check{\betav}_{b_n}(t)$ is asymptotically negligible under the condition of Theorem 1. Therefore, we have
\begin{eqnarray}\label{eq: col1_proof1}
	\max_{\substack{\lceil nb_n \rceil \le j \le n- \lceil nb_n \rceil, \\ j \notin \mathbb{A}}}\left|\sum_{i = \lceil nb_n \rceil}^j\mathbf{C}(\check{\betav}(t_i)-\betav(t_i))-\sum_{i =1}^j \mathbf{C} \Sigma^{-1}(t_i)\psi(e_i)\bx_i\right| = O_{\mathbb{P}}\left(\pi_n\sqrt{\frac{n}{b_n}}\right).
\end{eqnarray}
According to Lemma \ref{lem:3}, we have
\begin{eqnarray}\label{eq: col1_proof2}
	\left|\sum_{i = 1}^{\lceil nb_n \rceil - 1}\mathbf{C}(\check{\betav}(t_i) - \betav(t_i)) \right| \le \sum_{i = 1}^{\lceil nb_n \rceil - 1}\left|\mathbf{C}(\check{\betav}(t_i) - \betav(t_i))\right| = O_{\mathbb{P}}\left(\sqrt{nb_n}\log
	^4 n + nb_n^3\log n\right).
\end{eqnarray}
Similarly, observe that $\mathbb{A}$ includes at most $O(nb_n)$ elements, thus 
\begin{eqnarray}\label{eq: col1_proof3}
	\max_{\substack{\lceil nb_n \rceil \le j \le n- \lceil nb_n \rceil, \\ j \in \mathbb{A}}}\left|\sum_{i = \lceil nb_n \rceil}^j\mathbf{C}(\check{\betav}(t_i)-\betav(t_i))-\sum_{i =1}^j \mathbf{C} \Sigma^{-1}(t_i)\psi(e_i)\bx_i\right| = O_{\mathbb{P}}\left(\sqrt{nb_n}\log
	^4 n + nb_n^3\log n\right).
\end{eqnarray}
Combining \eqref{eq: col1_proof1}\eqref{eq: col1_proof2}\eqref{eq: col1_proof3}, we have
\begin{eqnarray}\label{eq:col_1}
    \max_{\lceil nb_n \rceil \leq j \leq n-\lceil nb_n \rceil}\left|\sum_{i=1}^{j} \mathbf{C}\left(\check{\betav}_{b_n}\left(t_{i}\right)-\betav\left(t_{i}\right)\right)-\sum_{i=1}^{j} \mathbf{C} \Sigma^{-1}\left(t_{i}\right) \psi\left(e_{i}\right) \mathbf{x}_{i}\right|=O_{\mathbb{P}}\left(\pi_{n} \sqrt{\frac{n}{b_{n}}}\right).
\end{eqnarray}
Recall that $\{\check{\Lambda}_{\mathbf{C}}(t)\}$ is the linear interpolation of $\{\check{\Lambda}_{\mathbf{C}}(t_j)\}_{j = 0}^n$, where $\check{\Lambda}_{\mathbf{C}}(t_j) = \sum_{i=1}^j \mathbf{C}\check{\betav}(t_i)/n$. Then for $t \in [t_j, t_{j+1})$, we have $\check{\Lambda}_{\mathbf{C}}(t) = \check{\Lambda}_{\mathbf{C}}(t_j) + n(t-t_j)[\check{\Lambda}_{\mathbf{C}}(t_{j+1}) - \check{\Lambda}_{\mathbf{C}}(t_j)]$, i.e.,  $\check{\Lambda}_{\mathbf{C}}(t) = \check{\Lambda}_{\mathbf{C}}(t_j) + (t-t_j)\mathbf{C}\check{\betav}(t_{j+1})$. Observe that
\begin{eqnarray*}
	\sqrt{n}(\check{\Lambda}_{\mathbf{C}}(t) - {\Lambda}_{\mathbf{C}}(t)) &=& \sqrt{n}[\check{\Lambda}_{\mathbf{C}}(t) - \check{\Lambda}_{\mathbf{C}}(t_j) + \check{\Lambda}_{\mathbf{C}}(t_j) - {\Lambda}_{\mathbf{C}}(t_j) + {\Lambda}_{\mathbf{C}}(t_j) - {\Lambda}_{\mathbf{C}}(t)] \crcr
	& =& \sqrt{n}(t-t_j)\mathbf{C}\check{\betav}(t_{j+1}) +  \frac{1}{\sqrt{n}}\sum_{i=1}^{j}\mathbf{C}(\check{\betav}(t_i) - \betav(t_i))+ O({\frac{1}{\sqrt{n}}}), 
	\end{eqnarray*}
we have 
\begin{eqnarray}\label{eq:col_2}
	\sup_{t \in [0,1]} \left|\sqrt{n}(\check{\Lambda}_{\mathbf{C}}(t) - {\Lambda}_{\mathbf{C}}(t)) - \frac{1}{\sqrt{n}}\sum_{i=1}^{\lfloor nt \rfloor}\mathbf{C}(\check{\betav}(t_i) - \betav(t_i))\right| = O_{\mathbb{P}}\left(\frac{1}{\sqrt{n}}\right).
\end{eqnarray}
The result follows in view of (\ref{eq:col_1}) and (\ref{eq:col_2}).
\end{proof} 

The following proposition follows from (A5)(A6) and Corollary 2 of \cite{wu2011gaussian}.
\begin{proposition}\label{prop: gaussian}
Suppose the conditions of Theorem \ref{thm:2} hold. Then for any $s\times p$, $s \le p$ full rank matrix $\mathbf{C}$, on a possibly richer probability space, there exist $i.i.d.$ $N(0, Id_s)$ random vectors $V_1, \cdots, V_n$, such that 
\begin{eqnarray*}
    \max_{1 \le j \le n}\left|\sum_{i=1}^j \mathbf{C}\Sigma^{-1}(t_i)\psi(e_i)\bx_i - \sum_{i=1}^j \Psi_{\mathbf{C}}(t_i)V_i\right| = o_{\mathbb{P}}\left(n^{\frac{1}{4}}\log^2 n\right).
\end{eqnarray*}
where $\Psi_{\mathbf{C}}(t) = (\mathbf{C}\Sigma^{-1}(t)\Psi(t)\Sigma^{-1}(t)\mathbf{C}^\top)^{\frac{1}{2}}$.
\end{proposition}

\begin{lemma}\label{lem:7}
Recall the random vectors $V_1, \cdots, V_n$ defined in Theorem 2, then under conditions of Theorem 2 and (A7), we have
\begin{eqnarray*}
	\left\{\tilde{\Psi}_{\lfloor t n\rfloor}\right\}_{0<t<1} \Rightarrow\left\{\Upsilon_{t}\right\}_{0<t<1},
\end{eqnarray*}
where $\tilde{\Psi}_t = \sum_{i=1}^t \Psi_{\mathbf{C}}(t_i) V_i/\sqrt{n}$ and $\left\{\Upsilon_{t}\right\}$ is a mean $0$ Gaussian process with covariance $\cov(\Upsilon_{t_1}, \Upsilon_{t_2}) = \int_{0}^{\min(t_1,t_2)}\Psi_{\mathbf{C}}(t)\Psi_{\mathbf{C}}^{\top}(t) \; dt$.
\end{lemma}
\begin{proof}
According to Corollary \ref{col:1}, by carefully checking Proposition \ref{prop: gaussian}, we know that on a possibly richer probability space, there exist $i.i.d.$ $N(0, Id_s)$ random vectors $V_1, \cdots, V_n$, such that 
\begin{eqnarray}\label{eq: thm2_proof2}
	\max_{\lceil nb_n \rceil \le j \le n- \lceil nb_n \rceil}\left|\sum_{i = 1}^j \mathbf{C}(\check{\betav}_{b_n}(t_i) - \betav(t_i)) - \sum_{i = 1}^j\Psi_{\mathbf{C}}(t_i)V_i\right| = O_{\mathbb{P}}\left(\pi_n \sqrt{\frac{n}{b_n}}\right).
\end{eqnarray}
Thus the finite-dimensional convergence is straightforward. We next prove tightness. For $0 < s_1 < s_2 <s_3 < 1$, if $\lfloor s_3n \rfloor - \lfloor s_1n \rfloor \ge 2$, we have
	\begin{eqnarray*}
		&&\E(|\tilde{\Psi}_{\lfloor s_3 n\rfloor} - \tilde{\Psi}_{\lfloor s_2 n\rfloor}|^2|\tilde{\Psi}_{\lfloor s_2 n
		\rfloor} - \tilde{\Psi}_{\lfloor s_2 n\rfloor}|^2 ) \cr
		&=& \dfrac{|tr(\mathbf{C}\Sigma_{\lfloor s_3 n\rfloor}\mathbf{C}^\top) - tr(\mathbf{C}\Sigma_{\lfloor s_2 n\rfloor}\mathbf{C}^\top)||tr(\mathbf{C}\Sigma_{\lfloor s_2 n\rfloor}\mathbf{C}^\top) - tr(\mathbf{C}\Sigma_{\lfloor s_1 n\rfloor}\mathbf{C}^\top)|}{n^2}\crcr
		&\le & \frac{1}{n^2}{|tr(\mathbf{C}\Sigma_{\lfloor s_3 n\rfloor}\mathbf{C}^\top) - tr(\mathbf{C}\Sigma_{\lfloor s_1 n\rfloor}\mathbf{C}^\top)|^2}	\crcr
		& \le &  \frac{L^2}{n^2}[tr(\mathbf{C}^\top\mathbf{C})]^2(\lfloor s_3 n\rfloor-\lfloor s_1 n\rfloor)^2 \crcr
		& \le & 2(\lfloor s_3 n\rfloor-\lfloor s_1 n\rfloor)^2.
		\end{eqnarray*}
		Additionally, if  $\lfloor s_3n \rfloor - \lfloor s_1n \rfloor = 0$ or $1$, then $|\tilde{\Psi}_{\lfloor s_3 n\rfloor} - \tilde{\Psi}_{\lfloor s_2 n\rfloor}|^2|\tilde{\Psi}_{\lfloor s_2 n
		\rfloor} - \tilde{\Psi}_{\lfloor s_2 n\rfloor}|^2 = 0$. The lemma follows from Theorem 13.5 of \cite{billingsley2013convergence}. 
\end{proof}

\subsubsection*{Proof of Theorem \ref{thm:2}}
\begin{proof}
Combining Corollary 2 with Proposition \ref{prop: gaussian}, we have 
\begin{eqnarray*}
    \sup_{t \in [b_n, 1-b_n]}\left|\sqrt{n}(\check{\Lambda}_{\mathbf{C}}(t) - {\Lambda}_{\mathbf{C}}(t)) - \sum_{i=1}^{\lfloor nt \rfloor}\Psi_{\mathbf{C}}(t_i)V_i/\sqrt{n}\right| = O_{\mathbb{P}}\left(\frac{\pi_n}{\sqrt{b_n}}\right),
\end{eqnarray*}
and the conclusion follows in view of Lemma \ref{lem:7}.
\end{proof}

Let $\mu$ defined as in Theorem 3, $\tilde{K}_{c_n}(t-t_i) := K^*_{c_n}({t - t_i - c_n}) + K^*_{c_n}({t - t_i + c_n}) - 2K^*_{c_n}({t - t_i})$, denote
\begin{eqnarray*}
	\mathbf{S}_i &=& \Sigma^{-1}(t_i + c_n)\sum_{k=1}^n \mathbf{x_k}\psi(e_k) K^*_{c_n}({t_k - t_i - c_n}) + \Sigma^{-1}(t_i - c_n)\sum_{k=1}^n \mathbf{x_k}\psi(e_k) K^*_{c_n}({t_k - t_i + c_n})\cr
	&-& 2 \Sigma^{-1}(t_i)\sum_{k=1}^n \mathbf{x_k}\psi(e_k) K^*_{c_n}({t_k - t_i}),
\end{eqnarray*}
and 
\begin{eqnarray*}
	 \mathbf{S}^{\Diamond}_i = \Sigma^{-1}(t_i)\sum_{k = 1}^n \mathbf{W}(t_i, \FF_k, \GG_k) [K^*_{c_n}({t_k - t_i - c_n}) + K^*_{c_n}({t_k - t_i + c_n}) - 2K^*_{c_n}({t_k - t_i})].
\end{eqnarray*}
\begin{lemma}\label{lem:8}
	For $\lceil 2n c_n\rceil \leq r \leq s \leq n-\lceil 2n c_n\rceil$, define $\Lambda_{r, s} = \frac{1}{n^2 c_n}\sum_{i=r}^s \mathbf{S}_i\mathbf{S}_i^{\top}$, then under condition (A5), we have
 $$
    \left \|\max _{\lceil 2n c_n\rceil  \leq r \leq s \leq n-\lceil 2n c_n\rceil}|\Lambda_{r, s} - \mathbb{E}(\Lambda_{r, s})| \right \| = O({\sqrt{c_n}}).
    $$
\end{lemma}
\begin{proof}
	Consider $\mathcal{F}_{i}=\left(\cdots, \eta_{i-1}, \eta_{i}\right)$, define $\mathcal{F}_{i, j}=\left(\cdots, \eta_{i-j-1}, \eta_{i-j}^{\prime}, \eta_{i-j+1}, \cdots, \eta_{i}\right)$ and $\GG_{i,j}$ similarly. Note that $\mathbf{S}_i$ is $(\mathcal{F}_{i+\left\lfloor {2n c_n} \right \rfloor}, \GG_{i+\left\lfloor {2n c_n} \right \rfloor})$ measurable and can be written as $f_i(\mathcal{F}_{i+\left\lfloor {2n c_n} \right \rfloor},\GG_{i+\left\lfloor {2n c_n} \right \rfloor})$, define $\mathbf{S}_{i,l} = f_i(\mathcal{F}_{i+\left\lfloor {2n c_n} \right \rfloor,l},\GG_{i+\left\lfloor {2n c_n} \right \rfloor,l})$. Further write $i_{*} = \max\{1, i-\left\lfloor {2n c_n} \right \rfloor\}$, $i^{*} = \min\{n, i+\left\lfloor {2n c_n} \right \rfloor\}$, then by summation-by-parts formula,
\begin{eqnarray*}
	\mathbf{S}_i &=& \Sigma^{-1}(t_i + c_n)\sum_{j=i_{*}}^{i^{*}-1} \sum_{k=i_{*}}^{j} \mathbf{x_k}\psi(e_k) \left[K^*_{c_n}({t_j - t_i - c_n})-K^*_{c_n}({t_{j+1} - t_i - c_n}))\right]\\&+&\Sigma^{-1}(t_i + c_n)\sum_{k=i_{*}}^{i^{*}}\mathbf{x_k}\psi(e_k) K^*_{c_n}({t_{i_{*}} - t_i - c_n}) \\ & +& \Sigma^{-1}(t_i - c_n)\sum_{j=i_{*}}^{i^{*}-1} \sum_{k=i_{*}}^{j} \mathbf{x_k}\psi(e_k) \left[K^*_{c_n}({t_j - t_i + c_n})-K^*_{c_n}({t_{j+1} - t_i + c_n}))\right]\\&+&\Sigma^{-1}(t_i - c_n)\sum_{k=i_{*}}^{i^{*}}\mathbf{x_k}\psi(e_k) K^*_{c_n}({t_{i_{*}} - t_i + c_n})\cr
	&-& 2\Sigma^{-1}(t_i)\sum_{j=i_{*}}^{i^{*}-1} \sum_{k=i_{*}}^{j} \mathbf{x_k}\psi(e_k) \left[K^*_{c_n}({t_j - t_i})-K^*_{c_n}({t_{j+1} - t_i}))\right]\cr
	&-& 2\Sigma^{-1}(t_i)\sum_{k=i_{*}}^{i^{*}}\mathbf{x_k}\psi(e_k) K^*_{c_n}({t_{i_{*}} - t_i}).
\end{eqnarray*}	
By Lemma 6(i) of \cite{zhou2013heteroscedasticity_sup} and (A5), 
$\left\| \mathbf{S}_i \right\|_{4} \le C\sqrt{2\left\lfloor 2n c_n \right\rfloor} = O(\sqrt{n c_n})$. Similarly,  $\left\| \mathbf{S}_{i,l} \right\|_{4} = O(\sqrt{n c_n})$.
By calculation,
\begin{eqnarray*}
	\left\| \mathbf{S}_i\mathbf{S}_i^{\top} -  \mathbf{S}_{i,l}\mathbf{S}_{i,l}^{\top} \right\|& =& \left\| \mathbf{S}_i(\mathbf{S}_i^{\top}-\mathbf{S}_{i,l}^{\top}) + (\mathbf{S}_i- \mathbf{S}_{i,l})\mathbf{S}_{i,l}^{\top} \right\| \crcr
 & \le& \left\|\mathbf{S}_i \right\|_4 \left\| \mathbf{S}_i^{\top}-\mathbf{S}_{i,l}^{\top} \right\|_4 + \left\| \mathbf{S}_i-\mathbf{S}_{i,l}  \right\|_4 \left\|\mathbf{S}_{i,l}^{\top} \right\|_4 \\&=&  O(\sqrt{n c_n}) \left\| \mathbf{S}_i-\mathbf{S}_{i,l}  \right\|_4 \\ &=  &O(\sqrt{n c_n}) \left(\sum_{j=l-2\left\lfloor 2n c_n \right \rfloor+1}^{l} \Delta_{\mathbf{W}}(j,4)\right).
\end{eqnarray*}
Consider the time series $\{f_i(\mathcal{F}_{i^{*}}, \GG_{i^{*}})f_i^{\top}(\mathcal{F}_{i^{*}},\GG_{i^{*}})\}_{i=\left\lfloor 2n c_n \right \rfloor}^{n-\left\lfloor 2n c_n \right \rfloor}$, then by Lemma 6(ii), we get that
\begin{eqnarray*}
   & &\left \|\max _{\lceil 2n c_n \rceil  \leq s \leq n-\lceil 2n c_n \rceil}\left|\sum_{j=\lceil 2n c_n \rceil}^s \mathbf{S}_j\mathbf{S}_j^{\top} - \mathbb{E}\left(\sum_{j=
\lceil 2n c_n \rceil}^s \mathbf{S}_j\mathbf{S}_j^{\top}\right)\right| \right \| \\ 
& \le&  O(\sqrt{n c_n})\cdot \sqrt{n} \cdot \sum_{l = 0}^{\infty}  \left(\sum_{j=l-2\left\lfloor 2n c_n \right \rfloor+1}^{l} \Delta_{\mathbf{W}}(j,4)\right)
\\&=&  O(\sqrt{n c_n})\cdot \sqrt{n}\cdot O(n c_n) = O(n^2 c_n^{3/2}).
\end{eqnarray*}
Therefore,
\begin{eqnarray*}
   && \left \|\max _{\lceil 2n c_n \rceil  \leq r \leq s \leq n-\lceil 2n c_n \rceil}\left|\sum_{j=r}^s \mathbf{S}_j\mathbf{S}_j^{\top} - \mathbb{E}\left(\sum_{j=r}^s \mathbf{S}_j\mathbf{S}
   _j^{\top}\right)\right| \right \|\\
   & \le& 2\left \|\max _{\lceil 2n c_n \rceil   \leq s \leq n-\lceil 2n c_n \rceil}\left|\sum_{j=\lceil 2n c_n \rceil}^s \mathbf{S}_j\mathbf{S}_j^{\top} - \mathbb{E}\left(\sum_{j=
\lceil 2n c_n \rceil}^s \mathbf{S}_j\mathbf{S}_j^{\top}\right)\right| \right \| =   O(n^2 c_n^{3/2}),
\end{eqnarray*}
which further suggests that 
\begin{eqnarray*}
    && \left\|\max_{\lceil 2n c_n\rceil  \leq r \leq s \leq n-\lceil 2n c_n\rceil}\left|\Lambda_{r, s} - \mathbb{E}(\Lambda_{r, s})\right| \right\| \cr & =& \frac{1}{n^2 c_n} \left\|\max_{\lceil 2n c_n\rceil  \leq r \leq s \leq n-\lceil 2n c_n\rceil}\left|\sum_{j=r}^s \mathbf{S}_j\mathbf{S}_j^{\top} - \mathbb{E}\left(\sum_{j=r}^s \mathbf{S}_j\mathbf{S}_j^{\top}\right)\right| \right\|= O({\sqrt{c_n}}).
\end{eqnarray*}
Similarly, if we consider the time series $\{f_i(\mathcal{F}_{i^{*}}, \GG_{i^{*}})\}_{i=\left\lfloor 2n c_n \right \rfloor}^{n-\left\lfloor 2n c_n \right \rfloor}$, we also get 
\begin{eqnarray*}
	\left \|\max _{\lceil 2n c_n \rceil  \leq s \leq n-\lceil 2n c_n \rceil}\left|\sum_{j=\lceil 2n c_n \rceil}^s \mathbf{S}_j\right| \right \| \le  \sqrt{n} \cdot \sum_{l = 0}^{\infty}  \left(\sum_{j=l-2\left\lfloor 2n c_n \right \rfloor+1}^{l} \Delta_{\mathbf{W}}(j,4)\right)=  \sqrt{n}\cdot O(n c_n) = O(n^{3/2} c_n).
\end{eqnarray*}
\end{proof}
Denote $\mathbb{B} = \{i: [t_i - 2c_n, t_i + 2c_n] \text{ contains a break point}\}$, then ${\mathbb{B}}$ contains at most $O(n c_n)$ elements. Let $\bar{\mathbb{B}}$ be the complement of ${\mathbb{B}}$ in $\{1,2,\cdots,n\}$.

\begin{lemma}\label{lem:9}
	Under (A2)(A5), $\max _{i \in \bar{\mathbb{B}}}\left|\mathbb{E}\left(\mathbf{S}_i\mathbf{S}_i^{\top}\right)-\mathbb{E}\left(\mathbf{S}_i^{\Diamond}{\mathbf{S}_i^{\Diamond}}^{\top}\right)\right|=O\left(n c_n^{3/2}\right).$
\end{lemma}
\begin{proof}
	Define $\mathbf{S}_i^{\circ} = \Sigma^{-1}(t_i)\sum_{k=1}^n \mathbf{x_k}\psi(e_k) \tilde{K}_{c_n}(t_k-t_i)$. When $i \in \bar{\mathbb{B}}$, by piecewise stochastic Lipstchiz condition and Lemma 6(i), we have  
\begin{eqnarray*}
	\left\|\mathbf{S}_i - \mathbf{S}_i^{\circ}\right\|_4  & \le & \left\|\left[\Sigma^{-1}(t_i)-\Sigma^{-1}(t_i + c_n)\right]\sum_{k = i_{*}}^{i^{*}} \mathbf{x_k}\psi(e_k) K^*_{c_n}({t_k - t_i - c_n})\right\|_4 \\& +& \left\|\left[\Sigma^{-1}(t_i)-\Sigma^{-1}(t_i - c_n)\right]\sum_{k = i_{*}}^{i^{*}} \mathbf{x_k}\psi(e_k) K^*_{c_n}({t_k - t_i + c_n})\right\|_4\\ & \le & O(c_n)O(\sqrt{n c_n}) = O\left(\sqrt{n c_n^3}\right).
\end{eqnarray*}
Also note that $\left\|\mathbf{S}_i^{\circ} \right\|_4  = O(\sqrt{nc_n})$ by similar argument of Lemma \ref{lem:8}, then
\begin{eqnarray*}
	\left|\mathbb{E}\left(\mathbf{S}_i{\mathbf{S}_i}^{\top}\right)-\mathbb{E}\left(\mathbf{S}_i^{\circ}{\mathbf{S}_i^{\circ}}^{\top}\right)\right|  & \le& \left\|\mathbf{S}_i \right\|_4 \left\| \mathbf{S}_i^{\top}-{\mathbf{S}_i^{\circ}}^{\top} \right\|_4 + \left\| \mathbf{S}_i-\mathbf{S}_i^{\circ}  \right\|_4 \left\|{\mathbf{S}_i^{\circ}}^{\top} \right\|_4 \\&=& O\left(\sqrt{n c_n}\right) \left\| \mathbf{S}_i-{\mathbf{S}_i^{\circ}}^{\top} \right\|_4 = O(n c_n^2).
\end{eqnarray*}
Next Consider the time series $\{\mathbf{W}(t_k, \FF_k, \GG_k)- \mathbf{W}(t_i, \FF_k, \GG_k)\}_{k = i_{*}}^{i^{*}}$ and denote it as $\{\mathbf{Z}_{i,k}\}_{k = i_{*}}^{i^{*}}$, note that $\mathbf{S}_i^{\circ} - \mathbf{S}_i^{\Diamond} = \Sigma^{-1}(t_i)\sum_{k=1}^n \mathbf{Z}_{i,k}\tilde{K}_{c_n} (t_k - t_i)$.
By summation-by-parts Formula, we obtain
\begin{eqnarray*}
\mathbf{S}_i^{\circ} - \mathbf{S}_i^{\Diamond} &=& \Sigma^{-1}(t_i)\sum_{k = i_{*}}^{i^{*}} \mathbf{Z}_{i,k} \tilde{K}_{c_n} (t_k - t_i)\\  &=&\Sigma^{-1}(t_i)\sum_{j=i_{*}}^{i^{*}-1} \sum_{k=i_{*}}^{j} \mathbf{Z}_{i,k} \left[\tilde{K}_{c_n} (t_j - t_i)-\tilde{K}_{c_n} (t_{j+1} - t_i)\right]+\Sigma^{-1}(t_i)\sum_{k=i_{*}}^{i^{*}} \mathbf{Z}_{i,k} \tilde{K}_{c_n} (t_{i^{*}} - t_i). 
\end{eqnarray*}
According to the piecewise stochastic Lipstchiz assumption, $\left\| \mathbf{Z}_{i,k}  \right \|_4 =  O(c_n)$. Additionally, $\left \|\mathcal{P}_{k-l}\mathbf{Z}_{i,k}\right\|_4 \le 2 \Delta_{\mathbf{W}}(l,4)$. Therefore, 
\begin{eqnarray*}
    \max _{1 \leq i \leq n} \max _{i_{*} \leq k \leq i^{*}}\left\|\Sigma^{-1}(t_i)\mathcal{P}_{k-l}\mathbf{Z}_{i,k}\right\|_4=O\left(\min \left\{c_n, \Delta_{\mathbf{W}}(l,4)\right\}\right).
\end{eqnarray*}
Hence we obtain by Lemma 6(i) of \cite{zhou2013heteroscedasticity_sup} that 
\begin{eqnarray*}
    \max _{i_{*} \leq j \leq i^{*}}\left\|\Sigma^{-1}(t_i)\sum_{k=i_{*}}^{j} \mathbf{Z}_{i,k}\right\|_4& \le &\sum_{l=0}^{\infty}\left\|\Sigma^{-1}(t_i)\sum_{k=i_{*}}^{j} \mathcal{P}_{k-l}\mathbf{Z}_{i,k}\right\|_4\\
    & =& \sum_{l=0}^{\infty} O(\sqrt{n c_n})\cdot\min \left\{c_n, \Delta_{\mathbf{W}}(l,4)\right\}\\
    & = &O(\sqrt{n c_n}\cdot \sqrt{c_n})=O\left(\sqrt{n }c_n\right),
\end{eqnarray*}
which implies that $\left\|\mathbf{S}_i^{\circ} - \mathbf{S}_i^{\Diamond}\right\|_4 = O\left(\sqrt{n }c_n\right)$. 
Then we get 
\begin{align*}
    \left|\mathbb{E}\left(\mathbf{S}_i^{\circ}{\mathbf{S}_i^{\circ}}^{\top}\right)-\mathbb{E}\left(\mathbf{S}_i^{\Diamond}{\mathbf{S}_i^{\Diamond}}^{\top}\right)\right| \le  O(\sqrt{n c_n}) \left\| \mathbf{S}_i^{\Diamond}-{\mathbf{S}_i^{\circ}}^{\top} \right\|_4 = O(n (c_n)^{3/2}).
\end{align*}
Finally, 
\begin{align*} 
 \max _{i \in \bar{\mathbb{B}}}\left|\mathbb{E}\left(\mathbf{S}_i\mathbf{S}_i^{\top}\right)-\mathbb{E}\left(\mathbf{S}_i^{\Diamond}{\mathbf{S}_i^{\Diamond}}^{\top}\right)\right| &\le \max _{i \in \bar{\mathbb{B}}} 	\left|\mathbb{E}\left(\mathbf{S}_i{\mathbf{S}_i}^{\top}\right)-\mathbb{E}\left(\mathbf{S}_i^{\circ}{\mathbf{S}_i^{\circ}}^{\top}\right)\right|  + \max _{i \in \bar{\mathbb{B}}}  \left|\mathbb{E}\left(\mathbf{S}_i^{\circ}{\mathbf{S}_i^{\circ}}^{\top}\right)-\mathbb{E}\left(\mathbf{S}_i^{\Diamond}{\mathbf{S}_i^{\Diamond}}^{\top}\right)\right|  \\& = O(n (c_n)^{3/2}).
\end{align*}
\end{proof}

\begin{lemma}\label{lem:10}
	Under conditions of Lemma \ref{lem:9}, we have $$\max _{i \in {\mathbb{B}}}\left|\mathbb{E}\left(\mathbf{S}_i\mathbf{S}_i^\top\right)-\mathbb{E}\left(\mathbf{S}_i^{\Diamond}{\mathbf{S}_i^{\Diamond}}^\top\right)\right| = O(n c_n).$$
\end{lemma}
\begin{proof}
	By Lemma 6(i) of \cite{zhou2013heteroscedasticity_sup},
$$
\max _{i \in {\mathbb{B}}}\left\|\mathbf{S}_i\right\| = O(\sqrt{n c_n}), \quad \max _{i \in {\mathbb{B}}}\left\|\mathbf{S}_i^{\Diamond}\right\| = O(\sqrt{n c_n}).
$$
The lemma follows immediately. 
\end{proof}

\begin{lemma}\label{lem:11}
Under conditions of Theorem 3, we have
	$$
   \max _{\lceil 2n c_n\rceil \leq i \leq n-\lceil 2n c_n\rceil}\left|\mathbb{E}\left(\mathbf{S}_i^{\Diamond}{\mathbf{S}_i^{\Diamond}}^\top\right)-n c_n \mu \cdot \Sigma^{-1}(t_i)\Psi(t_i)\Sigma^{-1}(t_i) \right|=O(1).
    $$
\end{lemma}
\begin{proof}
	Note that $\{\mathbf{W}(t_i, \FF_j, \GG_j)\}_{j = -\infty}^{\infty}$ is a stationary time series, let $\Gamma_i(k)$ be the $k$th auto covariance, then we have 
\begin{align*}
    \Psi(t_i) = \sum_{k=-\infty}^{\infty} \Gamma_i(k)
\end{align*}
By similar argument in Lemma 5 of \cite{zhou2010simultaneous_sup}, we have
$$
|\Gamma_{i}(k)| = O(\chi^{|k|}).
$$
Note that $\mathbf{S}_i^{\Diamond} = \Sigma^{-1}(t_i)\sum_{k = 1}^n \mathbf{W}(t_i, \FF_k, \GG_k) \tilde{K}_{c_n}(t_k - t_i )$, we have
\begin{eqnarray*}
	\mathbb{E}\left(\mathbf{S}_i^{\Diamond}{\mathbf{S}_i^{\Diamond}}^{\top}\right) & =& \Sigma^{-1}(t_i) \mathbb{E}\left[\left(\sum_{k = 1}^n \mathbf{W}(t_i, \FF_k, \GG_k)\tilde{K}_{c_n}(t_k-t_i)\right)\left(\sum_{l = 1}^n \mathbf{W}(t_i, \FF_l, \GG_l) \tilde{K}_{c_n}(t_l-t_i)\right)^{\top}\right]\Sigma^{-1}(t_i)\crcr
       &=& \sum_{k = {i_*}}^{i^{*}}\sum_{l = {i_*}}^{i^{*}}\tilde{K}_{c_n}(t_l-t_i)\tilde{K}_{c_n}(t_k-t_i)\Sigma^{-1}(t_i)\Gamma_{i}(l-k)\Sigma^{-1}(t_i).
\end{eqnarray*}
Note that $ \tilde{K}_{c_n}(t_k-t_i) = \tilde{K}_{c_n}(t_l-t_i) + \dot{\tilde{K}}_{c_n}(t^*_{k,l}-t_i)\frac{k-l}{n c_n}$, where $t^*_{k,l}$ is some value between $t_k$ and $t_l$, and
\begin{eqnarray*}
	\mu &=&  \int_{-2}^{2}[K^*(t-1)+K^*(t+1)-2K^*(t)]^2 \mathrm{d}t \crcr
      & =& \frac{1}{c_n} \int_{t_i-2 c_n}^{t_i+2 c_n}\left[K^*\left(\frac{t-t_i-c_n}{c_n}\right)+K^*\left(\frac{t-t_i+c_n}{c_n}\right)-2K^*\left(\dfrac{t-t_i}{c_n}\right)\right]^2 \mathrm{d}t\crcr
      & =& \frac{1}{c_n}\sum_{k = {i_*}}^{i^{*}}\tilde{K}_{c_n}^2({t_k-t_i})\cdot \frac{1}{n} + O\left(\frac{1}{n}\right), \qquad \text{for } \lceil 2nc_n \rceil \le i  \le n - \lceil 2nc_n \rceil,
\end{eqnarray*}
we can simplify $\mathbb{E}\left(\mathbf{S}_i^{\Diamond}{\mathbf{S}_i^{\Diamond}}^{\top}\right)$ by

\resizebox{1.05\linewidth}{!}{
\begin{minipage}{\linewidth}
\begin{align*}
&  \sum_{k = {i_*}}^{i^{*}}\sum_{l = {i_*}}^{i^{*}}\tilde{K}^2_{c_n}(t_l-t_i)\Sigma^{-1}(t_i)\Gamma_{i}(l-k)\Sigma^{-1}(t_i) + \sum_{k = {i_*}}^{i^{*}}\sum_{l = {i_*}}^{i^{*}}\tilde{K}_{c_n}(t_l-t_i)\Sigma^{-1}(t_i)\Gamma_{i}(l-k)\Sigma^{-1}(t_i)\cdot O\left(\frac{k-l}{n c_n}\right) \\
	&= \Sigma^{-1}(t_i)\left[\sum_{k=-\left\lfloor 2n c_n\right\rfloor}^{\left\lfloor 2n c_n\right\rfloor}\tilde{K}^2_{c_n}(t_k)\right]\Gamma_i(0)\Sigma^{-1}(t_i) + 2\Sigma^{-1}(t_i)\sum_{j = 1}^{2\left\lfloor 2n c_n\right\rfloor} \left[\sum_{k=-\left\lfloor 2n c_n\right\rfloor}^{\left\lfloor 2n c_n\right\rfloor-j}\tilde{K}^2_{c_n}(t_k)\right]\Gamma_i(j)\Sigma^{-1}(t_i)\\
	&+2 \Sigma^{-1}(t_i)\sum_{j = 1}^{2\left\lfloor 2n c_n\right\rfloor} \left[\sum_{k=-\left\lfloor 2n c_n\right\rfloor}^{\left\lfloor 2n c_n\right\rfloor-j}\tilde{K}_{c_n}(t_k)\right]\Gamma_i(j) O\left(\frac{j}{n c_n}\right) \Sigma^{-1}(t_i).
\end{align*}
  \end{minipage}
}
Therefore,

\resizebox{1.05\linewidth}{!}{
\begin{minipage}{\linewidth}
\begin{align*}
	&\left|\mathbb{E}\left(\mathbf{S}_i^{\Diamond}{\mathbf{S}_i^{\Diamond}}^{\top}\right)-n c_n \mu \cdot \Sigma^{-1}(t_i)\Psi(t_i)\Sigma^{-1}(t_i) \right| \\
	& \le  \left|\Sigma^{-1}(t_i)\right|^2 \left[\left|\sum_{k=-\left\lfloor 2n c_n\right\rfloor}^{\left\lfloor 2n c_n\right\rfloor}\tilde{K}^2_{c_n}(t_k) - n c_n \mu  \right|\Gamma_i(0) + 2\sum_{j = 1}^{2\left\lfloor 2n c_n\right\rfloor} \left|\sum_{k=-\left\lfloor 2n c_n\right\rfloor}^{\left\lfloor 2n c_n\right\rfloor-j}\tilde{K}^2_{c_n}(t_k)-n c_n \mu\right|\Gamma_i(j) \right]\\
    & + 2 \left|\Sigma^{-1}(t_i)\right|^2 \sum_{j > 2\left\lfloor 2n c_n\right\rfloor}n c_n \mu \Gamma_i(j) + 2 \left|\Sigma^{-1}(t_i)\right|^2\sum_{j = 1}^{2\left\lfloor 2n c_n\right\rfloor} \left|\sum_{k=-\left\lfloor 2n c_n\right\rfloor}^{\left\lfloor 2n c_n\right\rfloor-j}\tilde{K}_{c_n}(t_k)\right|\Gamma_i(j) O\left(\frac{j}{n c_n}\right)\crcr
    &= O(c_n) + \sum_{j = 1}^{2\left\lfloor 2n c_n\right\rfloor} O(j \cdot \chi^j) + \sum_{j > 2\left\lfloor 2n c_n\right\rfloor} n c_n \mu \cdot O(\chi^j) + \sum_{j = 1}^{2\left\lfloor 2n c_n\right\rfloor} O(j \cdot \chi^j) =O(1).
\end{align*}
\end{minipage}}
\end{proof}

\subsubsection*{Proof of Theorem \ref{thm:3}} 
\begin{proof}
According to the uniform Bahadur representation in Lemma \ref{lem:5}, we have
\begin{eqnarray}\label{eq:boot_diff}
	&&\max_{\lceil 2nc_n \rceil \le i \le n - \lceil 2nc_n \rceil }\left|\sqrt{\frac{c_n}{\mu}}(\check{\betav}(t_i + c_n) + \check{\betav}(t_i - c_n)-2\check{\betav}(t_i))-\frac{1}{\sqrt{\mu}}\frac{1}{n\sqrt{c_n}}\mathbf{S}_i\right| \crcr
		& =& O_{\mathbb{P}}(\pi_n/\sqrt{n})+\left|\sqrt{\frac{c_n}{\mu}}\left[{\betav}(t_i + c_n) +{\betav}(t_i - c_n)-2{\betav}(t_i)\right]\right|\crcr
		&=& O_{\mathbb{P}}\left(\frac{\pi_n}{\sqrt{n}}+c_n^{\frac{5}{2}}\right),
\end{eqnarray}
where $K(\cdot)$ is substituted for $K^*(\cdot)$ because of the Jackknife bias-corrected estimator.\\
Combining Lemma \ref{lem:8} to Lemma \ref{lem:11}, we have
\begin{eqnarray*}
	\left\| \max _{\lceil 2n c_n\rceil \leq r \leq s \leq n-\lceil 2n c_n\rceil}\left|\frac{1}{\mu} \Lambda_{r, s} -\int_{r / n}^{s / n} \Sigma^{-1}(u)\Psi(u)\Sigma^{-1}(u)\right|\right\| = O(\sqrt{c_n}).
\end{eqnarray*}
Thus for $\Delta$ defined in Theorem 3, we have 
\begin{equation}\label{eq:delta}
\begin{aligned}
&\left|\E\left(\Phi_{i}^{o}{\Phi_{j}^{o}}^\top \mid \{\mathbf{x}_i\psi(e_i)\}_{i = 1}^n\right) - \mathbf{U}(\min(t_i,t_j))\mathbf{C}\left[\int_{0}^{\min({t_i,t_j})}\Sigma^{-1}(u)\Psi(u)\Sigma^{-1}(u)\; du\right]\mathbf{C}^\top \mathbf{U}(\min(t_i,t_j))\right| \\
	& \le  O_{\mathbb{P}}\left(nc_n^5+\pi_n^2\right) +  \max_{\lceil 2nc_n \rceil \le i,j \le n - \lceil 2nc_n \rceil }\left|\sum_{k = \lceil 2nc_n \rceil}^{\min(i,j)}\frac{1}{\sqrt{\mu}}\frac{1}{n\sqrt{c_n}}\mathbf{S}_k\right|\cdot O_{\mathbb{P}}(c_n^{\frac{5}{2}}+\pi_n/\sqrt{n})\\ &+ \max_{\lceil 2nc_n \rceil \le i,j \le n - \lceil 2nc_n \rceil }\left|\dfrac{1}{\mu}\Lambda_{\lceil 2nc_n \rceil ,\min(i,j)} - \int_{0}^{\min({t_i,t_j})}\Sigma^{-1}(u)\Psi(u)\Sigma^{-1}(u)\; du\right| \\
	& = O_{\mathbb{P}}(\sqrt{c_n}+nc_n^5),
\end{aligned}
\end{equation}
which implies that $\Delta = O_{\mathbb{P}}(\sqrt{c_n}+nc_n^5)$. 
According to Theorem C.1 of \cite{zhou2020frequency_sup}, the result follows. 
\end{proof}


\subsubsection*{Proof of Theorem \ref{thm:test}} 
\begin{proof}
Since $f(\cdot)$ is continuously differentiable, we have
\begin{eqnarray*}
    && f(t, \{\check{\Lambda}_{\mathbf{C}}(v_i)\}_{i =1}^k) - f(t, \{{\Lambda}_{\mathbf{C}}(v_i)\}_{i =1}^k) \crcr
    &=& \sum_{j =1}^k\partial_j f(t, \{{\Lambda}_{\mathbf{C}}(v_i)\}_{i = 1}^k)(\check{\Lambda}_{\mathbf{C}}(v_j) - {\Lambda}_{\mathbf{C}}(v_j)) + o_{\mathbb{P}}(\max_{j =1}^k|\check{\Lambda}_{\mathbf{C}}(v_j) - {\Lambda}_{\mathbf{C}}(v_j)|).
\end{eqnarray*}
This implies that 
\begin{eqnarray*}
    &&\sqrt{n}(T_n(t) - T(t)) \crcr
    &=&  \sqrt{n}[\check{\Lambda}_{\mathbf{C}}(t) - {\Lambda}_{\mathbf{C}}(t)] 
     - \sqrt{n}[f(t, \{\check{\Lambda}_{\mathbf{C}}(v_i)\}_{i =1}^k) - f(t, \{{\Lambda}_{\mathbf{C}}(v_i)\}_{i =1}^k)]\crcr
     &=& \sqrt{n}[\check{\Lambda}_{\mathbf{C}}(t) - {\Lambda}_{\mathbf{C}}(t)] 
     - \sqrt{n}\sum_{j =1}^k\partial_j f(t, \{{\Lambda}_{\mathbf{C}}(v_i)\}_{i =1}^k)(\check{\Lambda}_{\mathbf{C}}(v_j) - {\Lambda}_{\mathbf{C}}(v_j)) + o_{\mathbb{P}}(1). 
\end{eqnarray*}
Therefore, \eqref{eq:Tn} follows from Theorem \ref{thm:2} and continuous mapping theorem. \\
Without loss of generality, we assume $\{v_1, v_2, \cdots, v_k\} \subseteq \{t_{i_*}, \cdots, t_{i^*}\}$ (otherwise we just need to modify the form of \eqref{eq:trans_boot} and \eqref{eq:trans_u} by adding additional entries, which doesn't affect the subsequent inference). 
Following the notations defined in Theorem \ref{thm:test}, we have 
\begin{equation}\label{eq:trans_boot}
    \resizebox{.9\hsize}{!}{$\underbrace{\begin{pmatrix}
        \check{L}_n(t_{i_*})\\
        \check{L}_n(t_{i_* + 1}) \\
        \vdots \\
        \check{L}_n(t_{i^*})
    \end{pmatrix}}_{\textstyle \check{\mathbf{L}}^B} 
    = \underbrace{\begin{pmatrix}
        I_s & \mathbf{0}_s & \cdots & -\partial_{1} f(t_{i_*}, \{\check{\Lambda}_{\mathbf{C}}(v_i)\}_{i =1}^k)& \cdots & -\partial_{k} f(t_{i_*}, \{\check{\Lambda}_{\mathbf{C}}(v_i)\}_{i =1}^k) & \cdots \\
        \mathbf{0}_s & I_s & \cdots & -\partial_{1} f(t_{i_* + 1}, \{\check{\Lambda}_{\mathbf{C}}(v_i)\}_{i =1}^k)& \cdots & -\partial_{k} f(t_{i_* + 1}, \{\check{\Lambda}_{\mathbf{C}}(v_i)\}_{i =1}^k) & \cdots \\
        \vdots & \vdots & \vdots & \vdots & \vdots & \vdots &\vdots \\
        \mathbf{0}_s & \mathbf{0}_s & \cdots & -\partial_{1} f(t_{i^*}, \{\check{\Lambda}_{\mathbf{C}}(v_i)\}_{i =1}^k)& \cdots & -\partial_{k} f(t_{i^*}, \{\check{\Lambda}_{\mathbf{C}}(v_i)\}_{i =1}^k) & I_s
    \end{pmatrix}}_{\textstyle :=A^B}
     \underbrace{\begin{pmatrix}
        \tilde{\mathbf{\Phi}}^\circ_n(t_{i_*})\\
        \tilde{\mathbf{\Phi}}^\circ_n(t_{i_*  +1}) \\
        \vdots \\
        \tilde{\mathbf{\Phi}}^\circ_n(t_{i^*})
    \end{pmatrix}}_{\textstyle \mathbf{\Theta}^B_{\text{short}}} $} 
\end{equation}
and
\begin{equation}\label{eq:trans_u}
    \resizebox{.9\hsize}{!}{$\underbrace{\begin{pmatrix}
        \tilde{\mathbf{U}}(t_{i_*})\\
        \tilde{\mathbf{U}}(t_{i_* + 1}) \\
        \vdots \\
        \tilde{\mathbf{U}}(t_{i^*})
    \end{pmatrix}}_{\textstyle \tilde{\mathbf{U}}} 
    = \underbrace{\begin{pmatrix}
        I_s & \mathbf{0}_s & \cdots & -\partial_{1} f(t_{i_*}, \{{\Lambda}_{\mathbf{C}}(v_i)\}_{i =1}^k)& \cdots & -\partial_{k} f(t_{i_*}, \{{\Lambda}_{\mathbf{C}}(v_i)\}_{i =1}^k) & \cdots \\
        \mathbf{0}_s & I_s & \cdots & -\partial_{1} f(t_{i_* + 1}, \{{\Lambda}_{\mathbf{C}}(v_i)\}_{i =1}^k)& \cdots & -\partial_{k} f(t_{i_* + 1}, \{{\Lambda}_{\mathbf{C}}(v_i)\}_{i =1}^k) & \cdots \\
        \vdots & \vdots & \vdots & \vdots & \vdots & \vdots &\vdots \\
        \mathbf{0}_s & \mathbf{0}_s & \cdots & -\partial_{1} f(t_{i^*}, \{{\Lambda}_{\mathbf{C}}(v_i)\}_{i =1}^k)& \cdots & -\partial_{k} f(t_{i^*}, \{{\Lambda}_{\mathbf{C}}(v_i)\}_{i =1}^k) & I_s
    \end{pmatrix}}_{\textstyle :=A}
     \underbrace{\begin{pmatrix}
        {\mathbf{U}}(t_{i_*})\\
        {\mathbf{U}}(t_{i_*  +1}) \\
        \vdots \\
        {\mathbf{U}}(t_{i^*})
    \end{pmatrix}}_{\textstyle \mathbf{U}_{\text{short}}}$}. 
\end{equation}
Recall the $ns$-dim vectors $\mathbf{\Theta}^B$ and $\mathbf{U}$ defined in Theorem \ref{thm:3}, we observe in \eqref{eq:trans_boot} and \eqref{eq:trans_u} that $[(\tilde{\mathbf{\Phi}}^\circ_n(t_{i_*}))^\top, (\tilde{\mathbf{\Phi}}^\circ_n(t_{i_*  +1}))^\top,\cdots, (\tilde{\mathbf{\Phi}}^\circ_n(t_{i^*}))^\top]^\top$ and $[({\mathbf{U}}(t_{i_*}))^\top, ({\mathbf{U}}(t_{i_*  +1}))^\top,\cdots, ({\mathbf{U}}(t_{i^*}))^\top]^\top$ are ``shortened" version of them, therefore we denote the terms as $\mathbf{\Theta}^B_{\text{short}}$ and $\mathbf{U}_{\text{short}}$, respectively.
Define the difference between the covariance structure of $\check{\mathbf{L}}^B$ and $\tilde{\mathbf{U}}$ as
\begin{equation*}
    \tilde{\Delta} := \max_{1 \le i,j \le (n - 2i_*)s}\left|[\cov(\tilde{\mathbf{U}}) - \cov(\check{\mathbf{L}}^B \mid \{(\bx_k, y_k)\}_{k=1}^n)]_{ij}\right|, 
\end{equation*}
then \eqref{eq:trans_boot} and \eqref{eq:trans_u} implies that
\begin{equation}
\begin{aligned}
    \tilde{\Delta} &= \max_{1 \le i,j \le (n - 2i_*)s}\left|[\cov(A{\mathbf{U}_{\text{short}}}) - \cov(A^B \mathbf{\Theta}^B_{\text{short}} \mid \{(\bx_k, y_k)\}_{k=1}^n)]_{ij}\right| \\
     & \le \max_{1 \le i,j \le (n - 2i_*)s}\left|[\cov(A{\mathbf{U}_{\text{short}}}) - \cov(A \mathbf{\Theta}^B_{\text{short}} \mid \{(\bx_k, y_k)\}_{k=1}^n)]_{ij}\right| \\
     & + \max_{1 \le i,j \le (n - 2i_*)s}\left|[\cov((A-A^B) \mathbf{\Theta}^B_{\text{short}} \mid \{(\bx_k, y_k)\}_{k=1}^n)]_{ij}\right| \\
     & \le \max_{1 \le i,j \le (n - 2i_*)s}\left|[A(\cov({\mathbf{U}_{\text{short}}}) - \cov( \mathbf{\Theta}^B_{\text{short}} \mid \{(\bx_k, y_k)\}_{k=1}^n))A^\top]_{ij}\right| \\
     & + \max_{1 \le i,j \le (n - 2i_*)s}\left|[\cov((A-A^B) \mathbf{\Theta}^B_{\text{short}} \mid \{(\bx_k, y_k)\}_{k=1}^n)]_{ij}\right| \\
     & := I_1 + I_2. 
\end{aligned}
\end{equation}
Note that $A$ is sparse due to the fact that $k$ is fixed and does not change with $n$, also the nonzero entries are bounded according to the conditions of Theorem \ref{thm:test}, it's straightforward to conclude that $I_1 = O_{\mathbb{P}}(\Delta)$. 
By Lipschitz continuity of $\partial_j f(t, \{\mathbf{s}_i\}_{i =1}^k)$, there exists a constant $C_k$, such that
\begin{eqnarray*}
    |\partial_j f(t, \{\check{\Lambda}_{\mathbf{C}}(v_i)\}_{i =1}^k) - \partial_j f(t, \{{\Lambda}_{\mathbf{C}}(v_i)\}_{i =1}^k)| \le C_k\max_{i =1}^k|\check{\Lambda}_{\mathbf{C}}(v_i) - {\Lambda}_{\mathbf{C}}(v_i)|, \forall j = 1, \cdots,k,
\end{eqnarray*}
which gives
\[\left|A- A^B\right|= O_{\mathbb{P}}(\max_{i =1}^k|\check{\Lambda}_{\mathbf{C}}(v_i) - {\Lambda}_{\mathbf{C}}(v_i)|) = O_{\mathbb{P}}(\frac{1}{\sqrt{n}}).\] 
Therefore we get $I_2 = O_{\mathbb{P}}(\frac{1}{{n}})$ in view of $\max_{1 \le i,j \le n - 2i_*}\left|[\cov(\mathbf{\Theta}^B_{\text{short}} \mid \{(\bx_k, y_k)\}_{k=1}^n)]_{ij}\right| = O_{\mathbb{P}}(1). $ Combining the results above, we get $\tilde{\Delta} = O_{\mathbb{P}}(\Delta) = O_{\mathbb{P}}(\sqrt{c_n} + nc_n^5)$. \\
Observe that $\check{\mathbf{L}}^B$ (conditional on $\{(\bx_k, y_k)\}_{k=1}^n$) and $\tilde{\mathbf{U}}$ are both Gaussian vectors, Theorem 4 follows from similar arguments as in Theorem \ref{thm:3} and Theorem C.1 of \cite{zhou2020frequency_sup}. 
\end{proof}

\subsubsection*{Proof of Proposition \ref{prop: typeIerror_LOFT}}
\begin{proof}
	Under the null hypothesis of the Lack-of-fit Test, $T(t) = 0$, then the result follows directly from Theorem \ref{thm:test}.
\end{proof}

\subsubsection*{Proof of Proposition \ref{prop: power_LOFT}}
\begin{proof}
According to Theorem \ref{thm:test},
\begin{eqnarray*}
    \lim_{n \to \infty}\lim_{B \to \infty}\mathbb{P}\left(\max_{i_* \le j \le i^*}|T_n(t_j) - T(t_j)|_\infty >  \hat{q}_{n,1-\alpha}/\sqrt{n}\right) \to \alpha.
\end{eqnarray*}
Since $\sup_{t \in (0,1)}|\Lambda_{\mathbf{C}}(t) - f(t, \{\Lambda_{\mathbf{C}}(v_i)\}_{i =1}^k)|_\infty \gg n^{-\frac{1}{2}}$, we have
\begin{eqnarray*}
    \max_{ i_* \le j \le i^*}\sqrt{n}|T(t_j)|_\infty \to \infty.
\end{eqnarray*}
Therefore,
\begin{eqnarray}\label{eq:divergent}
	 \lim_{n \to \infty}\lim_{B \to \infty}\mathbb{P}\left(\max_{i_* \le j \le i^*}\sqrt{n}|T_n(t_j)|_\infty > d_n\right) = 1
\end{eqnarray}
for any $d_n$ divergent arbitrarily slow. Meanwhile, by similar argument of Proposition 4.5 in \cite{zhou2020statistical}, since $\sqrt{n}\max_{i_* \le j \le i^*}|T_n(t_j) - T(t_j)|_\infty$ is asymptotically equivalent to the infinity norm of a $\frac{s(s-1)}{2}$ dimensional Gaussian random function and $s \le p$ is finite, we have $\hat{q}_{n, 1-\alpha} = O_{\mathbb{P}}(1)$. Therefore, the result follows from \eqref{eq:divergent} and $\hat{q}_{n, 1-\alpha} = O_{\mathbb{P}}(1)$.
\end{proof}

\subsubsection*{Proof of Proposition \ref{prop: qualitative_typeI}}
\begin{proof}
Under the null hypothesis, the true function $\Lambda_{\mathbf{C}}(t)$ must lie within the feasible region $\mathcal{N}_0$, which implies that 
\begin{eqnarray*}
		\sup_{t \in (0,1)}|{\tilde{\Lambda}_\mathbf{C}}(t) - {\check{\Lambda}_\mathbf{C}}(t)|_\infty \le \sup_{t \in (0,1)}|{\Lambda}_\mathbf{C}(t) - {\check{\Lambda}_\mathbf{C}}(t)|_\infty,
	\end{eqnarray*}
or equivalently in the finite sample, 
\begin{eqnarray*}
    \max_{i_* \le j \le i^*}|{\tilde{\Lambda}_\mathbf{C}}(t_j) - {\check{\Lambda}_\mathbf{C}}(t_j)|_\infty \le \max_{i_* \le j \le i^*}|{\Lambda}_\mathbf{C}(t_j) - {\check{\Lambda}_\mathbf{C}}(t_j)|_\infty.
\end{eqnarray*}
Therefore, 
\begin{equation*}
\begin{aligned}
     & \quad \lim_{n \to \infty} \lim_{B \to \infty} \mathbb{P}\left(\max_{i_* \le j \le i^*}|{\tilde{\Lambda}_\mathbf{C}}(t_j) - {\check{\Lambda}_\mathbf{C}}(t_j)|_\infty > \hat{q}_{n, 1-\alpha} /\sqrt{n}\right)\\  & \le \lim_{n \to \infty} \lim_{B \to \infty} \mathbb{P}\left(\max_{i_* \le j \le i^*}|{\Lambda}_\mathbf{C}(t_j) - {\check{\Lambda}_\mathbf{C}}(t_j)|_\infty> \hat{q}_{n, 1-\alpha} /\sqrt{n} \right) \\&= \alpha
\end{aligned}
\end{equation*}
in view of Proposition \ref{prop: typeIerror_LOFT}.
\end{proof}

\subsubsection*{Proof of Proposition \ref{prop:qualitative_power}}
\begin{proof}
Following the proof of Proposition \ref{prop: typeIerror_LOFT}, we have 
\begin{eqnarray*}
	\lim_{n \to \infty} \lim_{B \to \infty}\mathbb{P}\left(\max_{i_* \le j \le i^*}|{\Lambda}_\mathbf{C}(t_j) - {\check{\Lambda}_\mathbf{C}}(t_j)|_\infty> \hat{q}_{n, 1-\alpha} /\sqrt{n} \right) = \alpha.
\end{eqnarray*}
Under the alternative hypothesis that $\inf_{g \in \mathcal{N}_0}\sup_{t \in (0,1)}|{\Lambda}_{\mathbf{C}}(t)-g(t)|_\infty \gg n^{-\frac{1}{2}}$, it is straightforward to get
\begin{eqnarray*}
 \max_{ i_* \le j \le i^*}\sqrt{n}|{\Lambda}_{\mathbf{C}}(t_j)-\tilde{\Lambda}_{\mathbf{C}}(t_j)|_\infty \to \infty.
\end{eqnarray*}
Then we have
\begin{eqnarray*}
	\lim_{n \to \infty} \lim_{B \to \infty}\mathbb{P}\left(\max_{ i_* \le i \le i^*}\sqrt{n}|\check{\Lambda}_{\mathbf{C}}(t_i) - \tilde{\Lambda}_{\mathbf{C}}(t_i)|_\infty > d_n\right) = 1
\end{eqnarray*}
for any $d_n$ divergent arbitrarily slow. 

By similar argument in Proposition \ref{prop: power_LOFT}, the proposition follows in view of $\hat{q}_{n, 1-\alpha} = O_{\mathbb{P}}(1)$. 
\end{proof}

\setlength{\bibsep}{2pt}
\bibliographystyle{chicago}
{\refersize \bibliography{SuppPub_abbre}}

\makeatletter\@input{tmp.tex}\makeatother